\documentclass[%
 aip,
cp,  
 amsmath,amssymb,
 reprint,%
]{revtex4-2}

\usepackage[utf8]{inputenc}
\usepackage[T1]{fontenc}
\usepackage{bm}
\usepackage{dcolumn}

\usepackage{color}
\usepackage{graphicx}
\usepackage[caption=false]{subfig} 
\usepackage{overpic}
\usepackage[hidelinks]{hyperref}
\usepackage{xurl}
\usepackage{nameref}

\newsavebox{\PhaseBHPLeftBox}

\begin{document}

\title{Frustrated supermolecules: the high-pressure phases of crystalline methane}

\author{Marcin Kirsz}
\email[Corresponding author: ]{marcin.kirsz@ed.ac.uk}

\author{Miguel Martinez-Canales}
\email{miguel.martinez@ed.ac.uk}

\author{Ayobami Daramola}
\email{adaramo2@ed.ac.uk}

\author{John S. Loveday}
\email{J.Loveday@ed.ac.uk}

\author{Ciprian G. Pruteanu}
\email{cip.pruteanu@ed.ac.uk}

\author{Graeme J. Ackland}
\email{gjackland@ed.ac.uk}

\affiliation{Centre for Science at Extreme Conditions, School of Physics and Astronomy, The University of Edinburgh, Edinburgh EH9 3FD, United Kingdom}

\date{\today}

\begin{abstract}
Methane is the simplest hydrocarbon, yet it exhibits an extraordinarily complicated series of crystal phases.  Notably, the non-plastic phases have large unit cells with nearly, but not quite cubic symmetry.  Furthermore, although non-polar molecules interact very weakly, their reorganisation across phase transitions is very sluggish.
Here, we demonstrate that these complex structures can be understood as simple packing of near-spherical supermolecular clusters of methane molecules: the departure from cubic symmetry arising from the non-spherical nature of the molecules.
We use molecular dynamics based on density functional theory calculations to simulate the finite-temperature crystal structures of methane, finding that the complex Phase~A is based around a 13-molecule regular icosahedron, with 8 additional molecules forming the 21-molecule unit cell.  Similarly, Phase~B is based on a body-centred cubic $bcc$ packing of 17-molecule Z16 polyhedra, with the remaining 12 molecules per cell in tetrahedral interstices.  We demonstrate that the favored intermolecular separation depends sensitively on molecular orientation, leading to hindered rotation and suppressed entropy. The structures are determined by a trade-off between efficient packing and entropy.
\end{abstract}


\maketitle
\section{Introduction}
Methane is the smallest hydrocarbon and as such is widespread both on Earth and throughout extraterrestrial environments~\cite{tielens2013molecular}.
Despite being a simple and fundamental system, there is still a considerable gap in our understanding of methane's solid phases, which exist at low temperatures and high pressures.  

In the low pressure limit, the enthalpy ($H=U+PV$) is effectively synonymous with internal energy, and cohesion in methane arises from dispersion forces.  With an enthalpy of vaporisation of ~10~kJ/mol (about 100~meV per molecule), these interactions are relatively weak.  By contrast, at 5~GPa each molecule occupies about 30~\AA$^3$, so the PV term is close to 1~eV.  Thus, the density contribution is an order of magnitude greater than the dispersion energy.  At these pressures, the energy contribution is overwhelmingly dominated by short-range steric repulsion, and minimising the enthalpy can be thought of as finding efficient packing of tetrahedral molecules.  At room temperature, entropic effects are also significant in determining phase stability, since the six molecular degrees of freedom yield up to  $3k_{\mathrm{B}}T\approx75$~meV.

The current phase diagram has an ``onion-ring'' form~\cite{bini1997,MPA_CH4_onion_PD}. Upon cooling from the high-pressure melt, or compression at room temperature, four solid phases have been identified. Their phase boundaries all have a positive slope ($dP/dT>0$), indicating that the high-temperature phases are also favoured at lower pressures. 

The first phase to which methane crystallizes on cooling or pressurization is called Phase~I and has been determined~\cite{press1972structure} to be cubic, closed-packed space group $Fm\overline{3}m$, Strukturbericht A1.  The tetrahedral $T_d$ point group symmetry of the molecule is inconsistent with the cubic close-packed structure.  The symmetry is restored by rotational motion which results in a plastic crystal~\cite{PRESS1970253, press1972structure}.

On increasing pressure, methane transitions from Phase~I to Phase~A~\cite{bini1997}, which has a 21 molecule rhombohedral unit cell~\cite{nakahata1999structural}. The structure of this phase has been extensively studied using both X-ray single-crystal and neutron powder diffraction, which gave a full structural solution for Phase~A including hydrogen positions in the $R3$ space group~\cite{maynard2010}. The rhombohedral cell is very close to cubic ($\alpha=89.4^\circ$) and was reported as a distorted close packed structure.
A distinguishing feature of Phase~A is that the C--H bonds do not point toward one another and all hydrogen positions are well ordered. Despite the description as close packing, Phase A cannot be related to the cubic cell of Phase~I.

With further compression, Phase~A undergoes a slow transition to a cubic Phase~B~\cite{bini1995,umemoto2002x}.
To date, structural studies on Phase~B have identified the $I\overline{4}3m$ space group and determined the arrangement of the carbon sublattice~\cite{maynard2014}. However, a complete structural solution, including hydrogen positions, remains unresolved. The conventional unit cell of Phase~B contains 58 molecules, with carbon atoms arranged similarly to $\alpha$-Mn, across four distinct crystallographic sites consistent with the A12 Strukturbericht designation.

The transition between methane A and B has been extensively studied using Raman spectroscopy, both in hydrogenous~\cite{wu1995high, bini1997} and in deuterated methane~\cite{pruteanu2019effect}. The two phases are readily distinguished by the splitting of vibrational modes and by their distinct pressure-dependent evolution of the Raman spectra. The sluggish nature of the transformation implies significant hysteresis and uncertainty about the exact transformation pressure.

At even higher pressures, methane transforms to yet another phase, called HP~\cite{bini1997, hirai2008phase}. A recent study combining X-ray measurements and \textit{ab initio} calculations found Phase~HP to be stable between 20 and 73~GPa, and determined it to be a rhombohedral distortion of Phase~B (space group $R3$), with the unit cell containing 87 molecules (29 molecules in the primitive cell)~\cite{bykov2021structural}.

Ab initio structure search methods can determine possible minimum enthalpy ground state structures.  However, such methods cannot detect structures with rotating molecules which have entropic stabilisation, such as the experimentally reported $Fm\overline{3}m$ Phase~I. 

Ab initio, Born-Oppenheimer molecular dynamics (BOMD) offers a powerful route to identify symmetry-breaking and capture finite-temperature behaviour that static DFT inherently omits: molecular rotation, anharmonic vibrations, and entropy-driven stabilisation. BOMD has already provided key insights into the role of molecular rotation and reorientation in solid hydrogen's phase diagram, including calculation of non-harmonic Raman~\cite{kohanoff1997solid,magduau2013identification,magduau2017infrared,cooke2020raman,cooke2022calculating,pena2020quantitative,ackland2020structures}, as well as rotational melting in solid nitrogen~\cite{kirsz2024understanding}. Motivated by these successes, we employ extensive BOMD simulations of crystalline methane across relevant pressures and temperatures.

Molecular dynamics is also crucial in identifying the effective shape of a molecule: for example, rotations can cause a linear molecule such as N$_2$ to behave and pack as a disk or a sphere~\cite{kirsz2024understanding}.  Furthermore, rotating molecules can pack into ``supermolecules'', most notably the icosahedral (H$_2)_{13}$~\cite{ackland2018icosahedral,binns2018formation,carneiro2025emergence}.  This, in turn, is a component of density-driven packing in weakly bound or hard-sphere systems.~\cite{eldridge1993stability,schofield2001binary,schofield2005stability,bommineni2020spontaneous}

The philosophical approach to interpreting experiments is to assign the structure as the highest symmetry not ruled out by the data.   
By contrast, BOMD samples structures with no symmetry: symmetry emerges only as a time- and space-averaged property.  More recently, we have advocated that the candidates for the crystal structure should be the low enthalpy and stable structures found in calculations.  Such a structure should be ruled out if it is inconsistent with the data, but not on the grounds that a ``simpler'' structure fits the data.  Here, we pay attention to the relationship between calculation and data, leaving aside the experimentalist's interpretation.

In this paper, we quantify the role of rotational entropy in stabilising Phases I, A, and B; elucidate the origin of kinetic hysteresis in the A$\rightarrow$B transition; and reconcile static DFT predictions with experimental diffraction and spectroscopic data.

\section{Methods}

\subsection{Molecular Dynamics}

We performed several BOMD simulations for methane phases I, A, B, and HP at 300~K. The many-body electronic problem is solved using density functional theory (DFT) within the plane-wave pseudopotential approach, as implemented in CASTEP~\cite{CASTEP}.  The PBE~\cite{perdew1996} is used as the approximate exchange-correlation potential. This is consistent with other DFT studies of methane and hydrocarbons under pressure~\cite{bykov2021structural, gao2010, conway2019}. Selected configurations were recalculated with rSCAN~\cite{BartokRSCAN} and BLYP~\cite{BeckeXC,LYPXC}, and the key results are independent of the functional. For the BOMD simulations, we used CASTEP's ``MS\_PBE'' ultrasoft pseudopotentials with a 390~eV energy cutoff. In all cases, we chose a timestep of 0.5~fs, some 20 times smaller than the shortest vibrational periods (C--H stretches).

For each phase, the specific supercell sizes, pressure conditions, and ensemble choices are detailed in the corresponding Results subsections, where we tailor the setup to the experimentally relevant stability fields.
In all simulations, we monitor radial distribution functions (RDFs) and mean squared displacements (MSDs) to characterize structural order and atomic mobility.
We also carried out static calculations for methane, which are consistent with previous work~\cite{bykov2021structural}.

\subsection{X-ray and Neutron Diffraction Patterns}

Molecular dynamics simulations do not enforce symmetry, so rather than go through some intermediate model structure, we determine the diffraction patterns directly from the simulation in the $P1$ space group\cite{ackland2020structures}.
We used the BOMD trajectory to create an effective ``unit cell'' with atoms superposed from 50 independent snapshots. This gives us a direct sampling of the thermal broadening and the fractional occupation of particular C--H directions.  This 42000-atom cell was then run through GSAS~\cite{toby2001expgui} to generate the corresponding X-ray and neutron diffraction patterns. Provided the system is ergodic (i.e. time and space averages are equivalent), such patterns are suitable for direct comparison with experiment and, importantly, have no assumed symmetry.

\subsection{Rotational Free Energy}

The contribution to the free energy from rotational degrees of freedom is not conceptually straightforward. In classical physics, the freely rotating molecule has lowered free energy due to high entropy from equally sampling the solid angle $(\theta,\phi)$.  In quantum physics, it is the free rotor ground state $Y_{00}$ which samples all solid angles.  As a single quantum rotor state it has no entropy, rather obtaining its low free energy from its low kinetic energy.
The equivalence of these pictures is demonstrated in Hindered Rotor Theory.~\cite{cooke2020raman,cooke2022calculating}
Neither of these rotor free energies has a direct relationship to the phonon free energy routinely calculated from perturbation theory, assuming the motion to be harmonic.  However, rotating molecules usually have a shallow minimum in their potential energy surface, so lattice dynamics calculations are often fortuitously accurate.

To quantify the degree of orientational disorder for each molecule $i$ in our BOMD simulations, we compute unit vectors $\mathbf{u}_{i,j}(t)$ along the four C--H bonds ($j$). From their time series, we build the joint orientational probability density function (PDF) $P_i(\theta,\phi)$ on the unit sphere and compute the associated Shannon entropy,
\begin{equation}
  S_i^{\rm rot} \;=\; -\sum_{\theta,\phi} P_i(\theta,\phi)\,\ln P_i(\theta,\phi)\,\Delta \Omega,
\end{equation}
where $\Delta \Omega$ is the solid angle per bin. $S_i^{\rm rot}$ is then normalised by $\ln(4\pi)$ so that $0 \le S_i^{\rm rot} \le 1$. The sum runs over equal-area binned data. $S_i^{\rm rot}$ can be considered as a fraction of the maximum possible entropy $k_\mathrm{B}\ln(4\pi)$ for two angular degrees of freedom.

We use the per-molecule entropy to determine groups of atoms which are symmetry-related. It also allows us to distinguish between ordered and (partially) disordered sites.

To further aid identification of rotating molecules in the BOMD trajectories, we created a spherical heatmap of the $(\theta,\phi)$ directions from $P_i(\theta,\phi)$, shown as a Mollweide projection, oriented such that one bond lies at the centre of the projection.

\subsection{Crystal Structure and Symmetry Analysis}
Any snapshot taken from BOMD calculation will have no symmetry.  Over a long simulation, symmetry may emerge from the average positions of the atoms.  However, the atoms are identical, and with methane, this can be problematic if the molecules reorient or rotate. Since the molecules do not diffuse, taking the average position of the  carbon atoms is fine, but the mean position of a hydrogen in a rotating methane is in the centre of the carbon, which is not helpful.

We begin construction of the crystal symmetry by time averaging C-atom positions. 
For each molecule $i$, the H-atoms are then assigned based on the corresponding $P_i(\theta,\phi)$.
The symmetry of the structure is obtained using Spglib~\cite{spglib}. Together with previously-defined $S_i^{\rm rot}$ this allows us to identify symmetry-equivalent molecules and to determine the highest symmetry group consistent with the diffraction pattern.

\section{Results}

\begin{figure}[htbp]
    \centering
\includegraphics[width=0.48\textwidth]{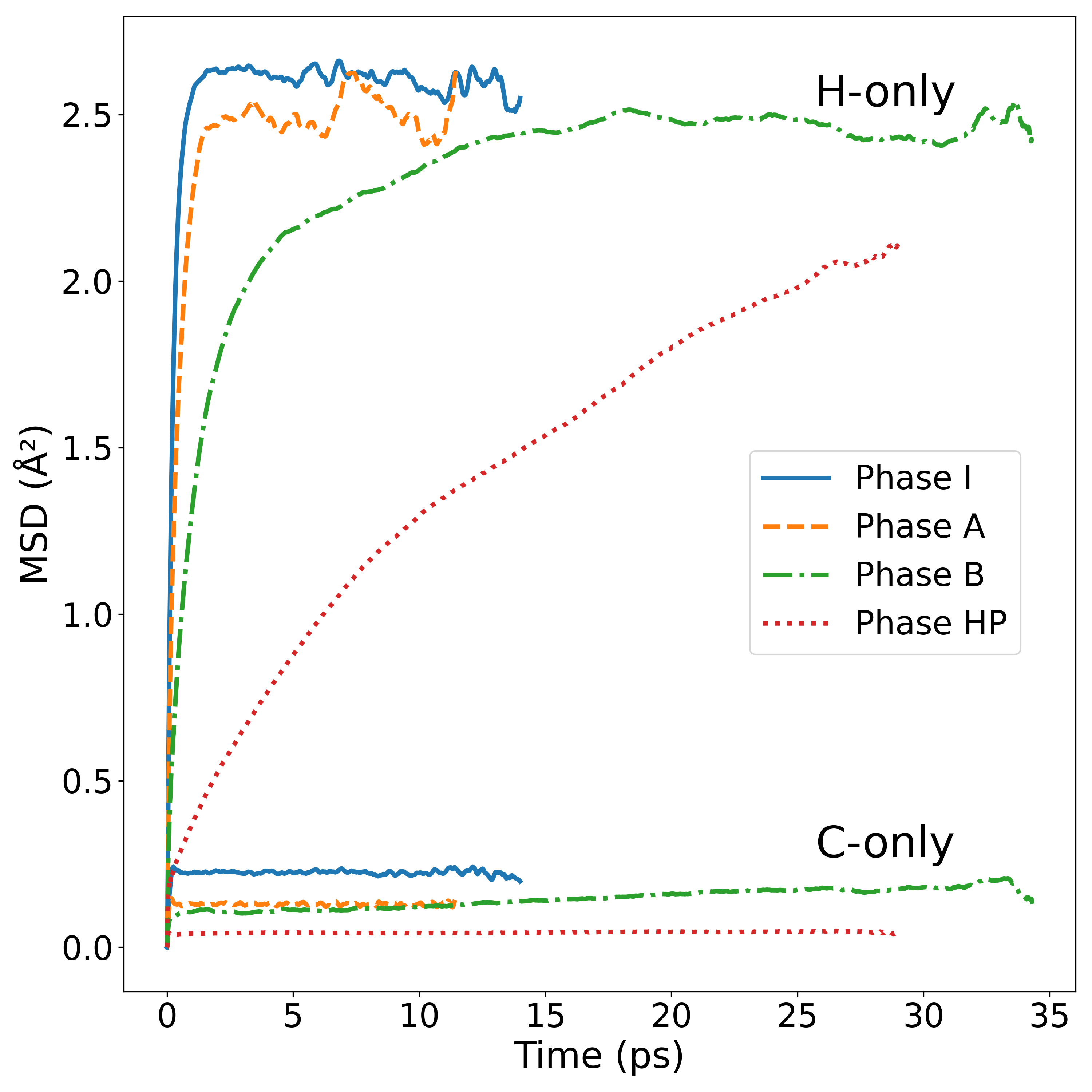} 
\includegraphics[width=0.48\textwidth]{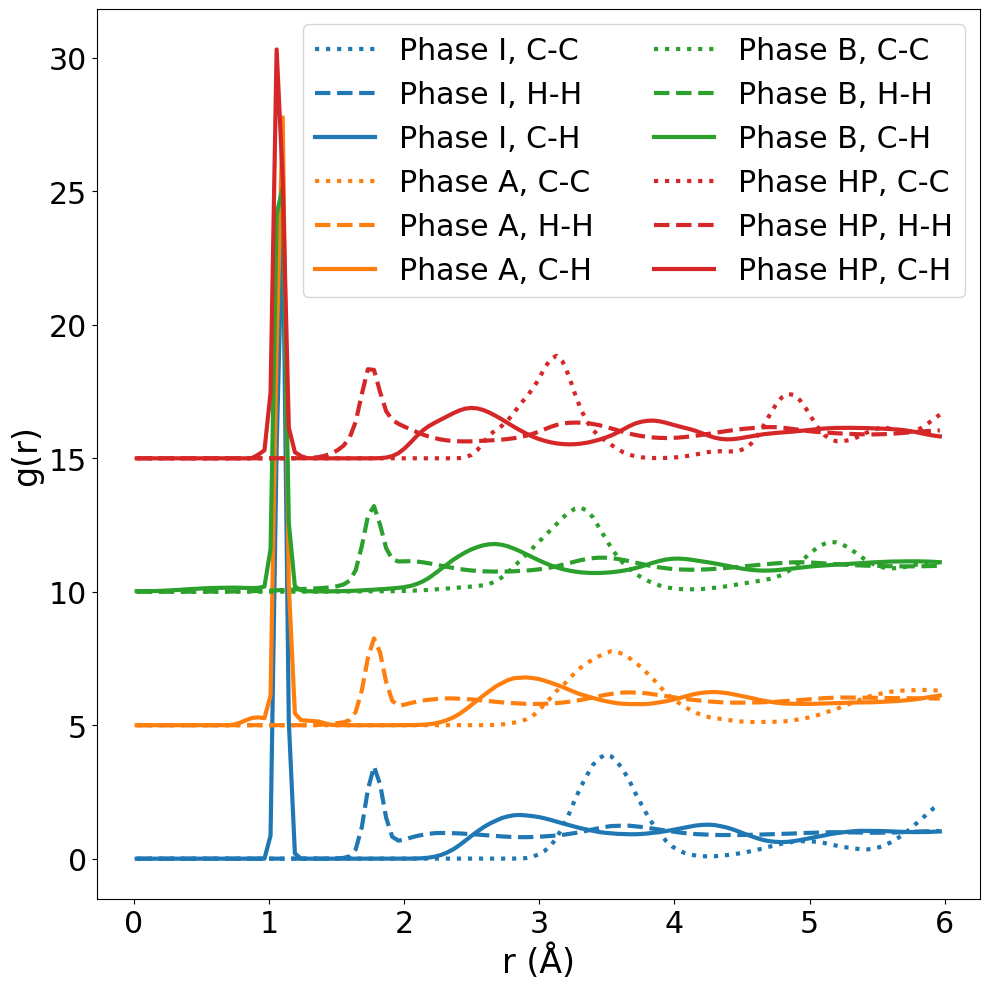} 
\caption{(Left) Mean squared displacement (MSD) of CH$_4$ in phases I, A, B and HP. For each phase, C- and H-only MSDs are shown. The high mobility of the H atoms compared to the nearly stationary C atoms is characteristic of plastic solid phases. (Right) Partial radial distribution functions for C--C, H--H, and C--H pairs.}
\label{fig:MSD+RDF}
\end{figure}

The MSDs become flat on a picosecond timescale, indicating non-diffusive crystalline behaviour without any C--H bond breaking (Fig.~\ref{fig:MSD+RDF}). Carbon atoms show low mobility, while hydrogen atoms display higher MSDs, consistent with molecules that either reorient or rotate. The C-C RDFs are a proxy for intermolecular spacings and for all phases they exhibit a well-defined first peak that drops to zero, confirming solid-like order with well-defined coordination numbers (Fig.~\ref{fig:MSD+RDF}).




\subsection{Phase~I}

Phase~I is widely reported as having the cubic close-packed structure $Fm\overline{3}m$. Static structure searches cannot find this structure because cubic close packing is incompatible with the $T_d$ point-group symmetry of the methane molecule. The closest approximant is a phase of symmetry $I\overline{4}2m$ with $c/a = 1.51$, which is competitive below 5~GPa.~\cite{bykov2021structural}

Spherical molecules in this structure could adopt the face-centred cubic ($fcc$) structure if $c/a=\sqrt{2}$, and this would be consistent with experiment. Therefore, we used $I\overline{4}2m$ with $c/a=\sqrt{2}$ as our initial condition for Phase~I molecular dynamics. We performed BOMD on a $2^3$ conventional supercell with 32 molecules (160 atoms), at 7~GPa and 300~K. We ran the simulation for a total of 15~ps. 

The structure remains solid, with reorienting but non-diffusive CH$_4$ molecules retaining their integrity and oscillating about their original positions. Fig.~\ref{fig:MSD+RDF} shows that the carbons have a low MSD of about 0.2~\AA$^2$, while hydrogens have an MSD converging to about 2.5~\AA$^2$. This is consistent with large-amplitude rotational motion of molecules with a C--H bond length of 1.09~\AA, rather than with translational diffusion or rigid orientational locking.

\begin{figure}
    \centering
    \begin{minipage}[c]{0.30\textwidth}
        \centering
        \includegraphics[width=\linewidth]{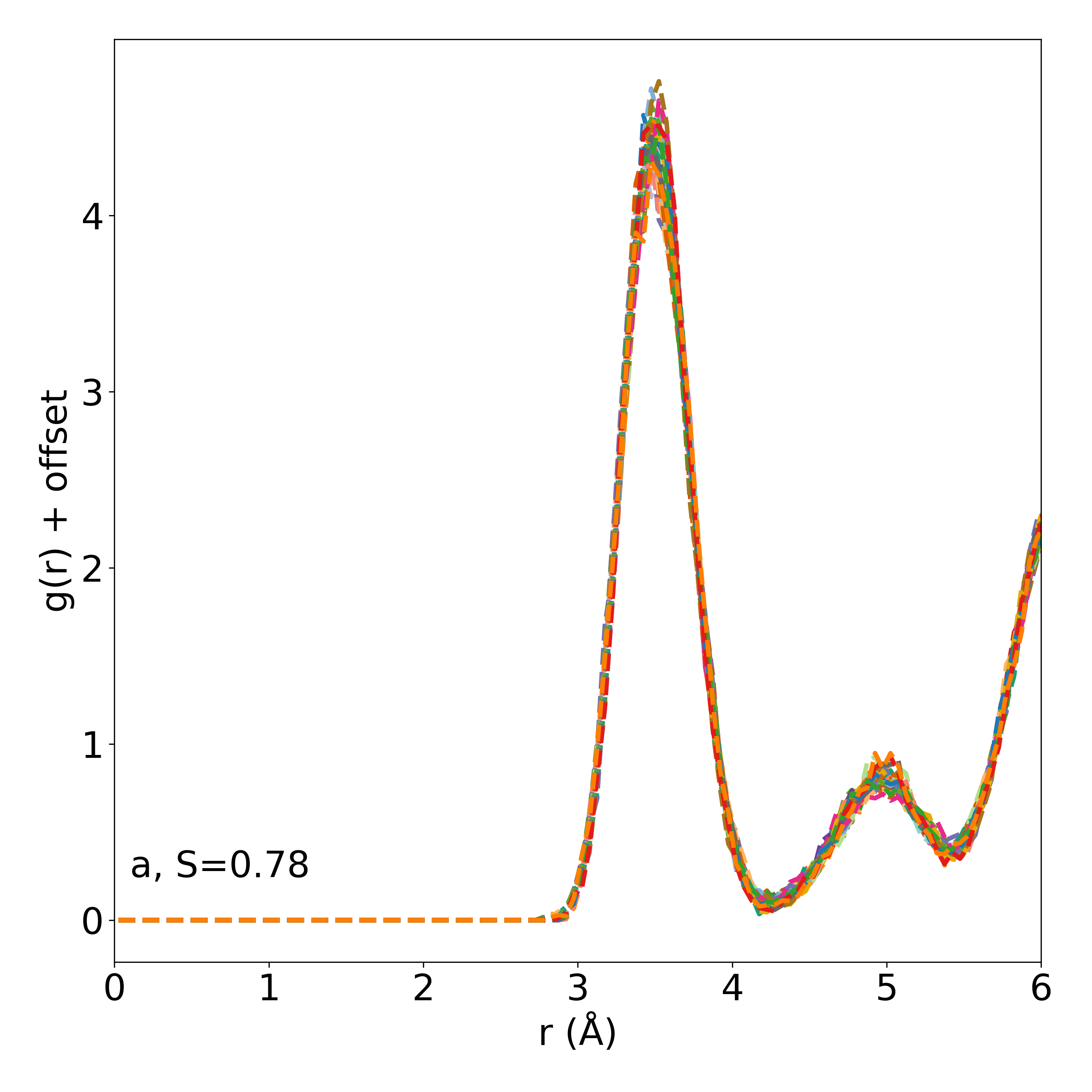} \\
        \includegraphics[width=\linewidth]{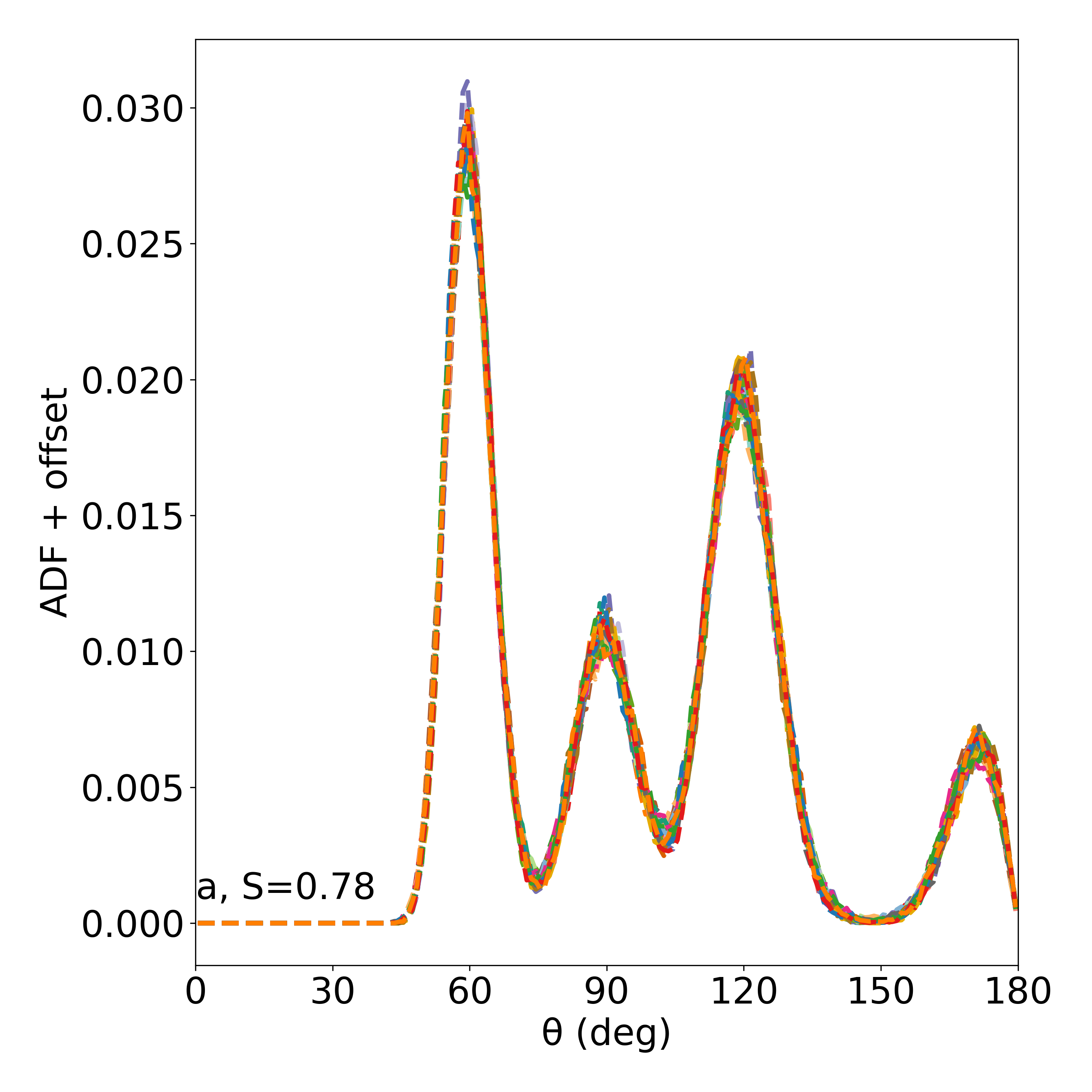}
    \end{minipage}
    \begin{minipage}[c]{0.50\textwidth}
        \centering
        \begin{overpic}[width=.9\linewidth]{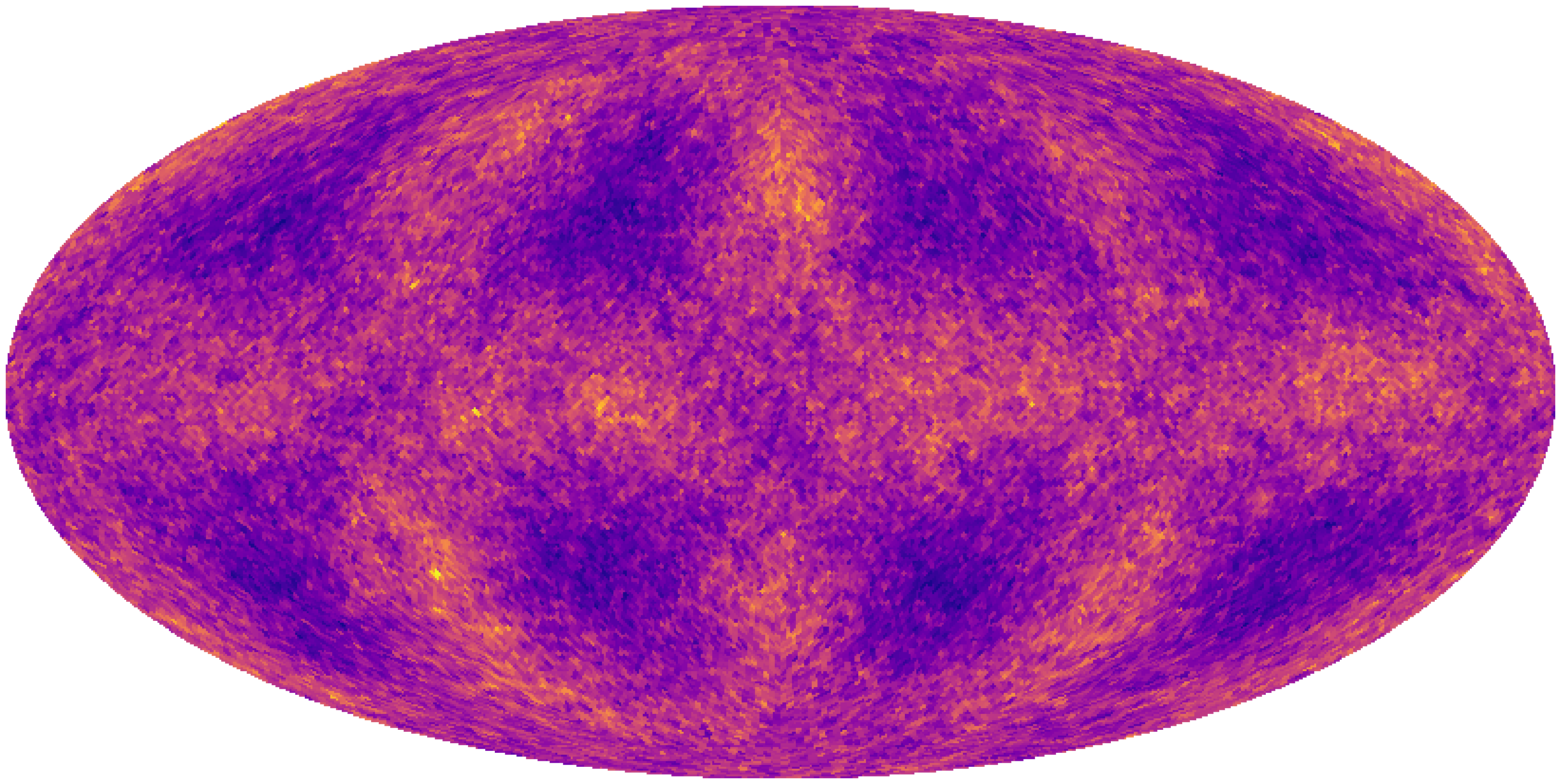}
            \put(10,45){\small\textbf{a}}
        \end{overpic}\par
        \vspace{2ex}
        \begin{overpic}[width=.75\linewidth]{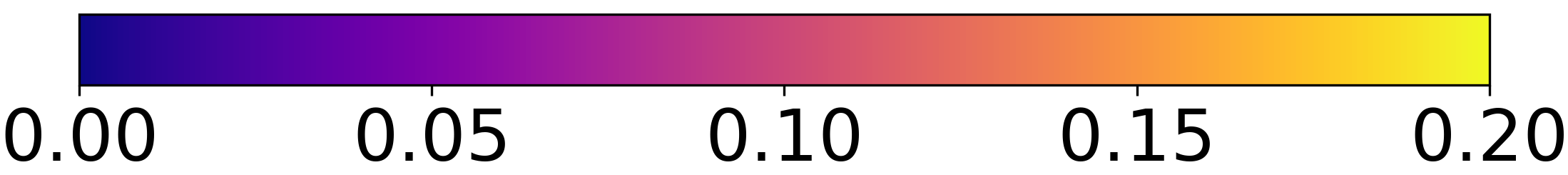}
        \end{overpic}\par
        \vspace{2ex}
        \begin{overpic}[width=.9\linewidth]{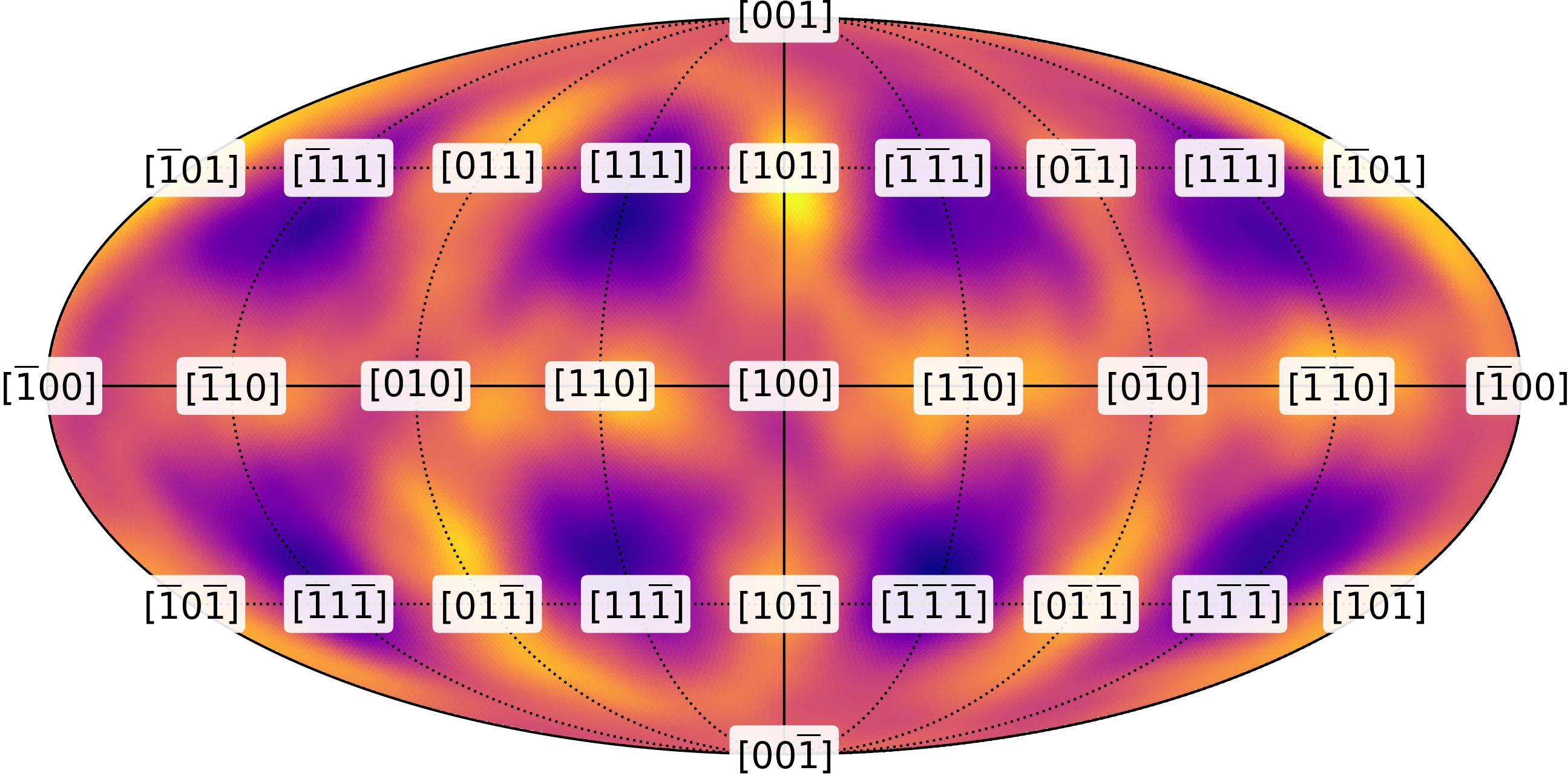}
        \end{overpic}
    \end{minipage}
    \caption{Top left and bottom left: site--site RDFs and ADFs for C--C pairs in CH$_4$ Phase~I ($Fm\overline{3}m$, single Wyckoff site a). RDF and ADF curves computed separately for each symmetrically equivalent carbon (all on the a site) lie on top of one another, coinciding with the corresponding ensemble-averaged distributions and confirming the expected $fcc$ positional order. Top right: Mollweide projection of the ensemble-averaged orientational PDF, showing pronounced preferred orientations and avoidance of eight distinct directions; the corresponding rotational entropy is $ S^\mathrm{rot} = 0.78 $ per molecule, indicating that the CH$_4$ molecules are not free rotors. Bottom right: the same PDF, smoothed with a Gaussian kernel with $ \mathrm{FWHM} = 10^\circ $ and labelled with the cubic crystallographic directions. C--H bonds avoid the cube body-diagonal $ \langle 111 \rangle $ directions, so C--H--C contacts with the 12 nearest neighbours are formed and broken along preferred, off-diagonal orientational pathways. This behaviour is consistent with CH$_4$ molecules acting as hindered rotors on the $fcc$ lattice sites, rather than freely sampling all orientations.}
    \label{fig:phaseI}
\end{figure}

The orientational probability density $ P(\theta,\phi) $ in Fig.~\ref{fig:phaseI} exhibits distinct maxima and clearly avoided directions, incompatible with an isotropic free-rotor model. The corresponding rotational entropy $ S^\mathrm{rot} = 0.78$ per molecule is about 80\% of the maximum classical value $ k_\mathrm{B}\ln(4\pi) $, directly quantifying the degree of rotational hindering. Thus, the methane molecules in Phase~I are neither rigidly locked nor freely rotors; they move freely while strongly avoiding the eight (111)-type orientations.  The free energy landscape is unusual, being largely flat, but with discrete peaks.

Despite this anisotropic instantaneous orientational distribution, the RDFs and ADFs for all  carbons lie on top of one another and coincide with the ensemble-averaged curves, confirming that the time-averaged carbon positions form an $ fcc $ lattice. The stress tensor becomes hydrostatic at $ c/a=\sqrt{2} $ (see the Supplementary Material for supporting analysis), and the trajectory shows a stable solid with no intermolecular proton transfer.

The $ I\overline{4}2m $ structure found in static searches is a plausible low-temperature orientationally ordered approximant, but at 300~K its free energy is exceeded by that of the orientationally disordered, hindered-rotor $ Fm\overline{3}m $ phase. Static enthalpy rankings alone are therefore insufficient to predict the observed Phase~I; inclusion of rotational degrees of freedom and their entropy is essential for capturing its stability.




\subsection{Phase~A}

\begin{figure}
    \centering

    \begin{minipage}[c]{0.32\textwidth}
        \centering
        \includegraphics[width=\linewidth]{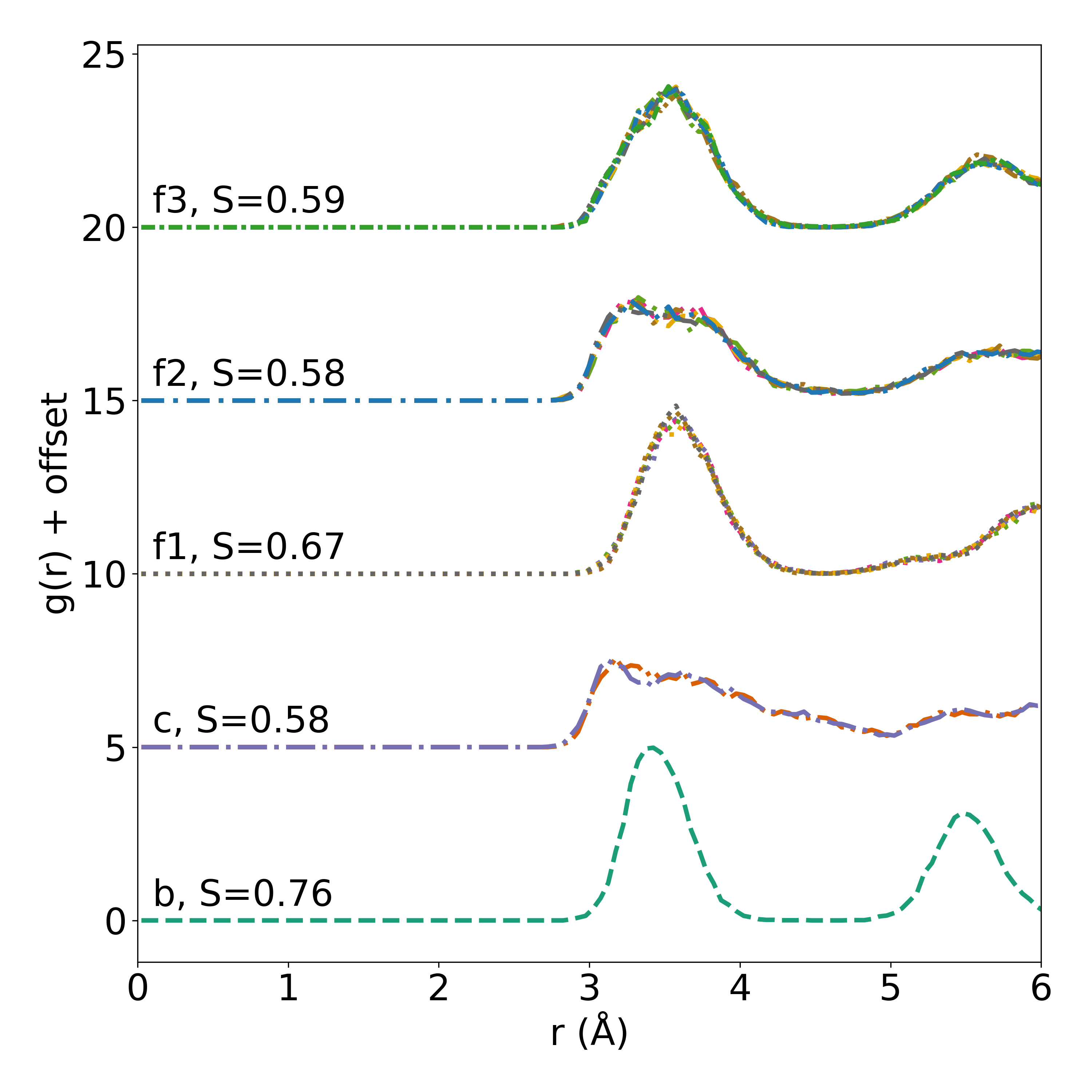}
    \end{minipage}
    \hfill
    \begin{minipage}[c]{0.32\textwidth}
        \centering
        \includegraphics[width=\linewidth]{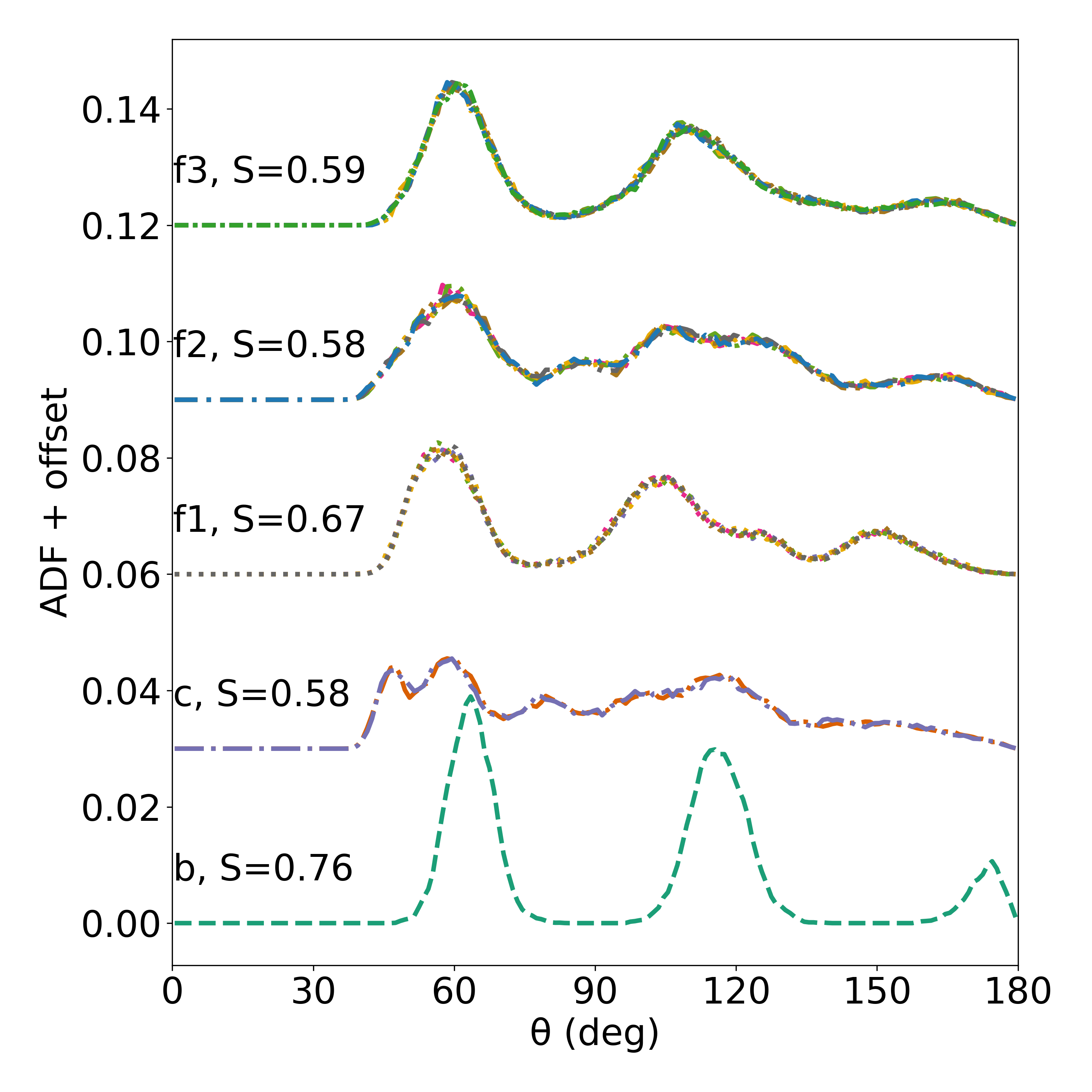}
    \end{minipage}
    \hfill
    \begin{minipage}[c]{0.34\textwidth}
        \centering
        \begin{minipage}[t]{0.48\linewidth}
            \centering
            \begin{overpic}[width=\linewidth]{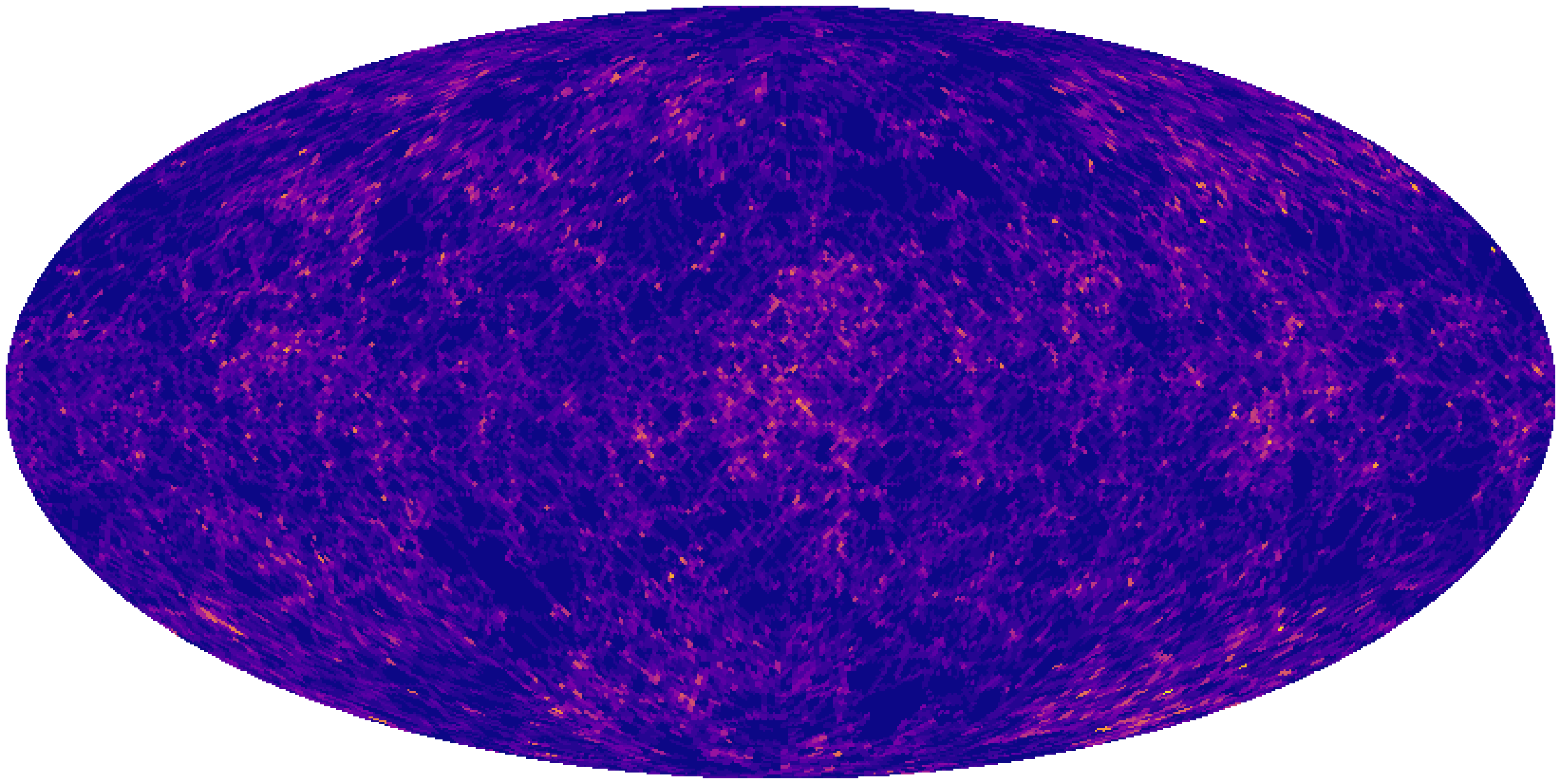}
                \put(0,0){\small\textbf{b}}
            \end{overpic}
        \end{minipage}\hfill
        \begin{minipage}[t]{0.48\linewidth}
            \centering
            \begin{overpic}[width=\linewidth]{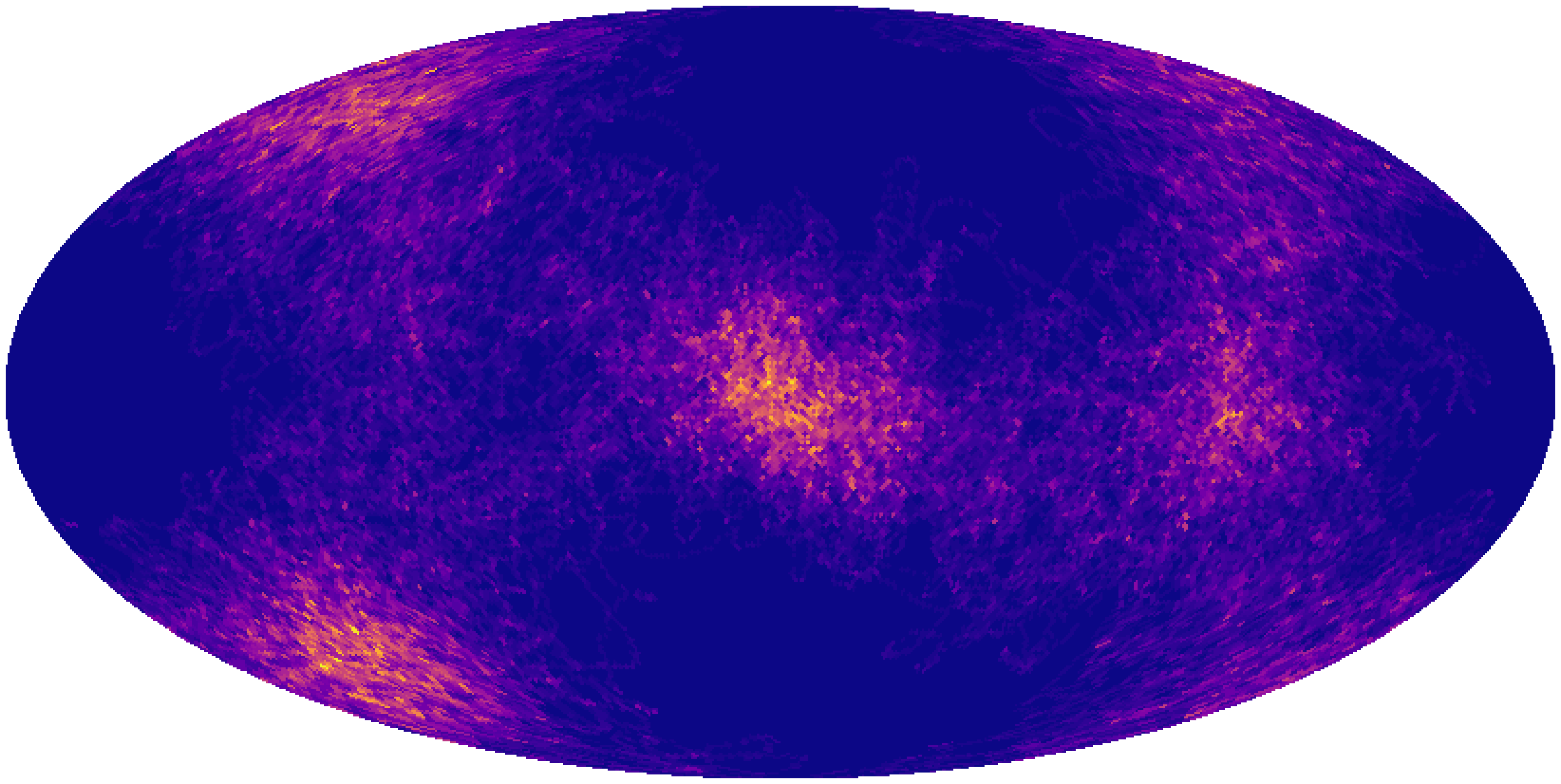}
                \put(0,0){\small\textbf{c}}
            \end{overpic}
        \end{minipage}
        \vspace{1ex}
        \begin{minipage}[t]{0.48\linewidth}
            \centering
            \begin{overpic}[width=\linewidth]{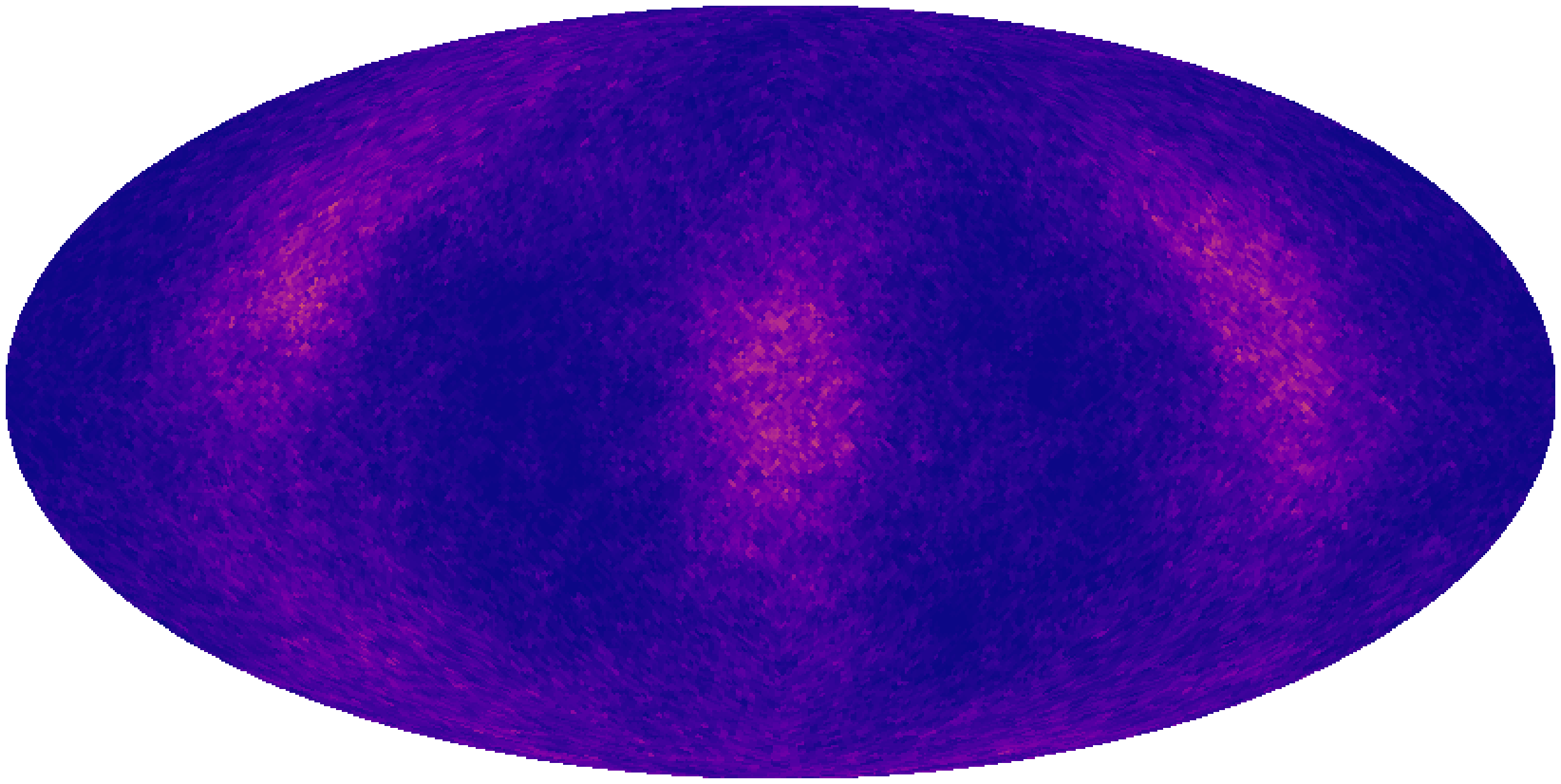}
                \put(0,0){\small\textbf{f1}}
            \end{overpic}
        \end{minipage}\hfill
        \begin{minipage}[t]{0.48\linewidth}
            \centering
            \begin{overpic}[width=\linewidth]{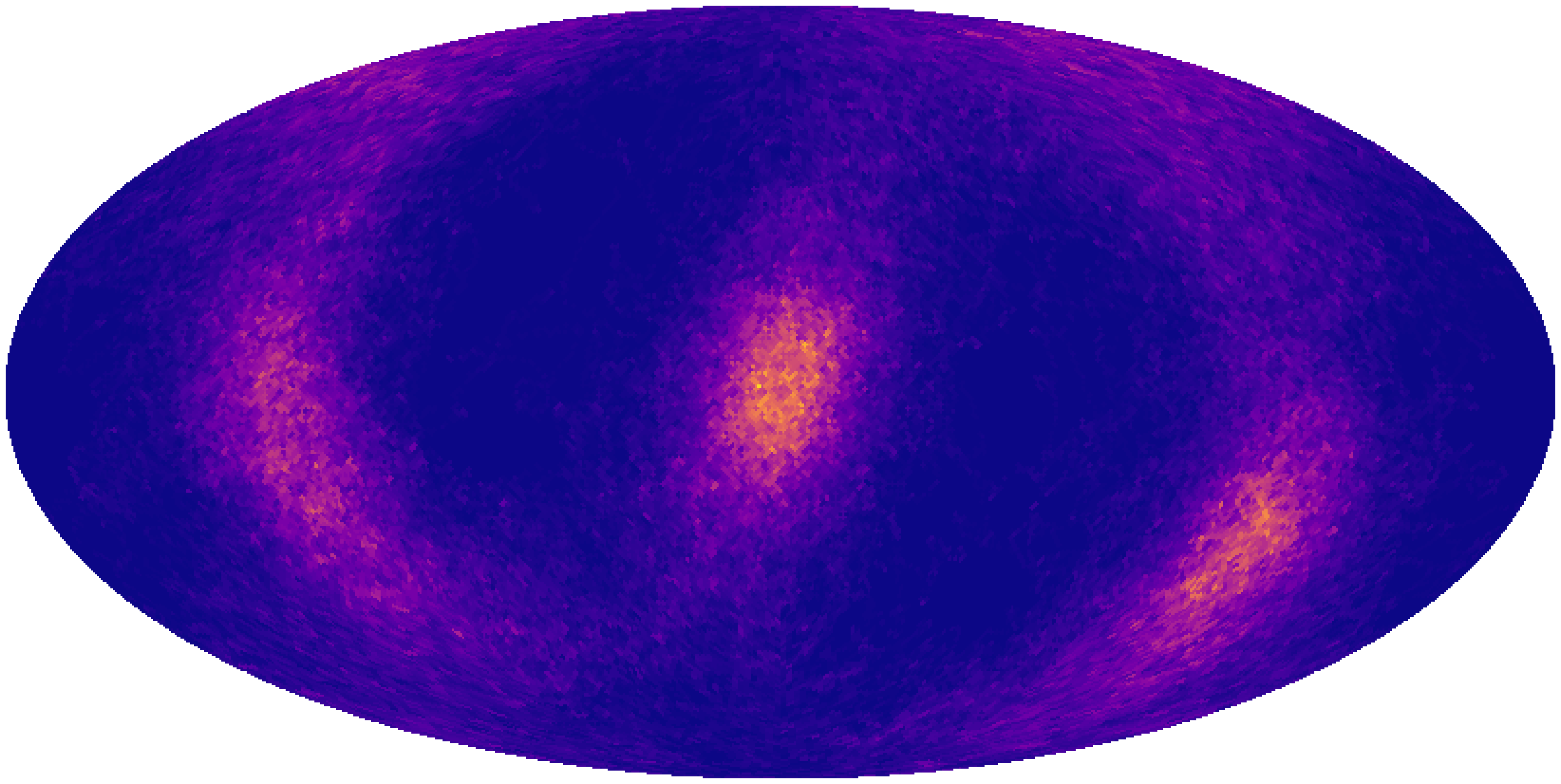}
                \put(0,0){\small\textbf{f2}}
            \end{overpic}
        \end{minipage}
        \vspace{1ex}
        \begin{minipage}[t]{0.48\linewidth}
            \centering
            \begin{overpic}[width=\linewidth]{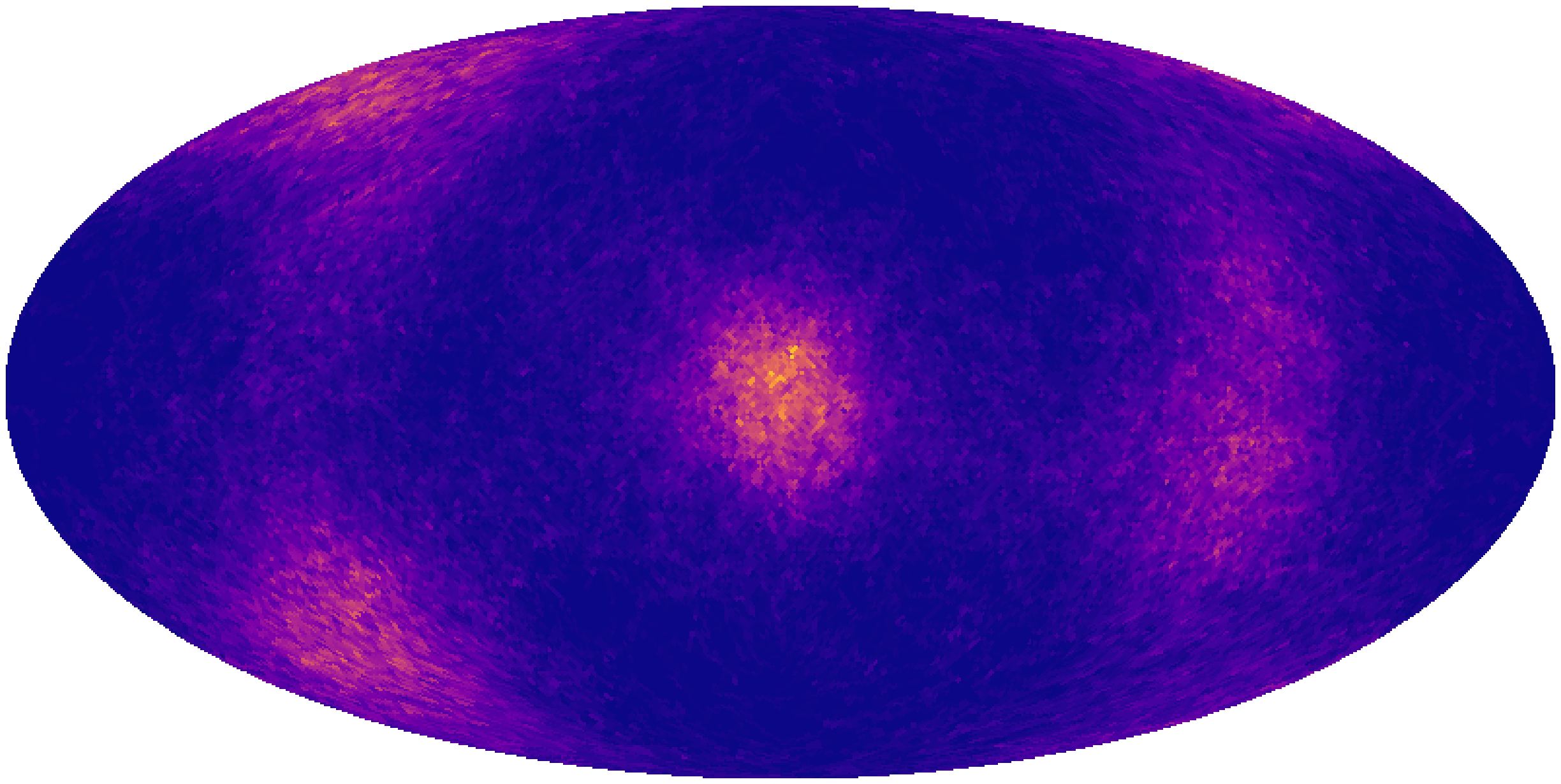}
                \put(0,0){\small\textbf{f3}}
            \end{overpic}
        \end{minipage}
        \vspace{1ex}
        \begin{minipage}[t]{\linewidth}
            \centering
            \begin{overpic}[width=.9\linewidth]{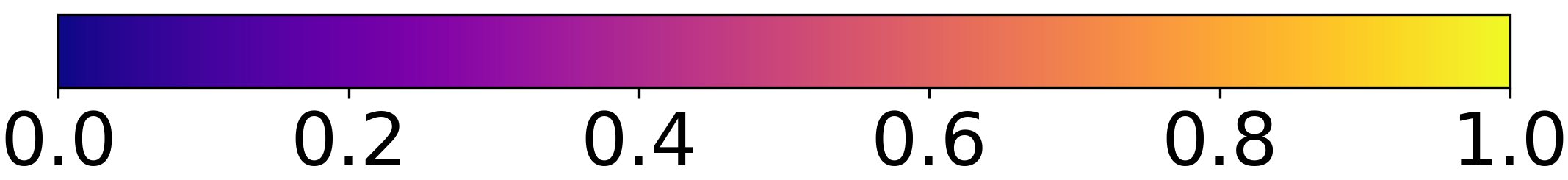}
            \end{overpic}
        \end{minipage}
    \end{minipage}
    \caption{C--C radial and angular distribution functions showing the five environments (Wyckoff sites) of molecules in methane Phase~A simulations. The rotating molecule on the 1b site has the largest near-neighbour separations. Molecule in the 1b site is hindered-rotor like; one 6f site, labelled f1, has notably higher entropy than the remaining f-sites. Note that PDF (right) colour scale is different than in Fig.~\ref{fig:phaseI}. None of the molecules can be considered as ``static''.}
    \label{fig:phaseA}
\end{figure}

\subsubsection{Phase~A Simulations}
Phase~A is regarded as a very unusual structure, insofar as it has a near-cubic primitive unit cell (rhombohedral angle between 89 and 90$^o$), with 21 molecules but with molecular positions having $R3$ symmetry without any obvious relationship to a cubic structure. 
Constrained DFT structure searches  based on 21 molecules per unit cell and enforced $R3$ symmetry are able to find the experimentally reported $R3$ structure~\cite{maynard2010}. 
However, static DFT calculations find this phase to be unstable with respect to rotation of the molecules: unless $R3$ symmetry is enforced, there is a symmetry breaking on relaxation to a triclinic structure (named A-$P1$), which has substantially lower enthalpy than $R3$.  In all cases  the unit cell is remarkably close to a cube, angles are still between 89 and 90$^o$ and the cell lengths within 0.5\%\, of equality.

Furthermore, unconstrained structure searches~\cite{bykov2021structural} revealed several structures with enthalpy even lower than the A-$P1$ structure in the pressure range where Phase~A is located. These structures will be discussed in section ~\nameref{sec:low_temp}.

For BOMD, we used the primitive cell containing 105 atoms, large enough that the two nearest-neighbour shells are fully independent molecules. Given the importance of steric effects and the fast-decaying nature of van der Waals interactions, we expect and observe this system size to be adequate to overcome orientational frustration.  We ran two different simulations in the NVT ensemble: in one case using a triclinic unit cell, based on the calculated A-$P1$ minimum enthalpy structure, and the other with a rhombohedral cell, as per experiment.  We ran both BOMD simulations in the NVT ensemble for 12~ps. Where applicable, the same simulation settings as for Phase~I, were chosen. We initiated both structures via static relaxation at 7~GPa.
The final pressure, taking into account the kinetic contribution, was around 8~GPa. To test finite size effects, we then repeated the simulation with a $2^3$ supercell containing 840 atoms.
All three simulations had equivalent radial and angular distribution functions, and produced equivalent simulated diffraction patterns (see SM).  We conclude that our results are independent of system size or supercell shape.

The molecular orientation shows frequent reorientation, such that after about 2~ps, molecule orientations are uncorrelated with initial conditions. After a long run, the average position of individual hydrogen atoms is in the middle of the molecule as discussed in the Methods section.

\subsubsection{Phase~A Structure}

The radial distribution function shows a very well-defined first peak, so we can easily assign coordination numbers to each molecule and generate the ADF (Figs. \ref{fig:MSD+RDF} and \ref{fig:phaseA}). On relaxation of the structure, this near-neighbour peak comprises 25 inequivalent ``bonds'' between various neighbour pairs, with a range of lengths between 3.1 and 4.2~\AA\,.  No other pairs are found less than 5~\AA~apart. 

We created averaged structures from the average molecular positions and tested their symmetry using SPGLIB.  Surprisingly, this is $R\overline{3}$, higher than either the initial cells or the experimentally reported values.

In the initial $R3$ structure setting, there were three molecules on Wyckoff 1a sites, i.e. on the body diagonal, and three molecules on each of the six independent 3b sites.  The emergent inversion symmetry which creates $R\overline{3}$ is about one of the 1a ($R3$) sites, which becomes 1b in $R\overline{3}$.  Inversion links the other two 1a sites, which become 2c in $R\overline{3}$,  and  pairs up the 3b ($R3$) sites into  6f ($R\overline{3}$) sites.

The RDF and ADF (Fig.~\ref{fig:phaseA}) demonstrate both the different environments of the molecules at different $R\overline{3}$ sites and the similarity of those at symmetry equivalent $R3$ Wyckoff sites  (see SM).  Notably, all $R{3}$ sites linked by the inversion symmetry to get $R\overline{3}$ have indistinguishable RDF and PDF.
 
The MSD (Fig. \ref{fig:MSD+RDF}) is consistent with either rotating or reorientating molecules, but the $S^{rot}$ entropy gives deeper insight into the different molecular behaviours. The $\theta,\phi$ PDF indicates that only the 1b molecule rotates nearly-freely ($S^{rot} = 0.76$), other molecules have a strongly preferred orientation, but reorient between the 12 permutations of the hydrogen atoms on a picosecond timescale. 
  
We note that the 1b site is the inversion centre, and so this molecule cannot be ordered without breaking $R\overline{3}$ symmetry.  This 1b site has a very simple RDF and ADF, the single C--C peak indicates a shell of 12 neighbours, and angles of 63, 115 and 180$^o$ reveal it to be the central site in an icosahedron. 

The 6f2 and 6f3 molecules, which form the icosahedron, are oriented such that they have C--H bonds pointing directly away from the central 1b site.  We propose that this facilitates the rotation of the central molecule. 
The fivefold symmetry means the icosahedron cannot be incorporated  into any space group, so the 6f2 and 6f3 sites are symmetrically inequivalent, although their RDF, ADF and entropy are similar.  Indeed the 6f2 and 6f3 sites have very close near-neighbours outside the octahedron, and show low rotation and entropy, whereas the 6f1 site has near-neighbour distances similar to 1b, and  exhibits high $S^{rot}$.

The shortest intermolecular spacings are between the 2c and  6f1 sites. Taken together, these eight molecules form a distorted cube centred on the origin.  A cube is not an efficient packing, even when distorted, but this seems to be compensated for by the very short intermolecular separations facilitated by the favorable orientations.  If one thinks of these as ``bonds'', their low enthalpy is primarily due to the shortness of the distance contributing to low volume structures. 


The structure found in the simulations is extremely close to that reported experimentally; the icosahedral supermolecule is present, if unnoticed.
All diffraction peaks are in the same place in simulation and experiment and correspond to either $R3$ or $R\overline{3}$, the maximum possible symmetry for an octahedron in a cube.  
The X-ray single-crystal data analysis, considering carbon atoms only, obtained a better fit to the intensities with R${3}$ symmetry while the time-averaged simulation is close to $R\overline{3}$ (with finite sampling, any computed average is strictly $P1$.). However the molecular positions in the two models are extremely close. 
The  neutron powder data was analysed using three models, all different from the simulation, and the best fit being $R3$ with molecules orientationally ordered.  The sample was somewhat textured , so the neutron data are thus less emphatic than are the x-ray data~\cite{maynard2014}.
The most notable difference is that, in ($R3$), the three molecules along the (111) rotation axis were assumed to be all pointing in the same direction~\cite{maynard2014}. In $R\overline{3}$ one of these molecules is the rotating one at the centre of the icosahedron, and in simulation another rotates to become symmetry-related to the third as 2c. An illustration of this is given in SM.
The experimental proposal for orientational order came from the best-fit to possible fixed molecular orientations with $T_d$ symmetry, the MD shows another possibility where one molecule rotates.  We note that the dynamic reorientation observed in the simulation simply permutes the hydrogen atoms, so has no impact on the diffraction analysis.

\begin{table}[htbp]
\centering
\begin{tabular}{|c|c|c|c|c|} 
 \hline
Cell &  Angles ($^\circ$) &Symmetry & Volume (\AA$^3$) & \\  
\hline
triclinic & 89.30, 89.67, 89.27& $R\overline{3}$ & 643 &\\
rhombohedral & 88.88& $R\overline{3}$ &645& \\
Expt. & 89.5 & $R3$& 616&\\
\hline
Site & Coordinates & Entropy   &  Coordination& r$_{min}$(\AA)\\
\hline
1b  & 0.50 0.50  0.50 & 0.76 & 12 & 3.36 \\ 
2c  & 0.12 0.12  0.12 & 0.58 & 13 & 3.12 \\ 
6f1 & 0.36 0.45  0.12 & 0.67 & 14 & 3.12 \\ 
6f2 & 0.64 0.13  0.01 & 0.58 & 13 & 3.39 \\ 
6f3 & 0.63 0.76  0.23 & 0.59 & 13 & 3.25 \\ 
\hline
\end{tabular}
\caption{ \label{tab:phaseA} Summary of the average molecular positions in the Phase~A simulations. Each frame has $P1$ symmetry: SPGLIB is used to find the symmetry for the time-averaged structure (8~GPa a=8.642~\AA, $\gamma=88.88^\circ$).  Experimental data (a=8.509~\AA, $\gamma=89.5^\circ$ at 9.1~GPa) is from~\cite{maynard2014}. Coordination is based on the first shell of neighbours between 3.1 and 4.2~\AA. $r_{min}$ denotes the nearest neighbour distance. The closest pairs not counted in the coordination shell are beyond 5.1~\AA.
The icosahedron is made of 6f2 and 6f3 sites.}
\end{table}

\subsubsection{Phase~A Simulated Diffraction Patterns}

\begin{figure}[htbp]
    \centering
        \includegraphics[width=0.36\textwidth]{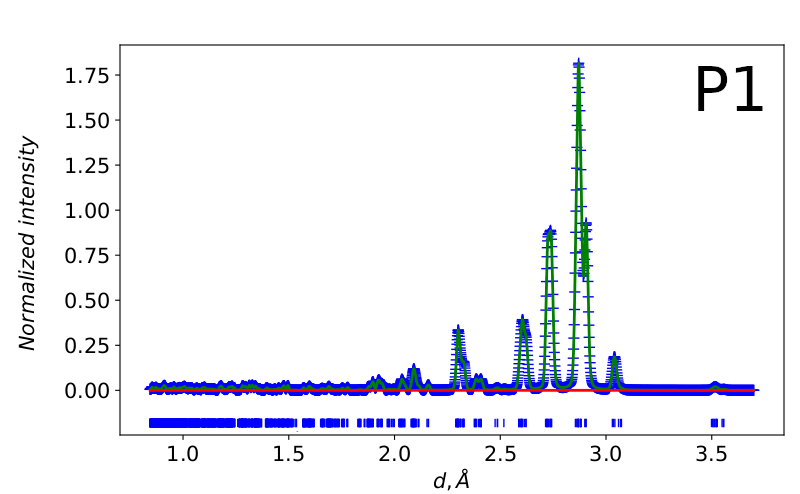}
        \includegraphics[width=0.26\textwidth]{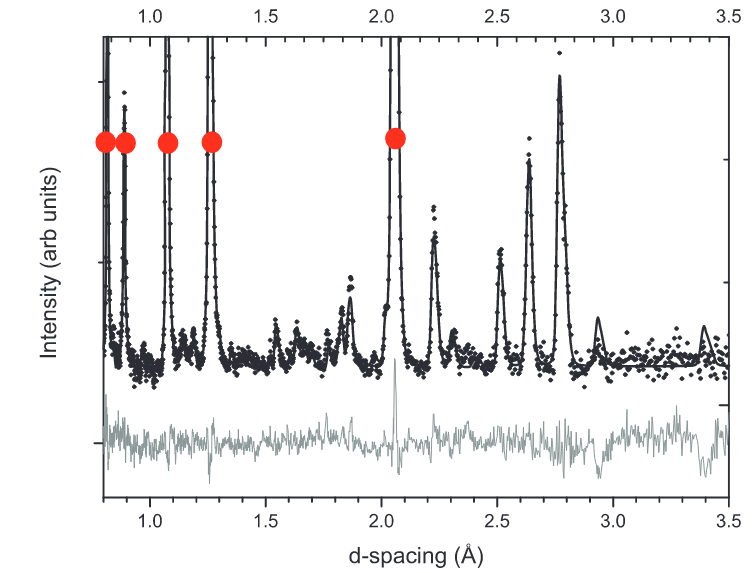}
        \includegraphics[width=0.36\textwidth]{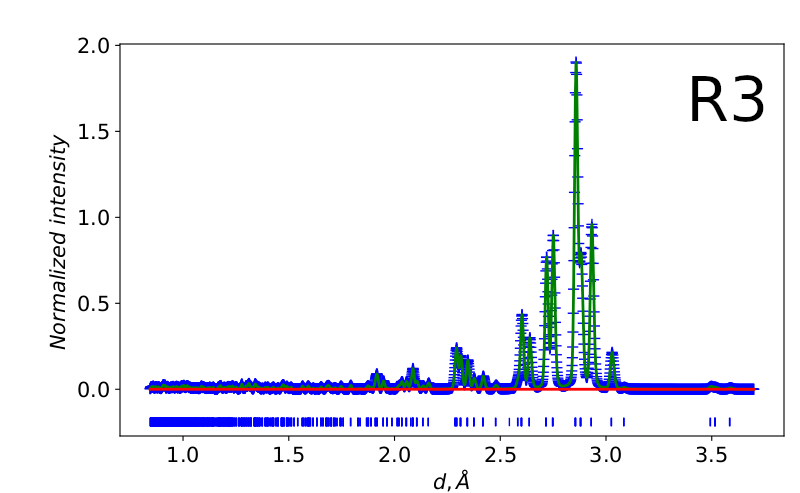}
    \caption{Phase~A at 8~GPa and 300~K, based on 50 uncorrelated frames from BOMD.
    (Left) Neutron diffraction simulation (all H replaced by D) for run with a triclinic cell. (Middle) Measured neutron diffraction pattern by Maynard-Casely \textit{et al.}~\cite{maynard2010}. Red circles mark diamond peaks coming from the anvils of the pressure cell. (Right) Simulated neutron diffraction for run with a rhombohedral cell.
    \label{fig:phaseA_stat}}
\end{figure}

We simulated time-of-flight (TOF) neutron diffraction patterns for each candidate phase using 50 frames from the BOMD simulation, which is enough to converge the pattern. Results are presented in Fig.~\ref{fig:phaseA_stat} (see SM for convergence tests and additional simulated patterns).

The cells are all close to cubic, which allows us to index the strong reflections into groups (see SM).  The small rhombohedral distortion splits these groups into well-defined peaks.  The agreement with the experimental data is good, given that there are no systematic absences: all peaks have non-zero intensity in theory, and the simulation gives very low intensity to those not observed in experiment.

Our best reconciliation of the simulation and experimental data is thus that phase A has a degree of orientational disorder of the molecules - a conclusion that is in accord with thermodynamic expectation given that phase A transforms on cooling.  The icosahedral supermolecule necessitates a reduction of symmetry from cubic to $R\overline{3}$, although the associated distortions from cubic are small. 
X-ray diffraction spot intensities~\cite{maynard2014} imply symmetry $R3$: the breaking of centrosymmetry is not explained by the simulations.  It could arise from  particular molecular orientations; methane does not have inversion symmetry, so $R\overline{3}$ requires disorder at the 1b site. The X-ray data was fitted  considering the carbon atoms only; ultimately, the final resolution of this question will require neutron single crystal data. 

\subsection{Phase~B and HP}

\begin{figure}
\centering

\begin{minipage}[c]{\textwidth}
  \begin{minipage}[c]{0.32\textwidth}
    \centering
    \includegraphics[width=\linewidth]{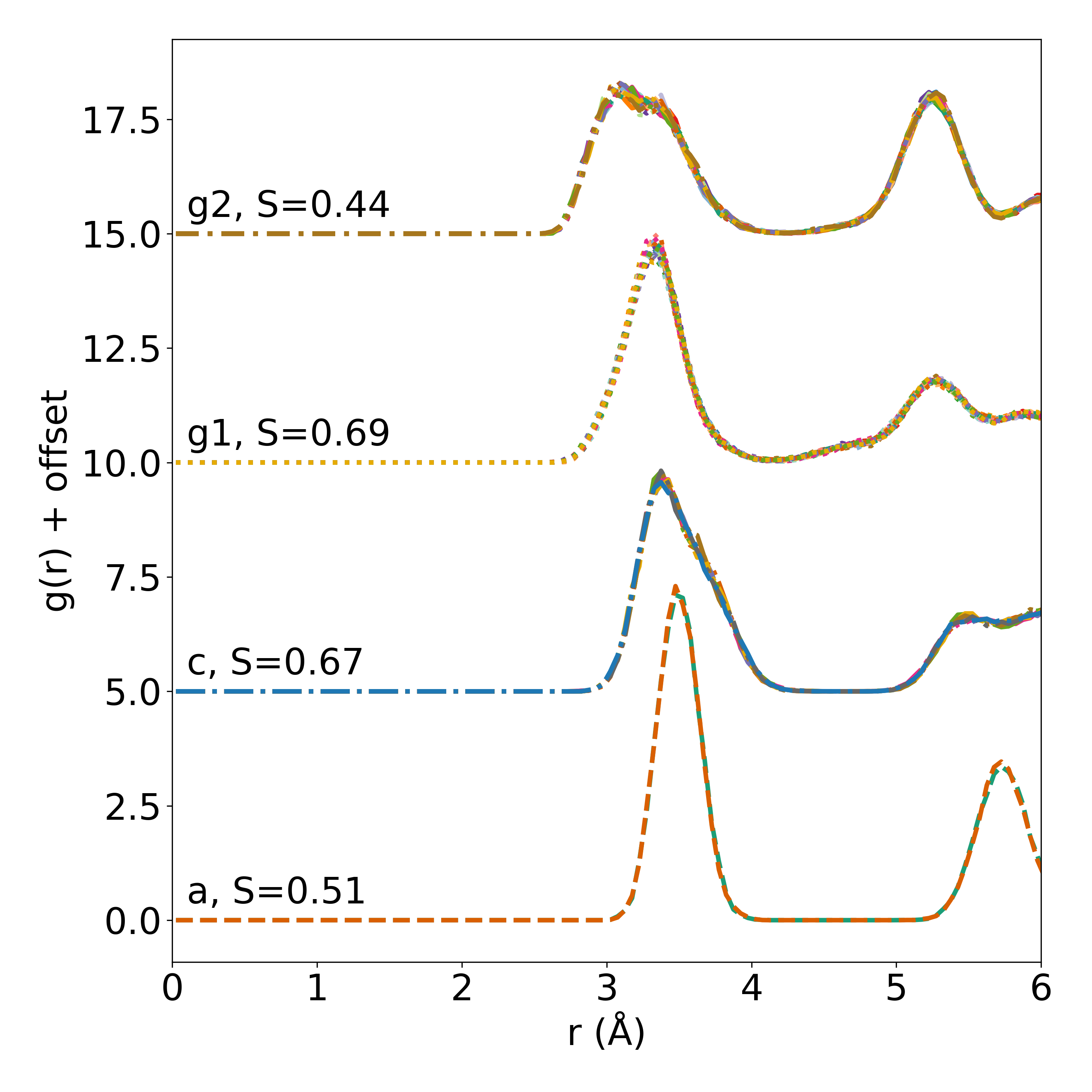}
  \end{minipage}\hfill
  \begin{minipage}[c]{0.32\textwidth}
    \centering
    \includegraphics[width=\linewidth]{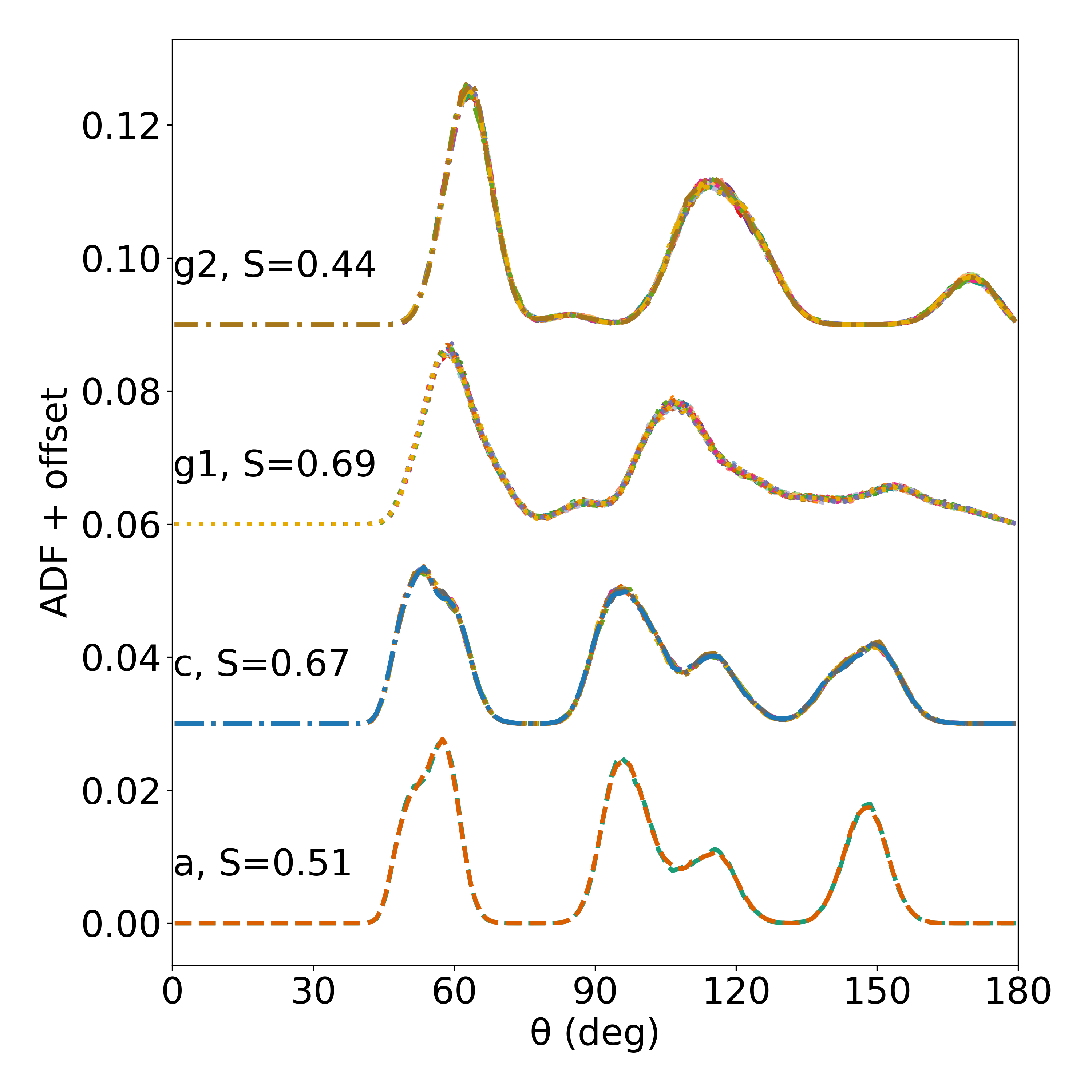}
  \end{minipage}\hfill
  \begin{minipage}[c]{0.34\textwidth}
    \centering
    \begin{minipage}[t]{0.48\linewidth} 
      \centering
      \begin{overpic}[width=\linewidth]{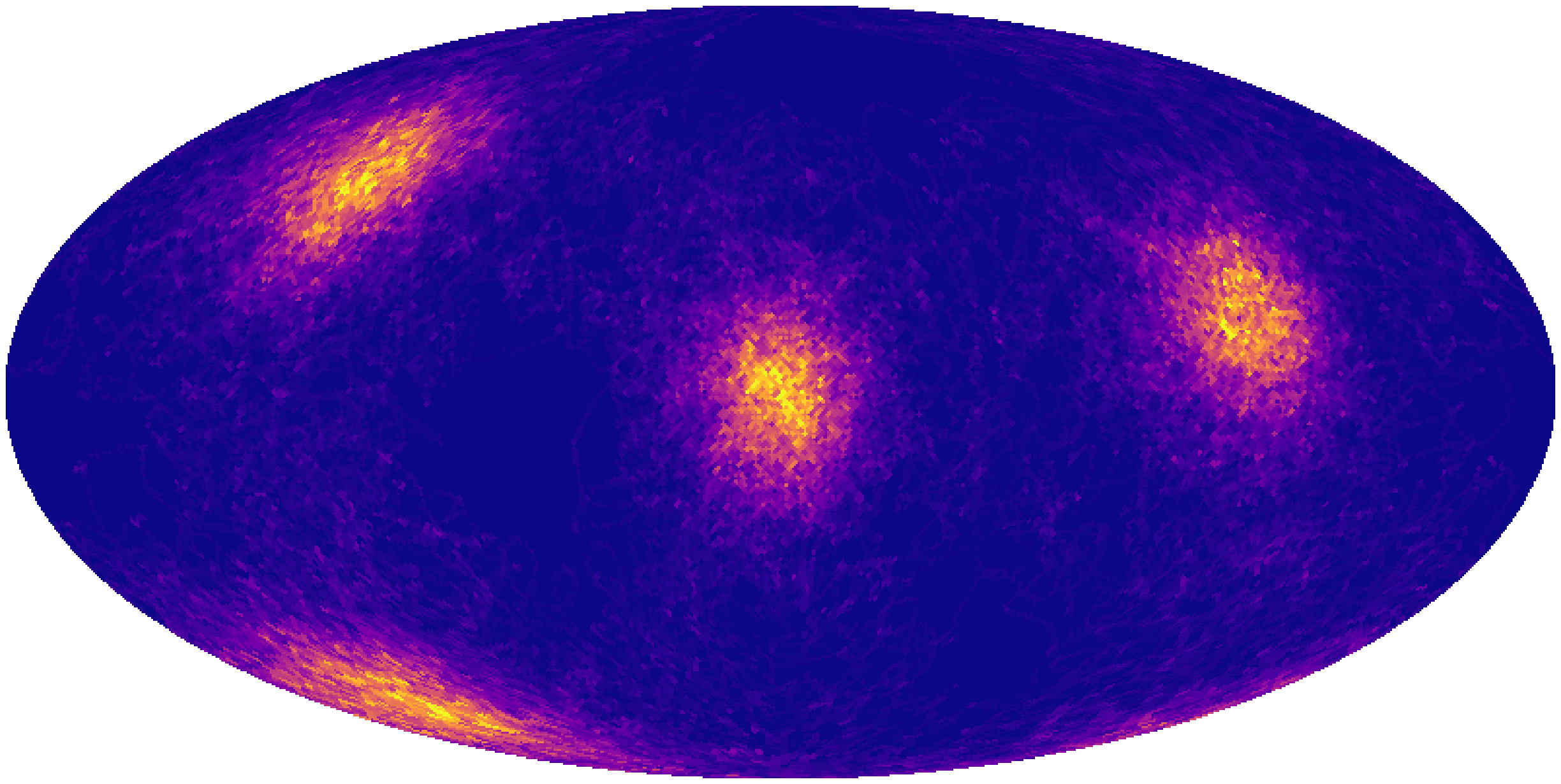}
        \put(-3,6){\small\textbf{a}}
      \end{overpic}     
    \end{minipage}\hfill
    \begin{minipage}[t]{0.48\linewidth}
      \centering
      \begin{overpic}[width=\linewidth]{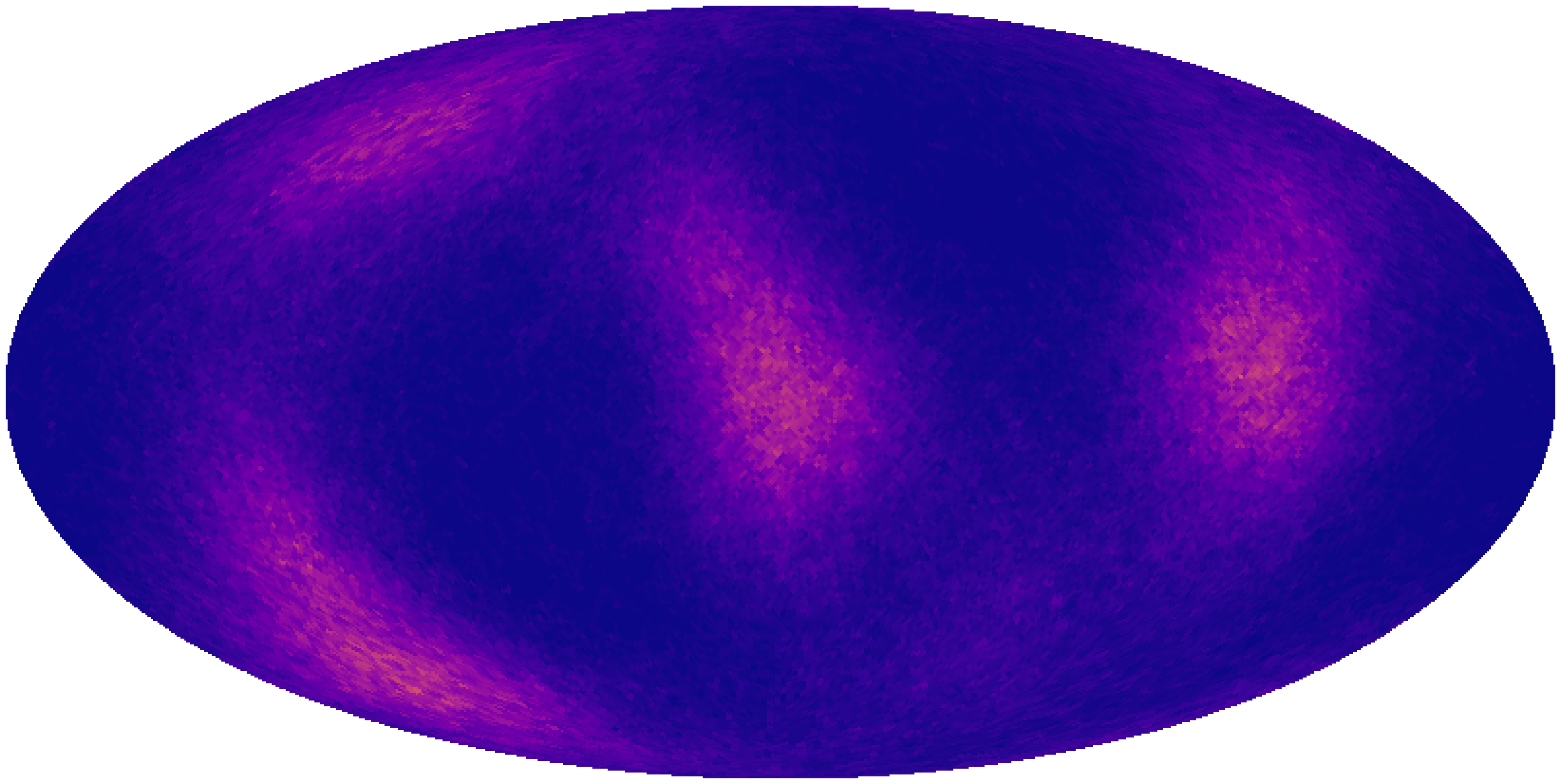}
        \put(-3,6){\small\textbf{c}}
      \end{overpic}
    \end{minipage}

    \vspace{1.5ex}

    \begin{minipage}[t]{0.48\linewidth}
      \centering
      \begin{overpic}[width=\linewidth]{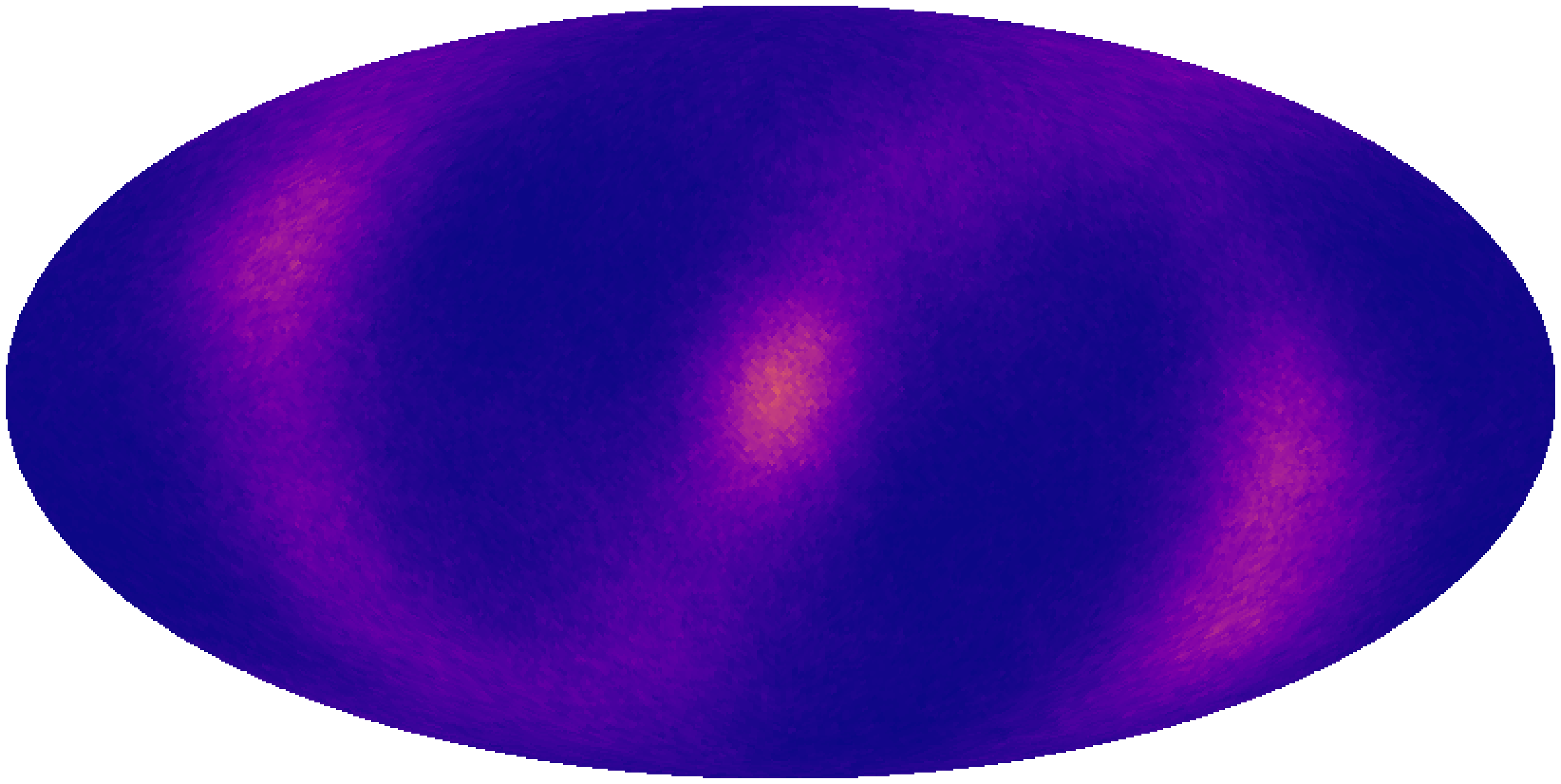}
        \put(-3,-1){\small\textbf{g1}}
      \end{overpic}
    \end{minipage}\hfill
    \begin{minipage}[t]{0.48\linewidth}
      \centering
      \begin{overpic}[width=\linewidth]{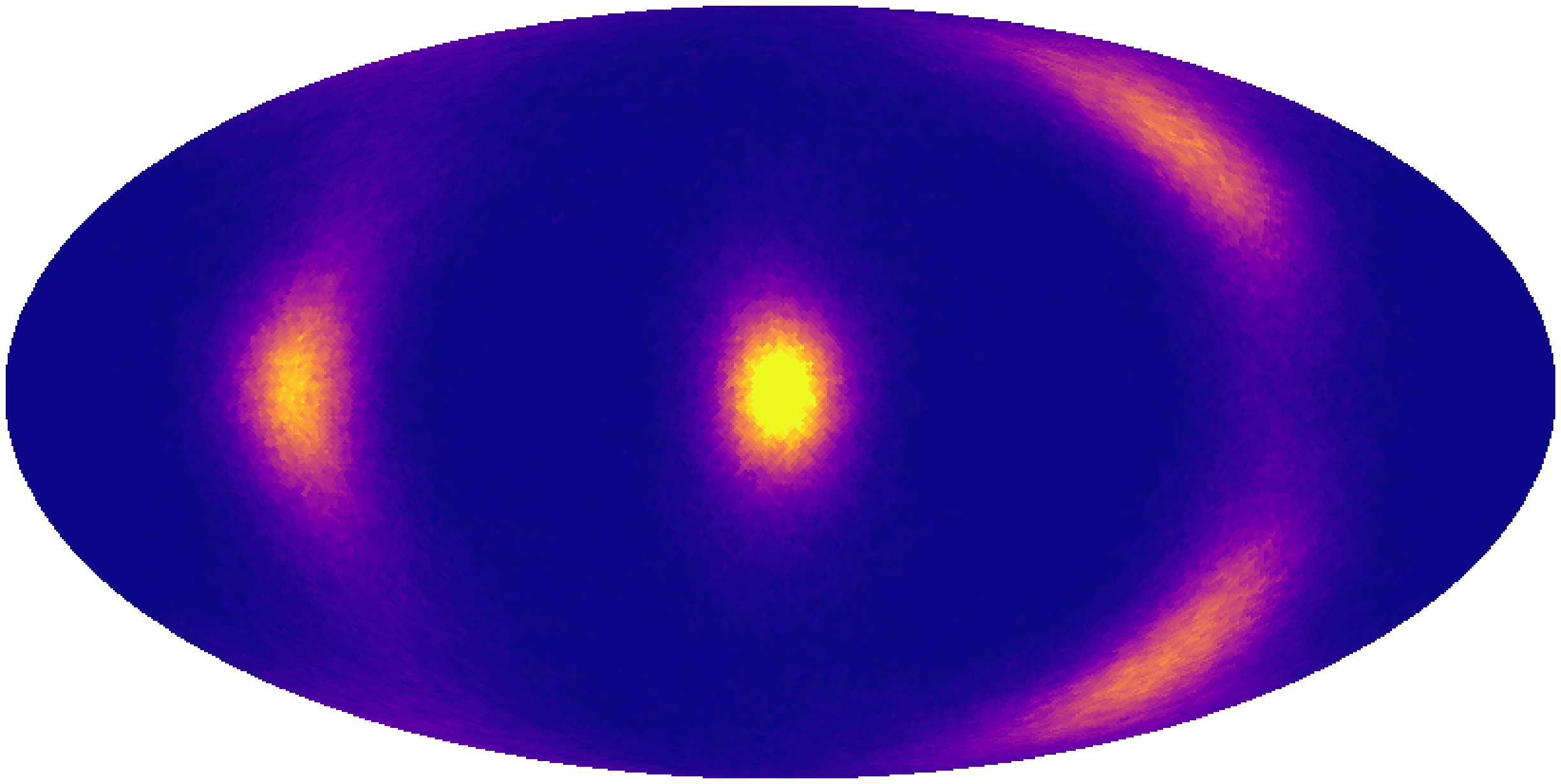}
        \put(-3,-1){\small\textbf{g2}}
      \end{overpic}
    \end{minipage}

    \vspace{6ex}

    \begin{minipage}[t]{\linewidth}
      \centering
      \begin{overpic}[width=.9\linewidth]{figures/opdfs/oPDF_colourbar_horizontal.png}
      \end{overpic}
    \end{minipage}
  \end{minipage}
\end{minipage}


\begin{minipage}[c]{\textwidth}
  \begin{minipage}[c]{0.32\textwidth}
    \centering
    \includegraphics[width=\linewidth]{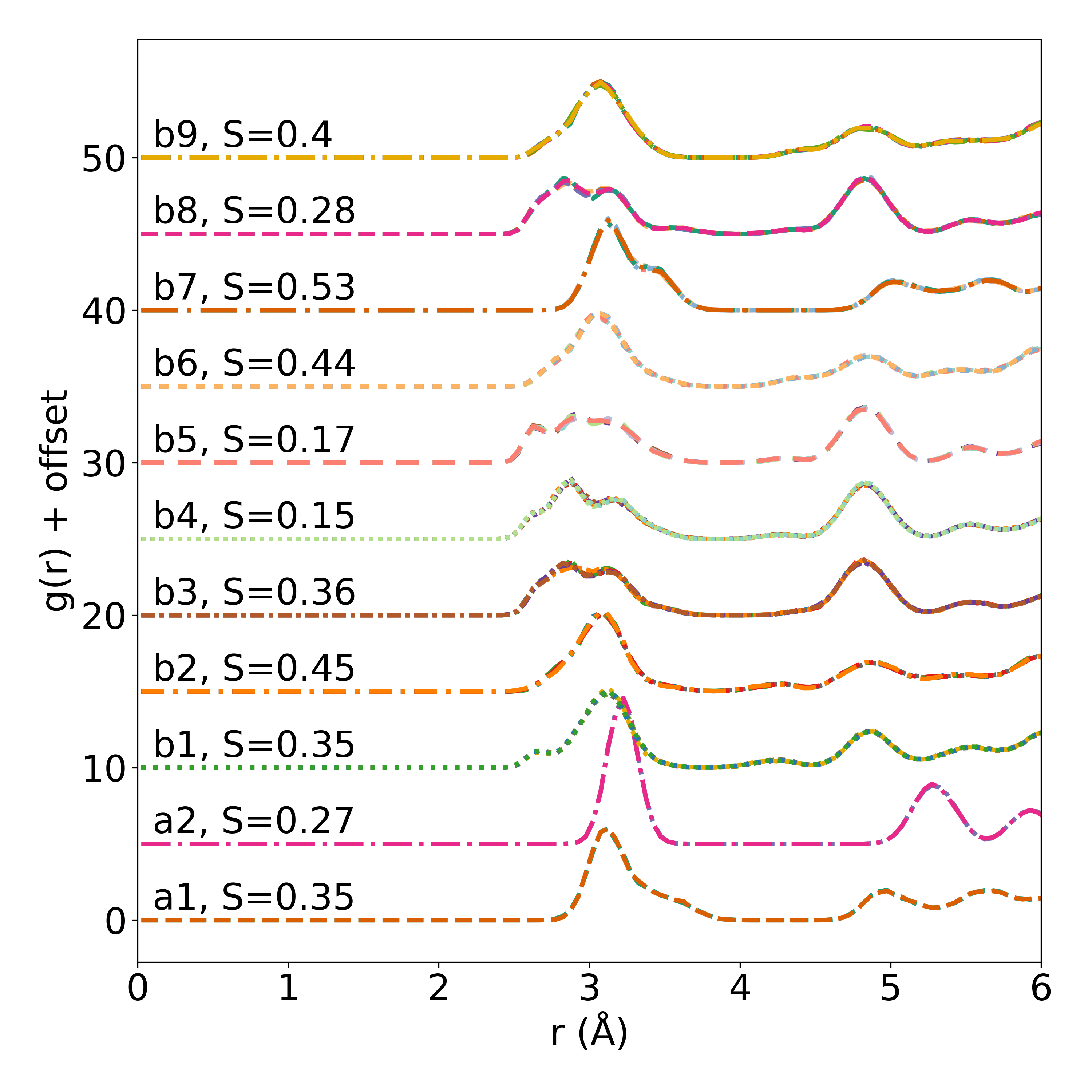}
  \end{minipage}\hfill
  \begin{minipage}[c]{0.32\textwidth}
    \centering
    \includegraphics[width=\linewidth]{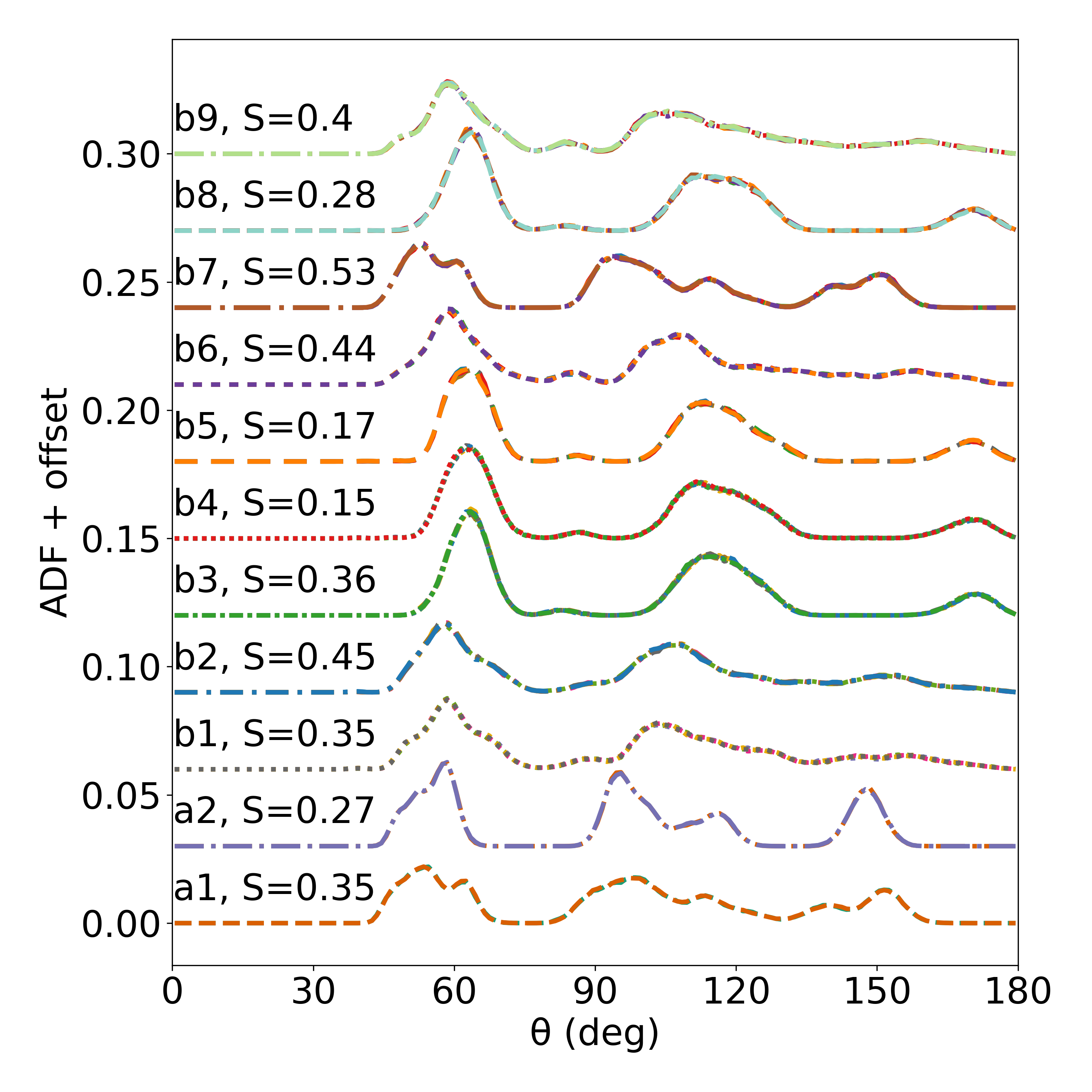}
  \end{minipage}\hfill
  \begin{minipage}[c]{0.34\textwidth}
    \centering

    \begin{minipage}[t]{0.48\linewidth}
      \centering
      \begin{overpic}[width=\linewidth]{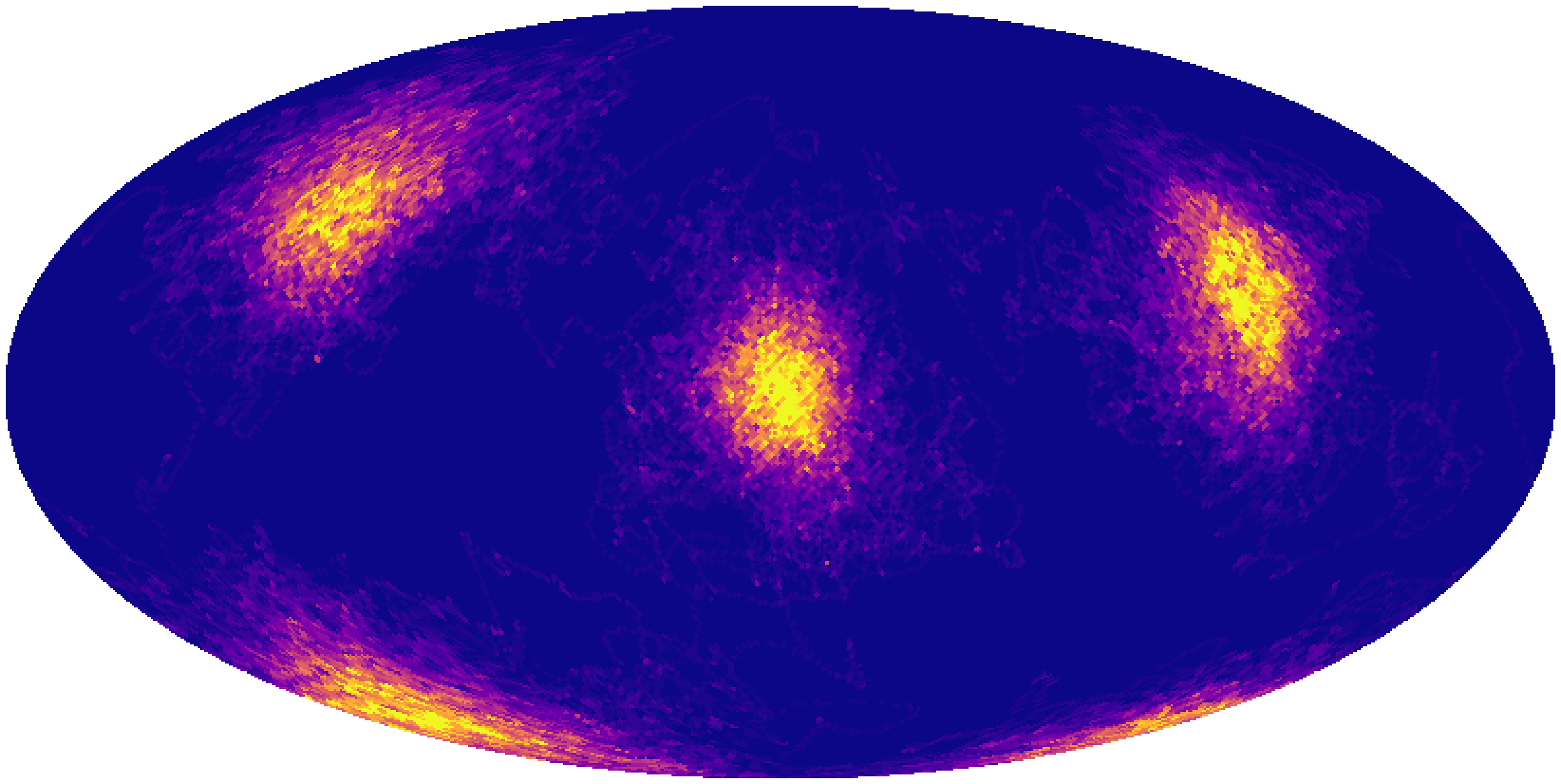}
        \put(-3,-1){\small\textbf{a1}}
      \end{overpic}
    \end{minipage}\hfill
    \begin{minipage}[t]{0.48\linewidth}
      \centering
      \begin{overpic}[width=\linewidth]{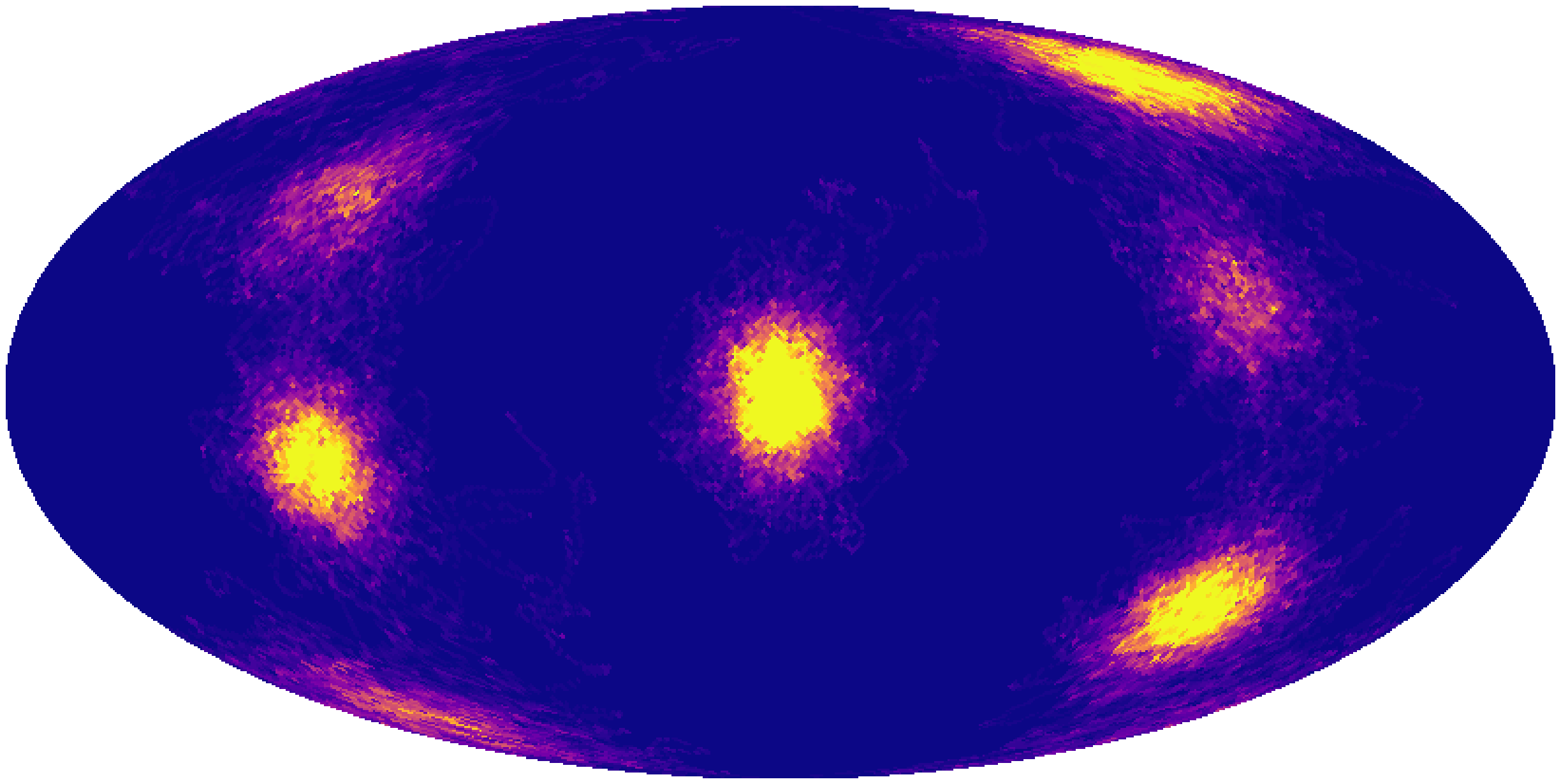}
        \put(-3,-1){\small\textbf{a2}}
      \end{overpic}
    \end{minipage}

    \vspace{1.5ex}

    \begin{minipage}[t]{0.31\linewidth}
      \centering
      \begin{overpic}[width=\linewidth]{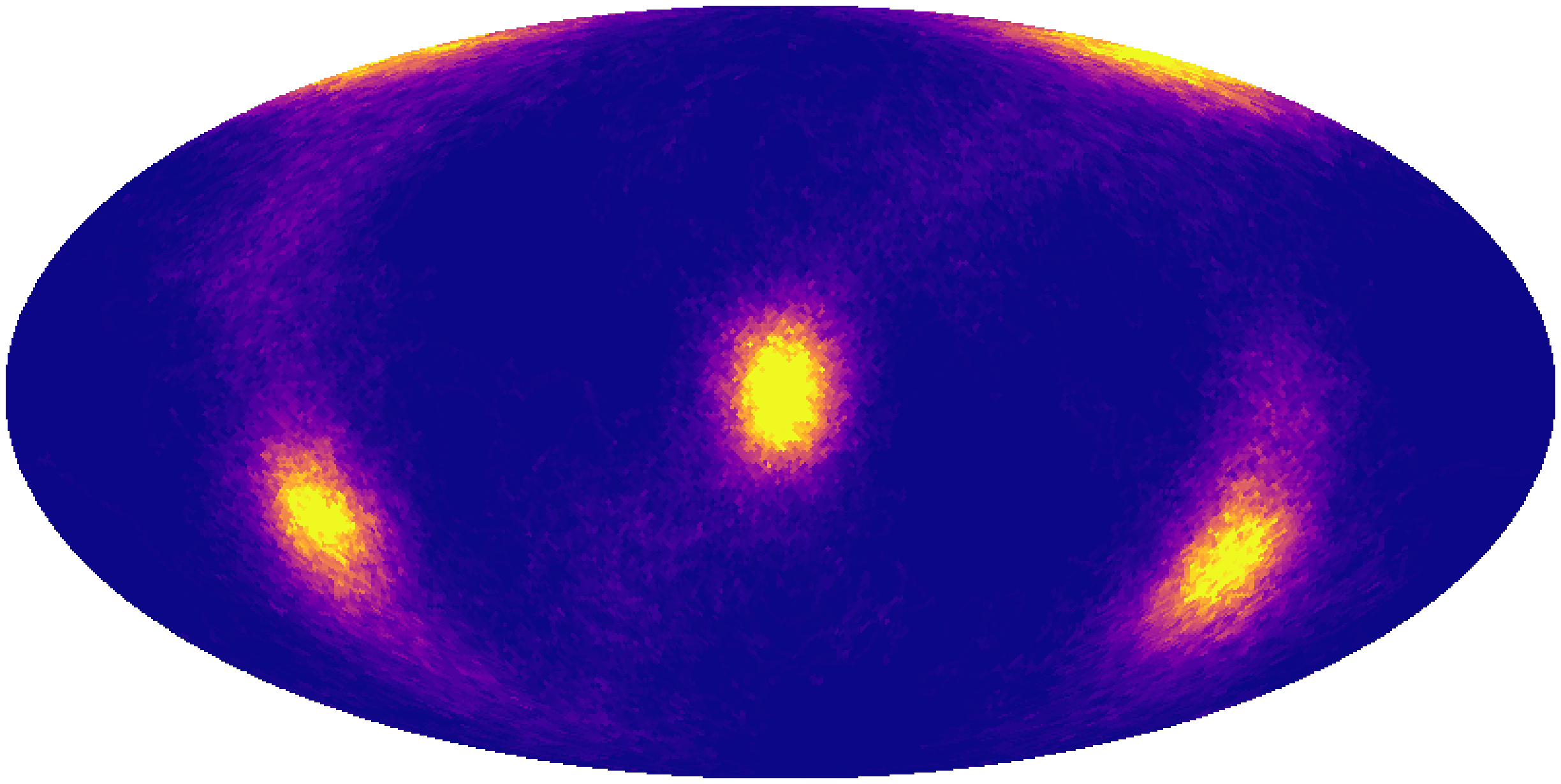}
        \put(-3,-12){\small\textbf{b1}}
      \end{overpic}
    \end{minipage}\hfill
    \begin{minipage}[t]{0.31\linewidth}
      \centering
      \begin{overpic}[width=\linewidth]{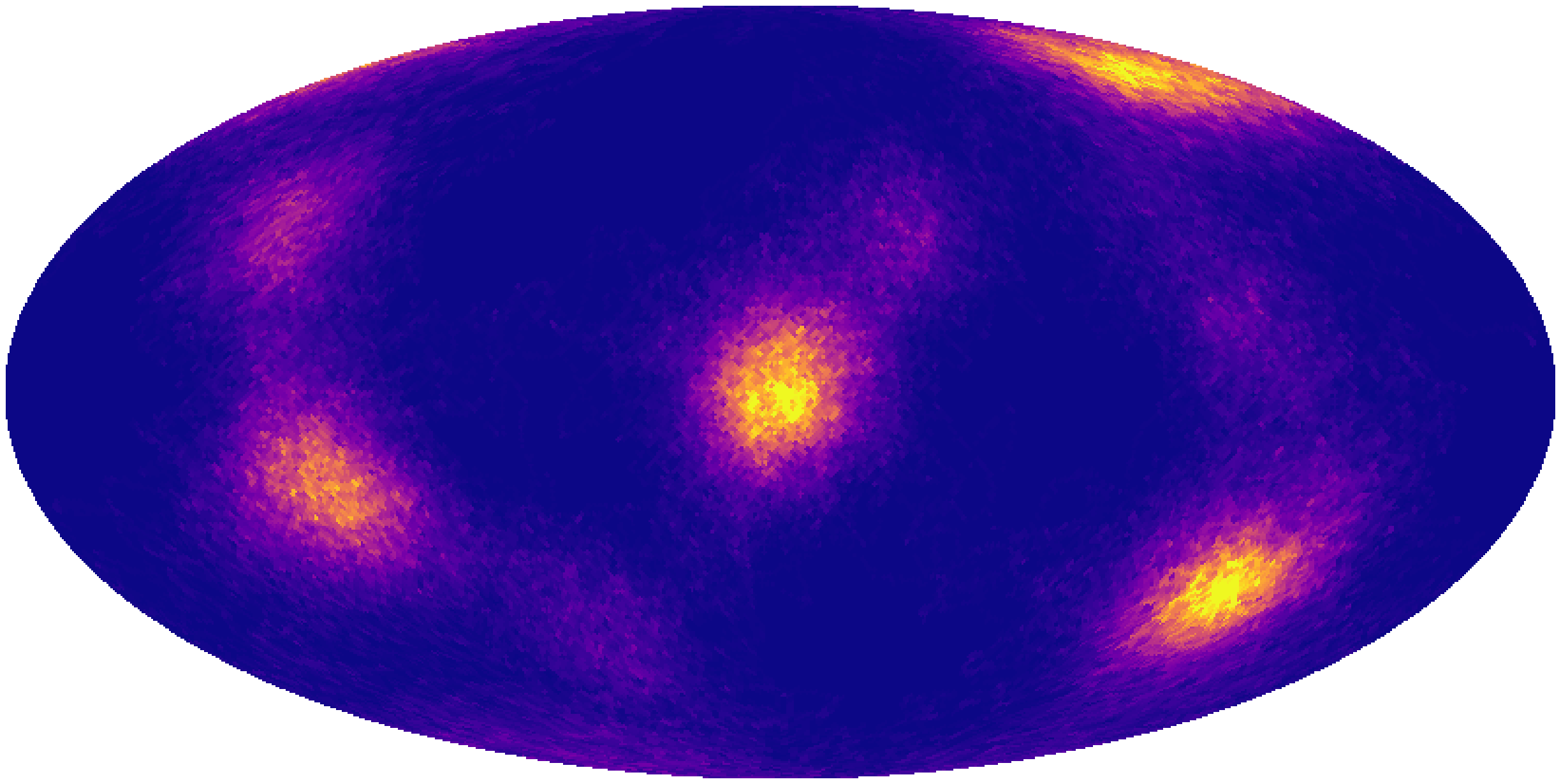}
        \put(-3,-12){\small\textbf{b2}}
      \end{overpic}
    \end{minipage}\hfill
    \begin{minipage}[t]{0.31\linewidth}
      \centering
      \begin{overpic}[width=\linewidth]{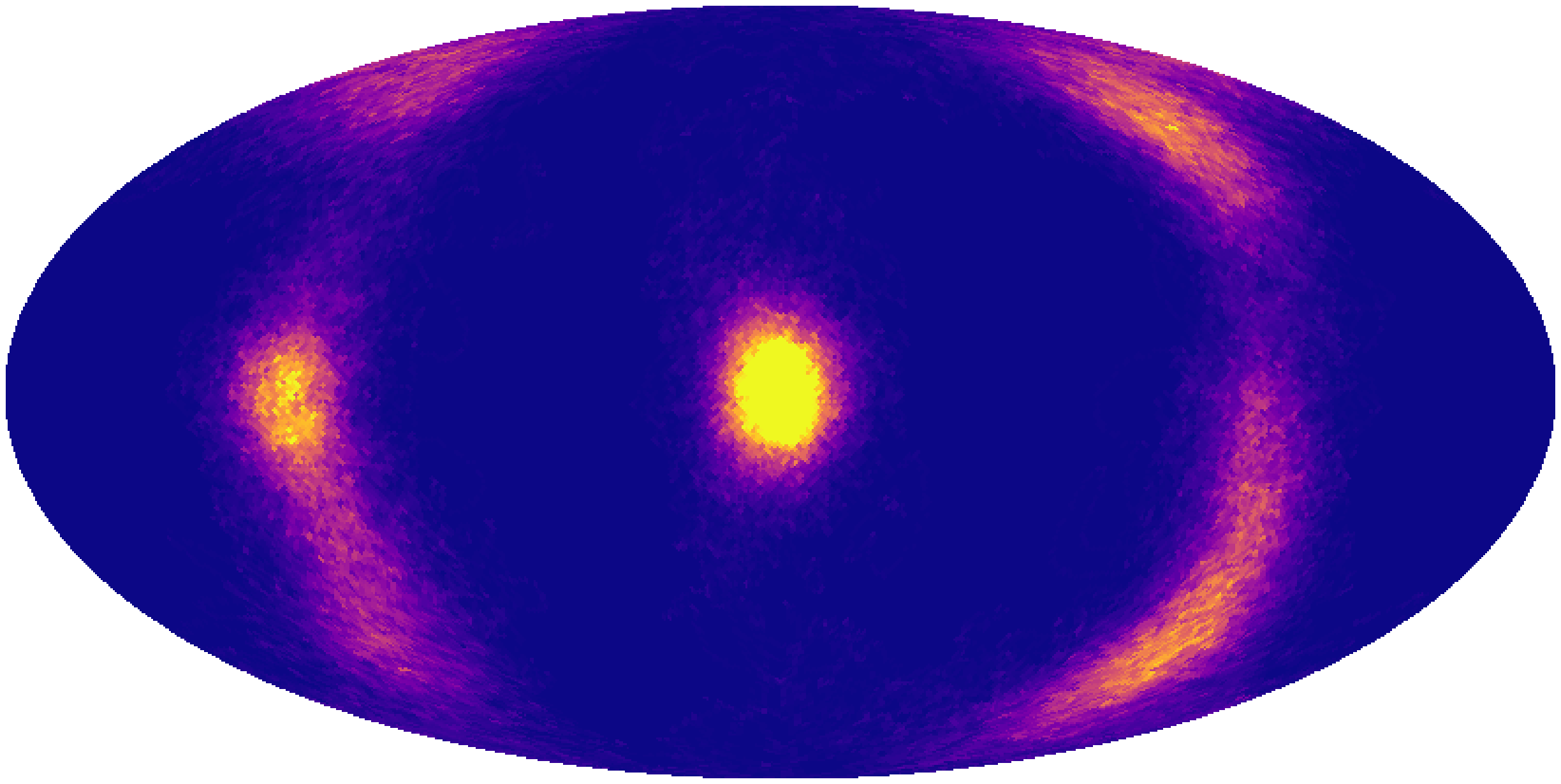}
        \put(-3,-12){\small\textbf{b3}}
      \end{overpic}
    \end{minipage}

    \vspace{1.5ex}

    \begin{minipage}[t]{0.31\linewidth}
      \centering
      \begin{overpic}[width=\linewidth]{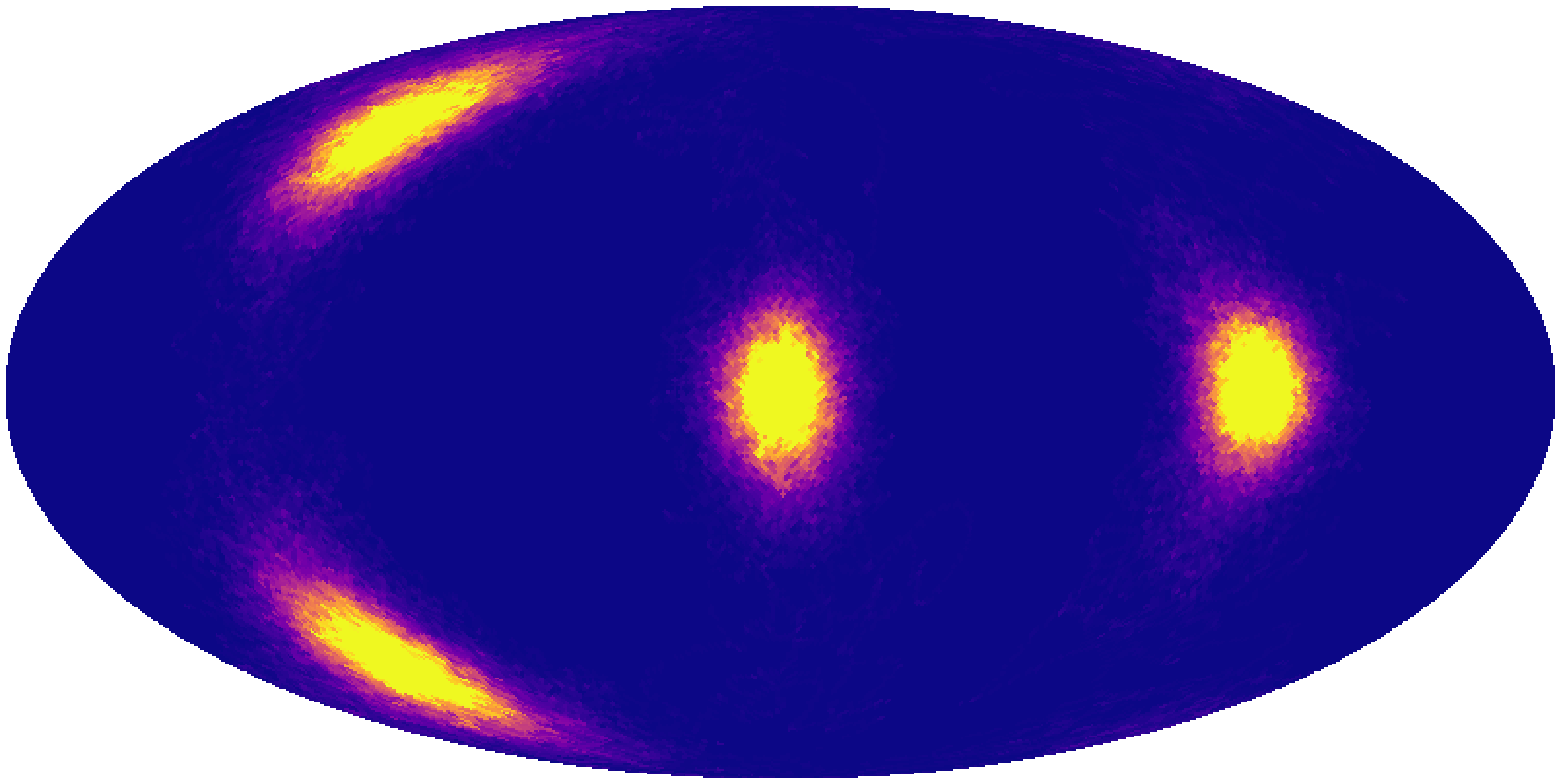}
        \put(-3,-12){\small\textbf{b4}}
      \end{overpic}
    \end{minipage}\hfill
    \begin{minipage}[t]{0.31\linewidth}
      \centering
      \begin{overpic}[width=\linewidth]{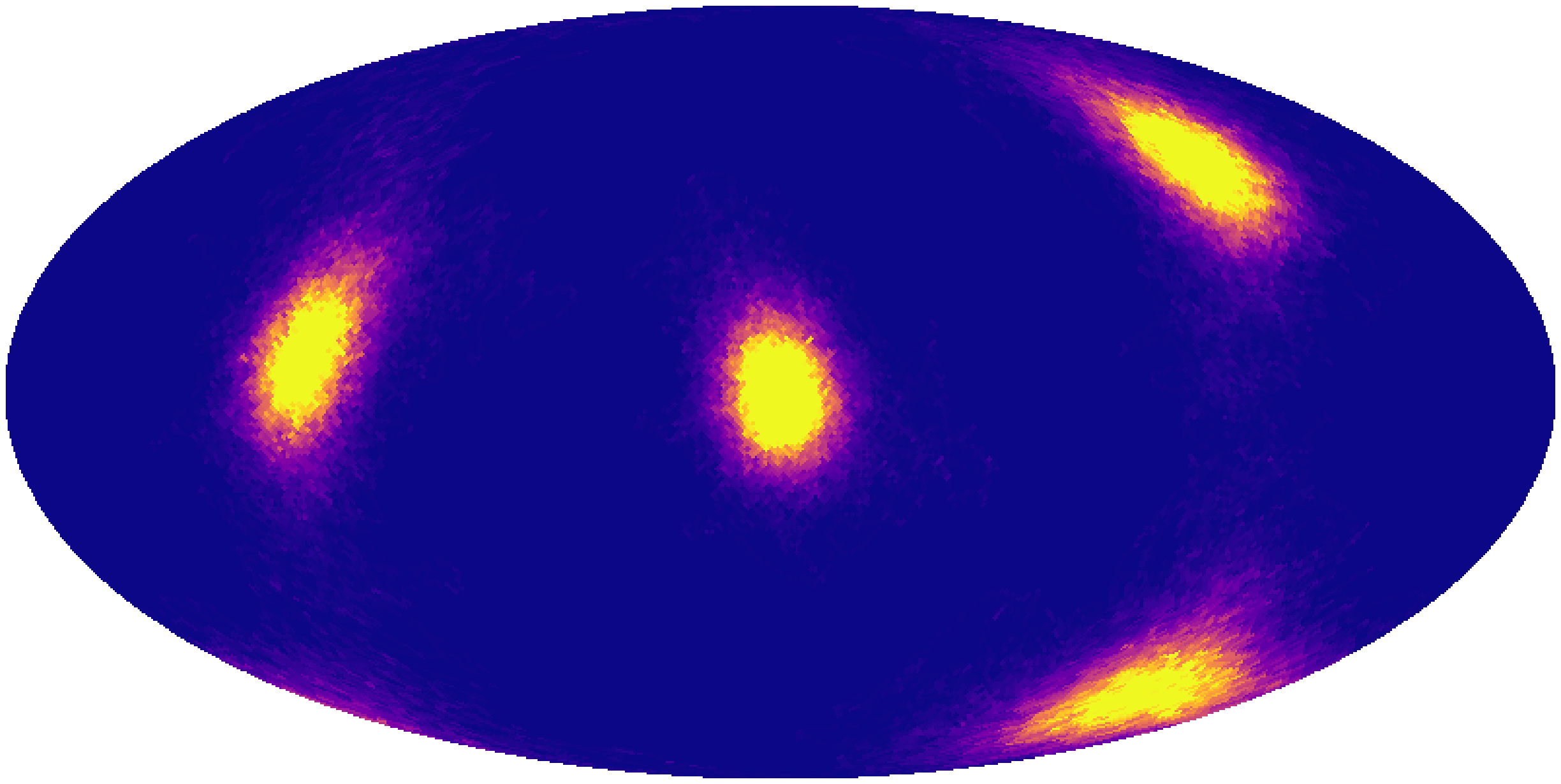}
        \put(-3,-12){\small\textbf{b5}}
      \end{overpic}
    \end{minipage}\hfill
    \begin{minipage}[t]{0.31\linewidth}
      \centering
      \begin{overpic}[width=\linewidth]{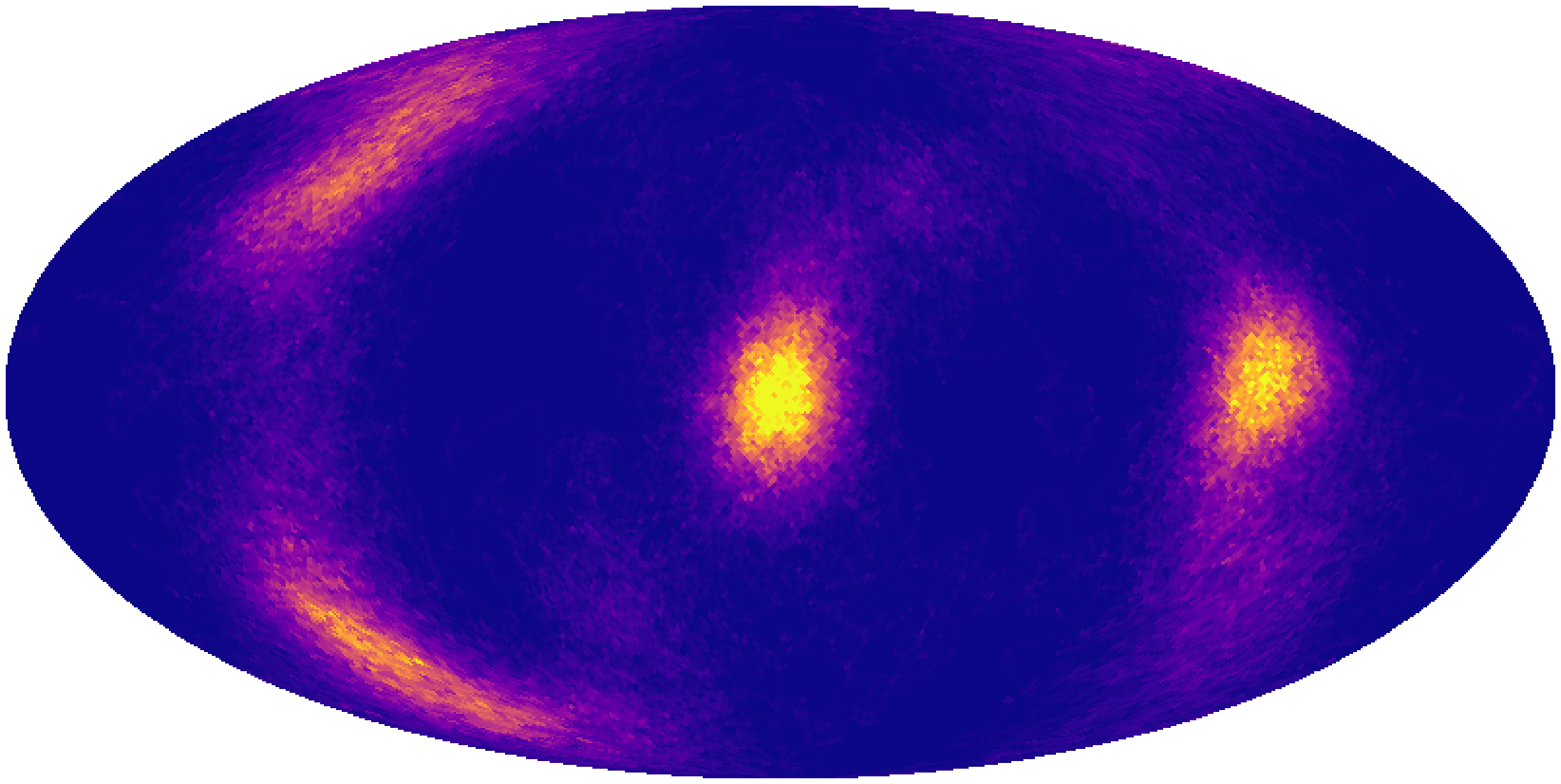}
        \put(-3,-12){\small\textbf{b6}}
      \end{overpic}
    \end{minipage}

    \vspace{1.5ex}

    \begin{minipage}[t]{0.31\linewidth}
      \centering
      \begin{overpic}[width=\linewidth]{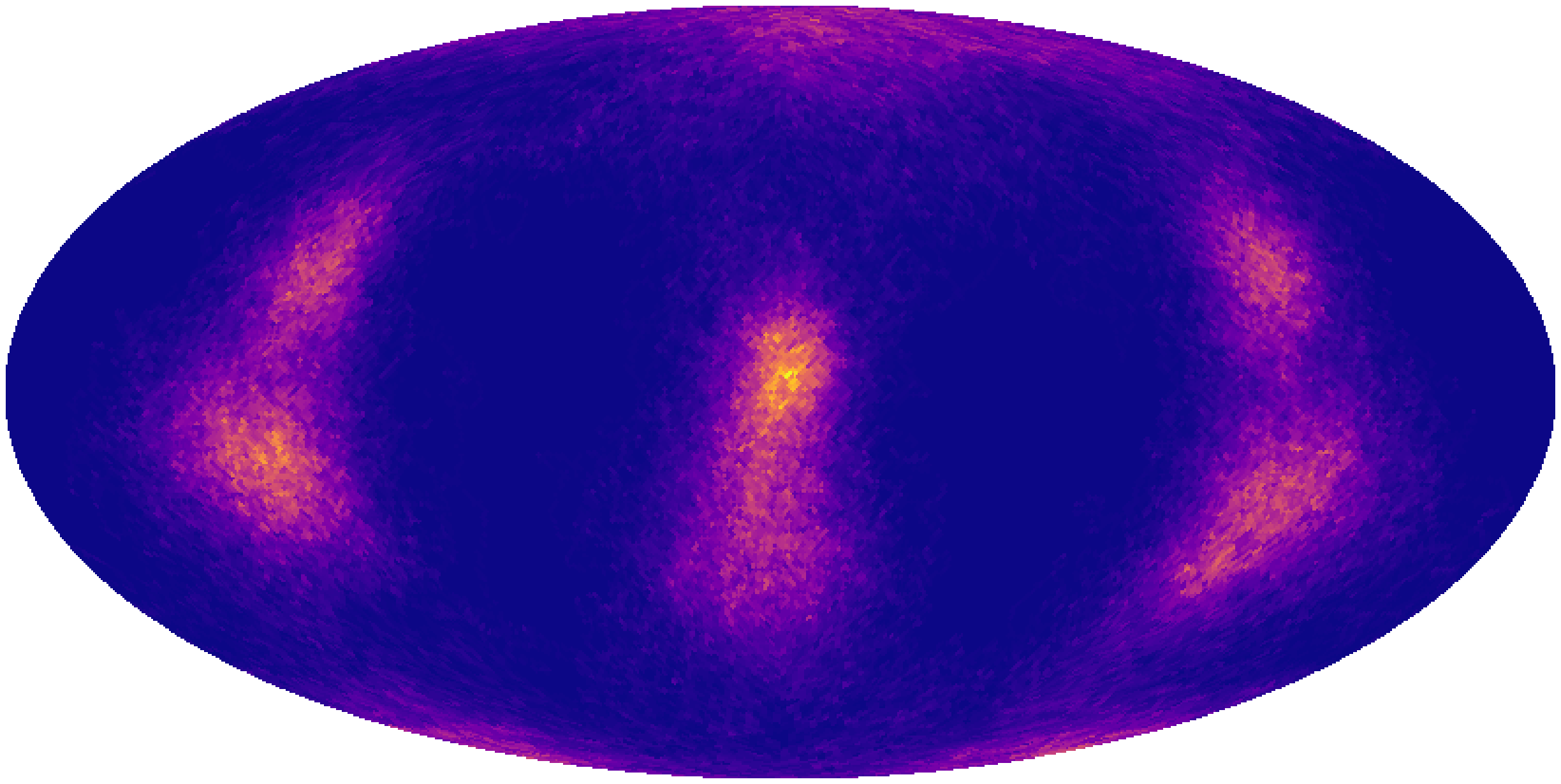}
        \put(-3,-12){\small\textbf{b7}}
      \end{overpic}
    \end{minipage}\hfill
    \begin{minipage}[t]{0.31\linewidth}
      \centering
      \begin{overpic}[width=\linewidth]{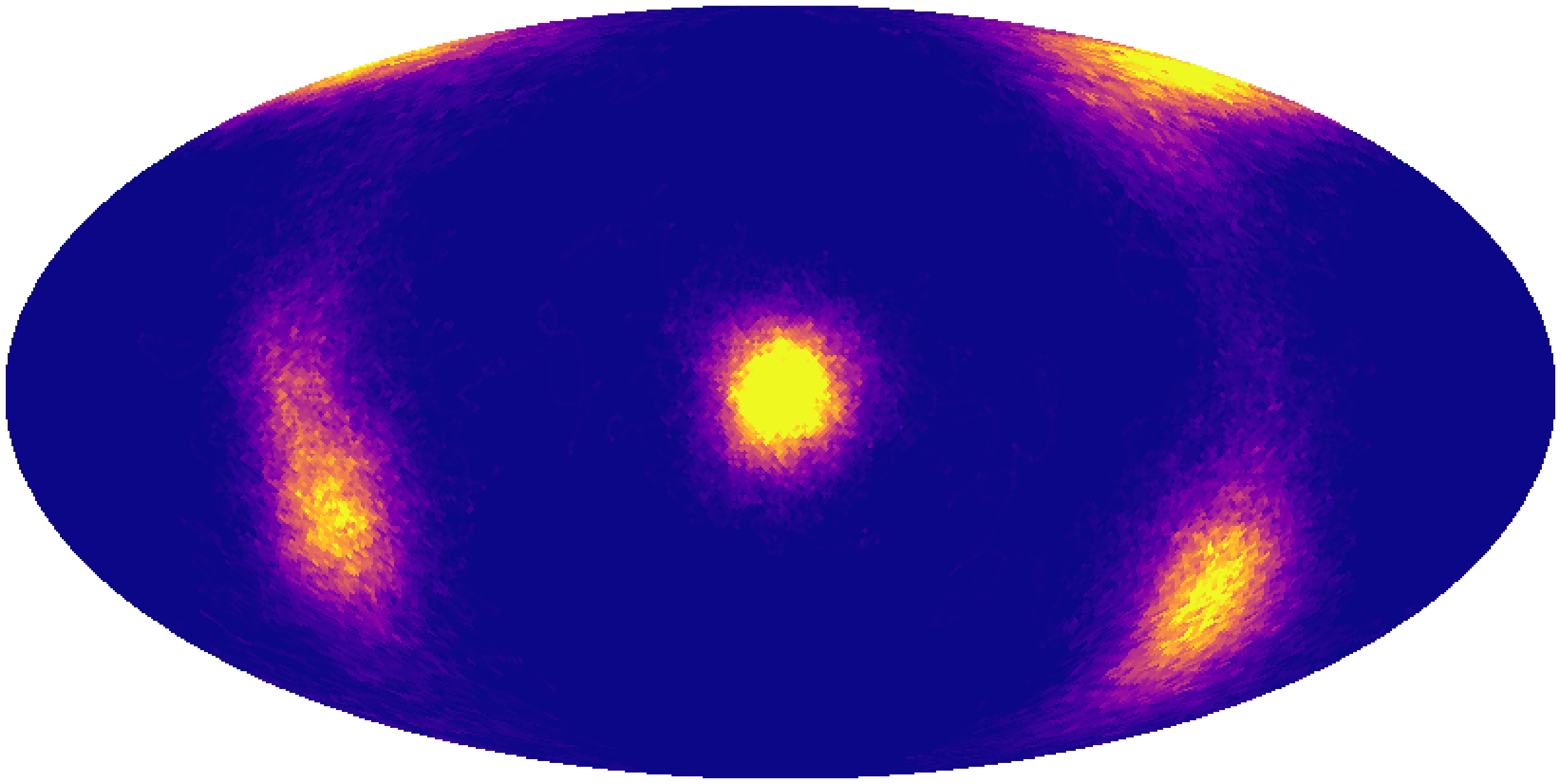}
        \put(-3,-12){\small\textbf{b8}}
      \end{overpic}
    \end{minipage}\hfill
    \begin{minipage}[t]{0.31\linewidth}
      \centering
      \begin{overpic}[width=\linewidth]{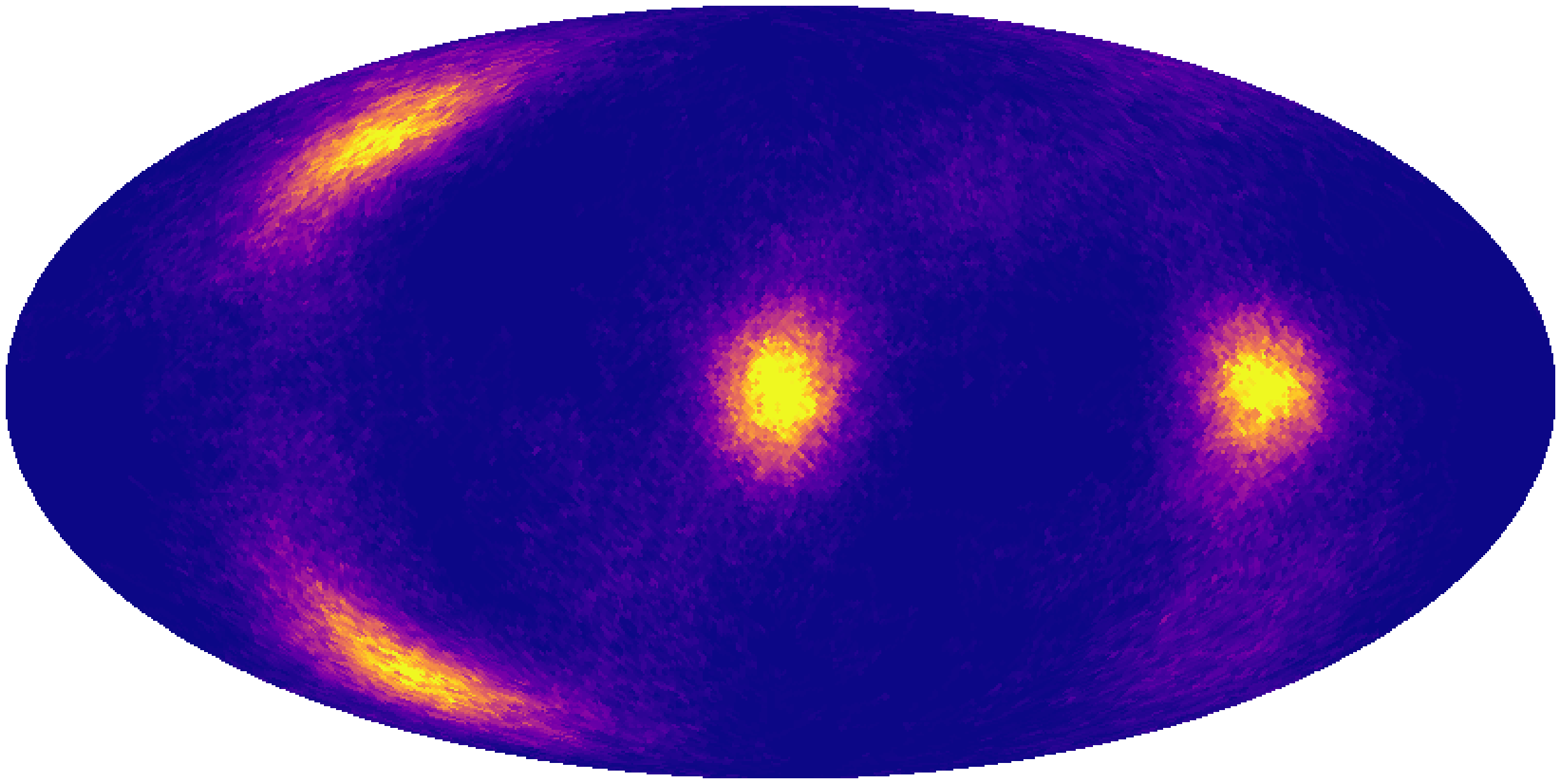}
        \put(-3,-12){\small\textbf{b9}}
      \end{overpic}
    \end{minipage}
  \end{minipage}
\end{minipage}

\caption{Phase~B (top row) and HP (bottom row). The symmetry breaking in B-HP relates $\text{2a} \rightarrow \text{a2}$; $\text{8c} \rightarrow \text{a1, b7}$; $\text{24g1} \rightarrow \text{b1, b2, b6, b9}$; $\text{24g2} \rightarrow \text{b3, b4, b5, b8}$.  Phase~B molecules on the c and g1 sites show some reorientational behavior, other sites are largely fixed.}
\label{fig:phaseB+HP}
\end{figure}

Phase~B has been reported to have the A12 $\alpha$-Mn structure, with $I\overline{4}3m$ cubic symmetry.  This structure has 58 molecules in the unit cell, on four symmetry-inequivalent molecular sites (Wyckoff 2a, 8c, 24g1 and 24g2) with coordination ranging from 12 to 16~\cite{bradley1927crystal}. All these sites have $T_d$ symmetry or lower, so a methane molecule can be placed there without breaking symmetry.  However, all $I\overline{4}3m$  structures found in constrained DFT structure searches have imaginary phonons associated with rotation on molecules on the 24g1 sites.  

The HP phase observed on compressing Phase~B is a small rhombohedral distortion from  cubic $\alpha$-Mn, to $R3$.  Although this structure has high enthalpy compared to structures discussed in~\nameref{sec:low_temp}, a combination of low ZPE and high entropy act to stabilise it at high temperatures~\cite{bykov2021structural}, and it is the most stable structure at 300~K and 25~GPa, according to quasiharmonic DFT calculation.
  
For the BOMD, we use a 290 atom quasicubic supercell of Methane HP, at 20~GPa as our initial configuration, with a rhombohedral cell with $\alpha=89.5^\circ$ angle.

We ran BOMD simulations first in the NVT ensemble for 10~ps, for equilibration. The system then evolved for 15~ps in the NPT ensemble. We used the Parrinello--Rahman barostat, allowing cell shape changes. The resulting MSD is shown in Fig.~\ref{fig:MSD+RDF}.  As for Methane I and A, the hydrogen MSD tends to the 2.5~\AA$^2$ value expected for hydrogen reorientations, however, this limit is only reached after 30~ps: the reorientation is an order of magnitude slower than phases I and A. 
An analysis of the mean hydrogen positions and the PDF in Fig.~\ref{fig:phaseB+HP} shows a preferred axis of rotation for the 24g2 site: consequently, the hydrogen MSD reaches 2~\AA$^2$ within 5~ps as three of the four C--H bonds exchange positions, but the final reorientations take an order of magnitude longer.


We have also monitored the evolution of lattice parameters during the 20~ps NPT simulation (see SM).  The ensemble-averaged  lattice parameters with time are cubic, in agreement with experiment for Phase~B.

\begin{table}[htbp]
\centering
\begin{tabular}{|c|c|c|c|c|} 
\hline
Site & Coordinates & Entropy  &  Coordination& r$_{range}$(\AA)\\
\hline
2a & 0 0 0& 0.51 &16  & 3.8 \\ 
8c &  0.317 & 0.67 & 16 & 3.63-4.12\\  
24g-1 & 0.087 0.931 0.713 & 0.69 & 13 &3.3-4.02 \\
24g-2& 0.145 0.145  0.528 & 0.44 & 12 & 3.11-4.11 \\ 
\hline
\end{tabular}
\caption{Summary of the average positions in the Phase~B simulations: each frame has $P1$ symmetry, averaged positions in $I\overline{4}3m$.  The rotational Shannon entropy, coordination and range of neighbour distances at each site are also given.}
\label{tab:phaseB}
\end{table}

\subsubsection{Phase~B and HP Structure Molecular Orientations}

Phase~B and HP have essentially identical molecular positions, differing only in the molecular orientation.  A notable feature of Phase~B (and HP) are the two Z16 Frank--Kasper (FK) supermoleculal  clusters, typical of Laves phases. These contains 17 molecules, centred on the 2a sites, with the neighbours on  24g2 and 8c sites.  The C--H bonds of the central molecule point directly towards the 8c sites, while each 24g2 molecule has a C--H bond pointing towards the centre.  As we saw in the dimer calculations, this implies a relatively long intermolecular distance compared with the shortest distances on the surface of the cluster. 
These supermolecules are arranged in a $bcc$ structure. The central site has $T_d$ symmetry, and that molecule has strong preferred orientation, as do the 24g2.

The orientation plot for Phase~B (Fig.~\ref{fig:phaseB+HP}) shows a clear difference between the various sites.  Significant rotation is observed for the 24g1  sites and the 8c sites (S=0.68$\pm$1) while the other sites remain firmly in their preferred orientation, (S=0.47$\pm$4).

The Z16 supermolecule is a feature in all of these $\alpha$-Mn type structures (Phase B, HP and the fixed-molecule $I\bar{4}3m$ structure).  The difference can be attributed to the 24g1 molecules.  In those sites, one of the C--H bonds is fixed, there are several behaviours for the remaining three:

\begin{itemize}
\item {\it Molecules are fixed in directions obeying the $I\overline{4}3m$ symmetry.}  This is possible, but high energy and not seen in BOMD, even on average.
\item {\it Molecules are freely rotating.}  This also allows $I\overline{4}3m$ symmetry, but is not observed.
\item {\it Molecules are disordered, oriented in one of two directions}: This is  Phase~B: $I\overline{4}3m$ symmetry is recovered on average by the disorder.  
\item {\it Molecules are ordered, alternating between four symmetry-breaking states.}:  This is Phase~HP.  The ordering to $R3$ breaks the 24g1 sites into four 6c sites.
\end{itemize}

\subsubsection{Bond Enthalpies, Energies  and Frustration}

In order to understand the physics of the structures, 
we consider the enthalpy of methane to be made up of contributions from neighbouring bonds. This is possible because, in all structures, there is a clear separation between the shell of nearest neighbours and the second neighbour shell. We consider enthalpy rather than energy because the different bondlengths significantly affect the density.

\begin{figure}
    \centering  \includegraphics[width=0.75\linewidth]{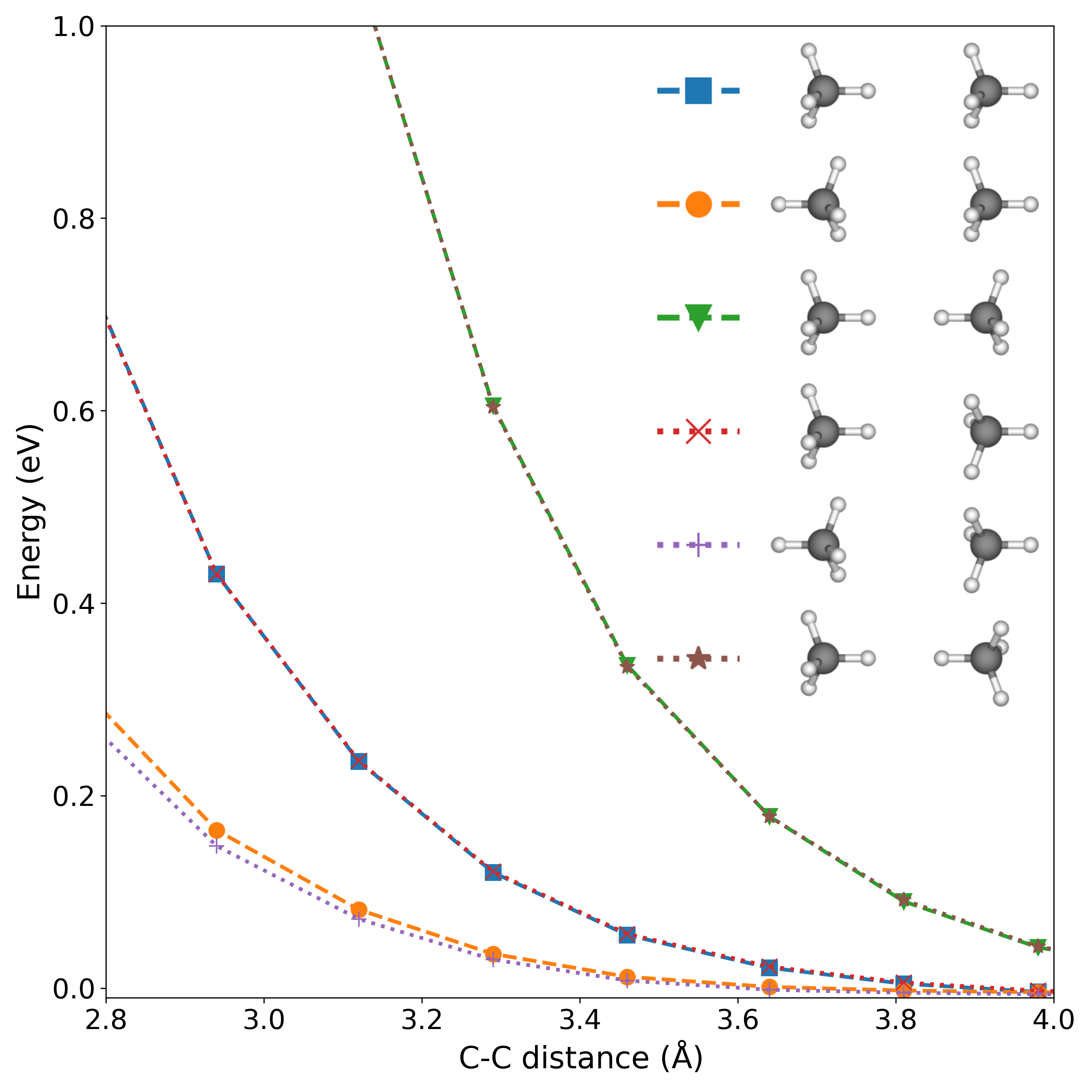}
\caption{Energy of a pair of methane molecules in vacuum as a function of the C--C separation. For the first three pairs, the molecules are eclipsed (mirrored); for the remaining three pairs, the molecules adopt a staggered configuration corresponding to a $ 60^\circ $ rotation about the separation axis. Calculations were performed with PBE in a cubic cell with side length 10~\AA{}.}

    \label{fig:dimer-msd}
\end{figure}

We calculated the energy of a methane dimer (Fig.~\ref{fig:dimer-msd}) with different orientations. At the observed separations, all ``bonds'' are repulsive. For a given a separation, there is a large energy difference between more and less favourable orientation.  For example, at 3.3~\AA\, there is a 90meV difference between the lowest energy arrangement, with molecules pointing away from each other, and a similar arrangement with one molecule pointing towards its neighbour.  The arrangement with both molecules pointing at each other is 350meV  
less stable.   Differences between staggered and eclipsed arrangements are small.

Calculating $PV$ and dividing by the bonds per cell, we find that each bond has approximately 100meV enthalpy contribution from density. Shorter bonds are assigned  lower enthalpy and the main way that bonds can be shortened is by favourable orientation of the molecules.  However, it is impossible to be favourably oriented with respect to all 12-16 neighbouring molecules, and we can expect all bonds to have similar energy,  so there is frustration and a range of bondlengths.

\begin{table}[htbp]
\centering
\begin{tabular}{|c|c|c|c|c|c|c|c|} 
 \hline
Symmetry & Atoms &   Name & Phase  & P (GPa) & T (K) & Notes & Enthalpy (eV/atom)  \\  
\hline

$I\overline{4}3m$  & 145 &  $\alpha$-Mn  & B & 20 & 0 & PBE &-6341.09617 \\
relaxed  & 145 &  d-$\alpha$-Mn  & B & 20 & 0 & PBE & -6342.35000\\ 
MD  & 145 &  d-$\alpha$-Mn  & B & 20 & 70 & PBE/0.1~ps & -6342.55259 \\ 
R3 & 145 &  HP-phase  & HP & 20 & 0 & PBE & -6342.81853 \\ 
\hline
\end{tabular}
\caption{Summary of 29-molecule structure calculation at 20GPa for $\alpha$-Mn type structure $I\overline{4}3m$ is relaxation with the full symmetry enforced - the structure with imaginary phonon modes.
``Relaxed'' relaxes the molecules along those modes to find a disordered local minimum, while ``MD'', enthalpy is relaxation of an BOMD snapshot with the unit cell and all atoms allowed to relax to a disordered minimum; $R3$ is the ordered arrangement seen in the HP phase. }
\label{hkls-p1}
\end{table}

\subsubsection{Phase~B Simulated Diffraction Patterns}

To assess which candidate structures are consistent with the reported Phase~B diffraction data, we simulated powder X-ray diffraction patterns directly from BOMD trajectories, following the same workflow as for Phase~A. We compare the generated patterns to the experimental X-ray pattern measured at 8~GPa and 300~K~\cite{maynard2014}.

\begin{figure}[htbp]
    \centering
    \includegraphics[width=0.48\columnwidth]{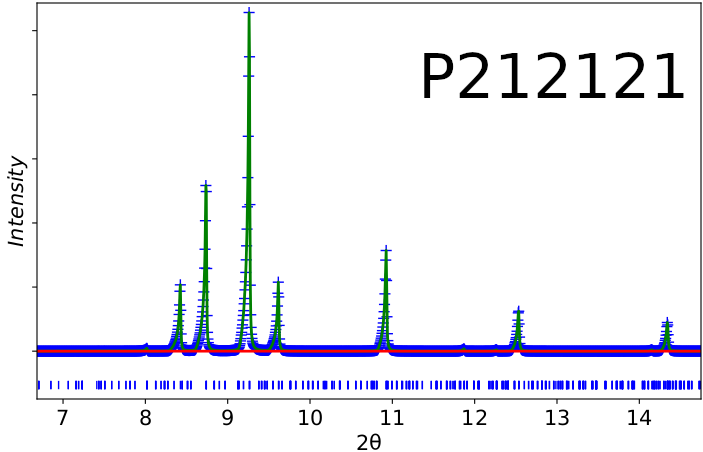}
    \includegraphics[width=0.48\columnwidth]{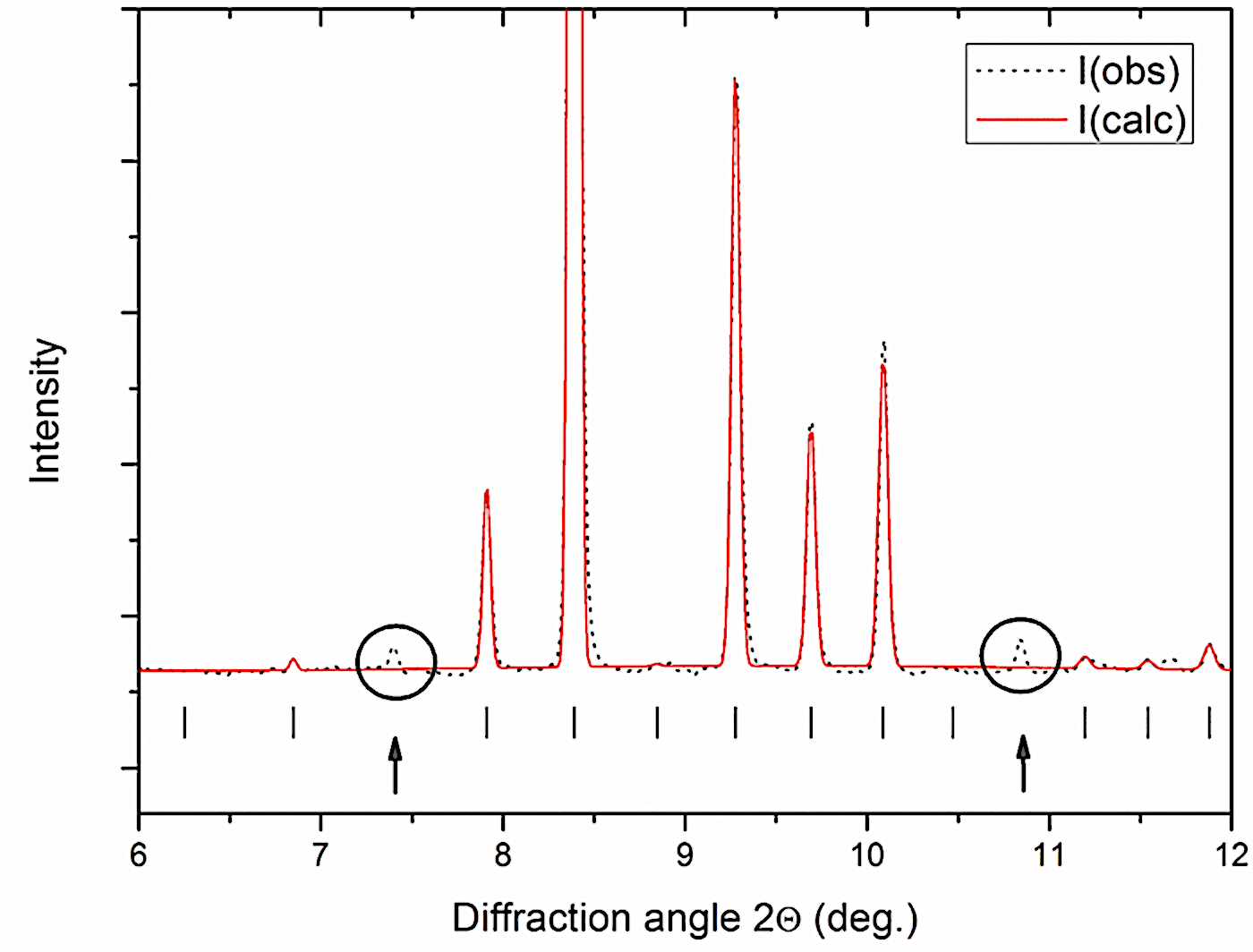}
    \includegraphics[width=0.48\columnwidth]{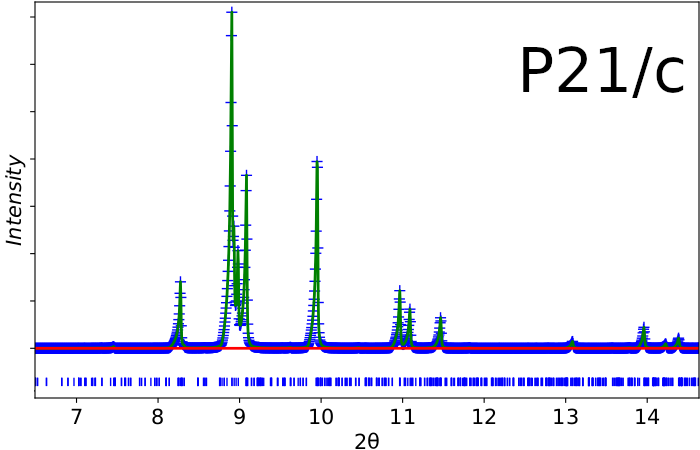}
    \includegraphics[width=0.48\columnwidth]{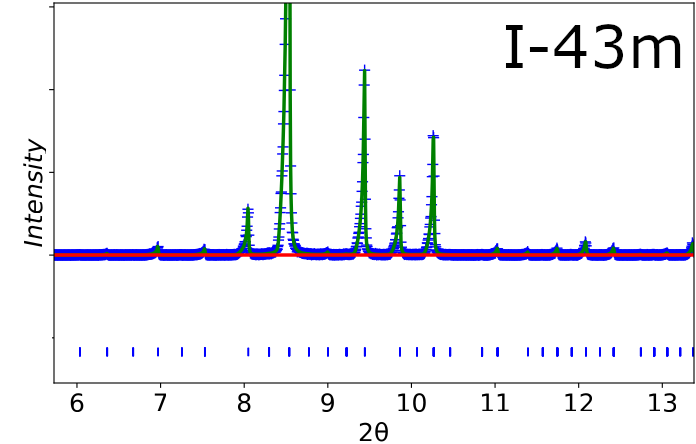}
    \caption{Comparison of simulated and experimental X-ray diffraction patterns for methane Phase~B over a short $2\theta$ range selected to highlight discriminating reflections (full range in SM). Simulations are generated from 50 BOMD snapshots combined into an effective $P1$ cell and processed with GSAS. Top left: $P2_1 2_1 2_1$ candidate~\cite{umemoto2002x}. Top right: experimental pattern at 8~GPa and 300~K from Maynard-Casely \textit{et al.}~\cite{maynard2014}. Bottom left: $P2_1/c$ candidate~\cite{umemoto2002x}. Bottom right: $I\overline{4}3m$ ($\alpha$-Mn/A12) model, which provides the closest overall match in this region.}

    \label{fig:phaseBshort}
\end{figure}

Figure~\ref{fig:phaseBshort} shows a restricted $2\theta$ interval chosen because it contains several reflections that are particularly sensitive to the underlying carbon sublattice in methane Phase~B. Alongside the $I\overline{4}3m$ (A12, $\alpha$-Mn type) model, we include two low-symmetry candidate structures previously proposed as low-temperature alternatives~\cite{umemoto2002x}, $P2_1 2_1 2_1$ and $P2_1/c$, simulated here under the same finite-temperature BOMD protocol.

Within this range, the $I\overline{4}3m$ model reproduces the experimental pattern significantly better than either low-symmetry candidate. In particular, it captures the observed peak positions and their grouping in a way that is consistent with a near-cubic cell, whereas the $P2_1 2_1 2_1$ and $P2_1/c$ candidates produce additional features and intensity redistribution that are not supported by the data (see \nameref{sec:low_temp} for simulation details). These discrepancies are not subtle: they appear as distinct extra reflections or peak splittings in the simulated patterns that would be readily detectable in the measured trace over the same angular range.

\subsection{Supermolecules: Quasicubic phases with odd molecule counts}

Although Phase~A does not have cubic symmetry, the rhombohedral angle is only about 1$^\circ$ away from 90.  Phase~B is cubic, but with a large prime number of molecules in the unit cell, and distorts by 1-2$^\circ$ to rhombohedral in the HP phase.  
These features can be resolved by considering supermolecular clusters.  

Phase~A has a 13 atom supermolecular cluster: the 1b atom has 12 neighbours in an icosahedral arrangement.  
Molecules on the surface of the icosahedron have C--H bonds pointing directly away from the central site, consequently intermolecular distance from the centre are smaller than those on the surface, which facilitates packing 12 neighbours.

Phase~B, has a 17 molecule cluster with $T_d$ symmetry centred on the a site which points directly at the four c sites. The 12 neighbours on the g2 site have C--H bonds pointing to the centre. Consequently, the intermolecular distances from the centre are longer than those on the surface, which facilitates packing 16 neighbours.

In Phase~A, the supermolecules form a cubic lattice while
the remaining eight molecules (2c and 6f1)  forming a distorted cube.  One can then approximate Methane-A as supermolecules in the Strukturbericht B2 (CsCl) structure, with a nearly regular icosahedron of 13 molecules at the cube centre, and a distorted cube of 8 molecules at the corner.  The $R\overline{3}$ is the highest  symmetry which can be obtained with a regular icosahedron in a cube: a threefold axis along (111) and an inversion centre.  The  cube distorts into $R\overline{3}$  with the two molecules on the threefold axis being different from those on the 6f1 site.

Turning now to Phase~B, we see that the 2a site has 16 near neighbours. This forms a 17 molecule cluster with $T_d$ symmetry centred on the 1a site and its 16 neighbours.  Such a structure is ``topologically close packed'' and is a well-known efficient packing with two different sized atoms.   Unlike the Laves phases, in Methane B all molecules are the same type: the a range of bondlengths introduces different sizes.  All intermolecular distances involving the central Z16 molecule have a C--H bond pointing directly along them; consequently, these distances are longer than the ones between Z16-surface molecules.  The central molecule is effectively bigger, which facilitates packing with the higher coordination number. 

The Z16 polyhedron can be visualised as four triangles and four hexagons,  capped with an additional molecule.  The C--H bonds  point to the cap of the hexagons. Furthermore, the Z16 clusters are organised on a $bcc$ structure (A2). 
The remaining 12 molecules (g2 sites) form a network of corner-sharing tetrahedra in the 12 tetrahedral interstices of the $bcc$ lattice $ (0, \frac{1}{4}, \frac{1}{2}) $.  

Thus the mysterious 21 molecule Phase~A is revealed as 13+8:
the B2 structure with Me$_{13}$Me$_8$. The
29-molecule cell is revealed as 17+12: the A2 structure with Me$_{17}$ with 12 filled tetrahedral interstitial sites.
Curiously, the small distortions from cubic in Phase A are because the supermolecule in incompatible with cubic symmetry, whereas in Phase HP it is the interstitial molecules which are unstable in their high-symmetry configuration.

\begin{figure}[htbp]
\centering

\begin{minipage}[c]{0.22\textwidth}
\centering
\begin{overpic}[width=\linewidth]{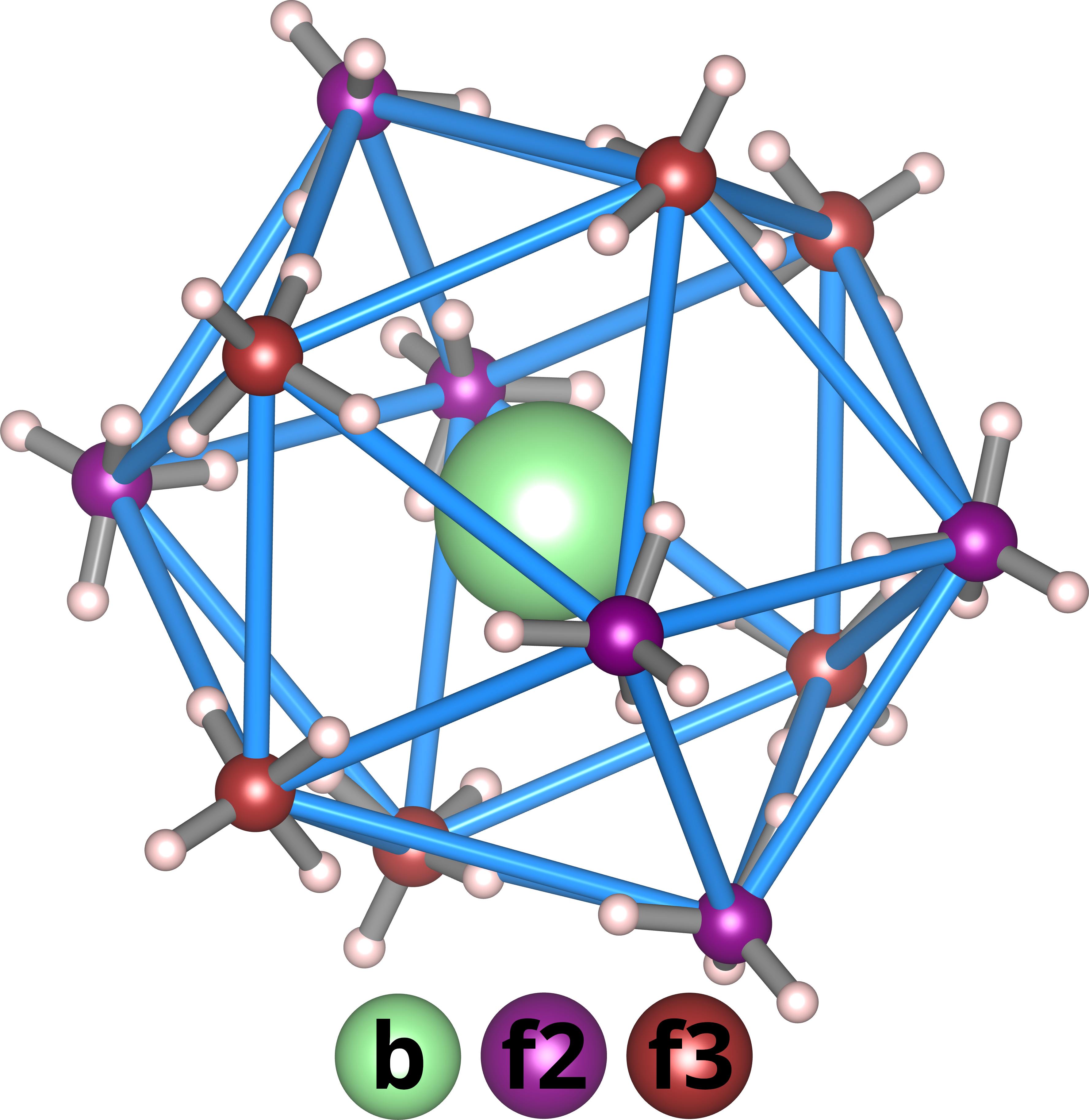}
\put(0,90){\small\textbf{A}}
\end{overpic}
\end{minipage}\hfill
\begin{minipage}[c]{0.22\textwidth}
\centering
\begin{overpic}[width=\linewidth]{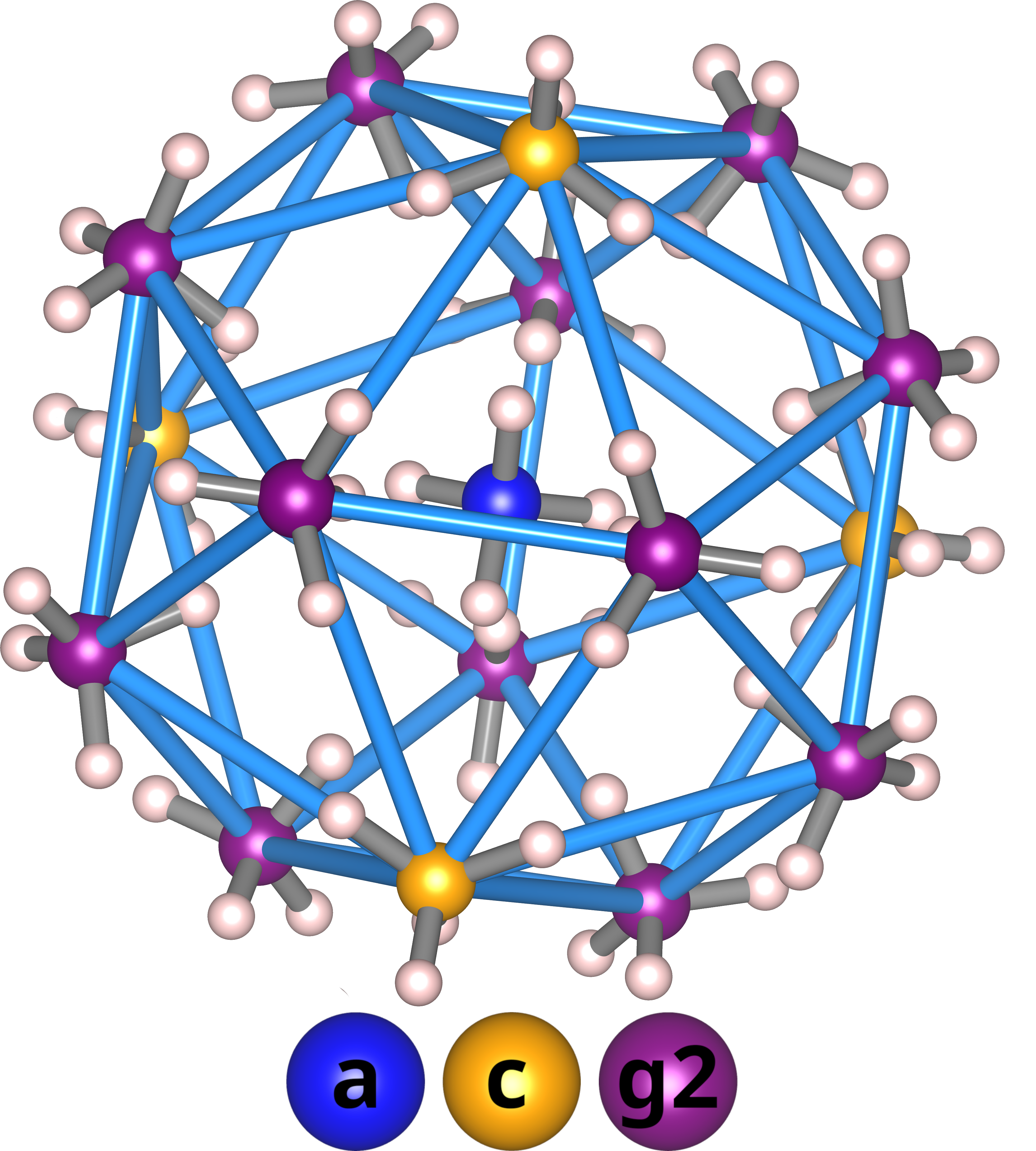}
\put(0,90){\small\textbf{B}}
\end{overpic}
\end{minipage}\hfill
\begin{minipage}[c]{0.22\textwidth}
\centering
\begin{overpic}[width=\linewidth]{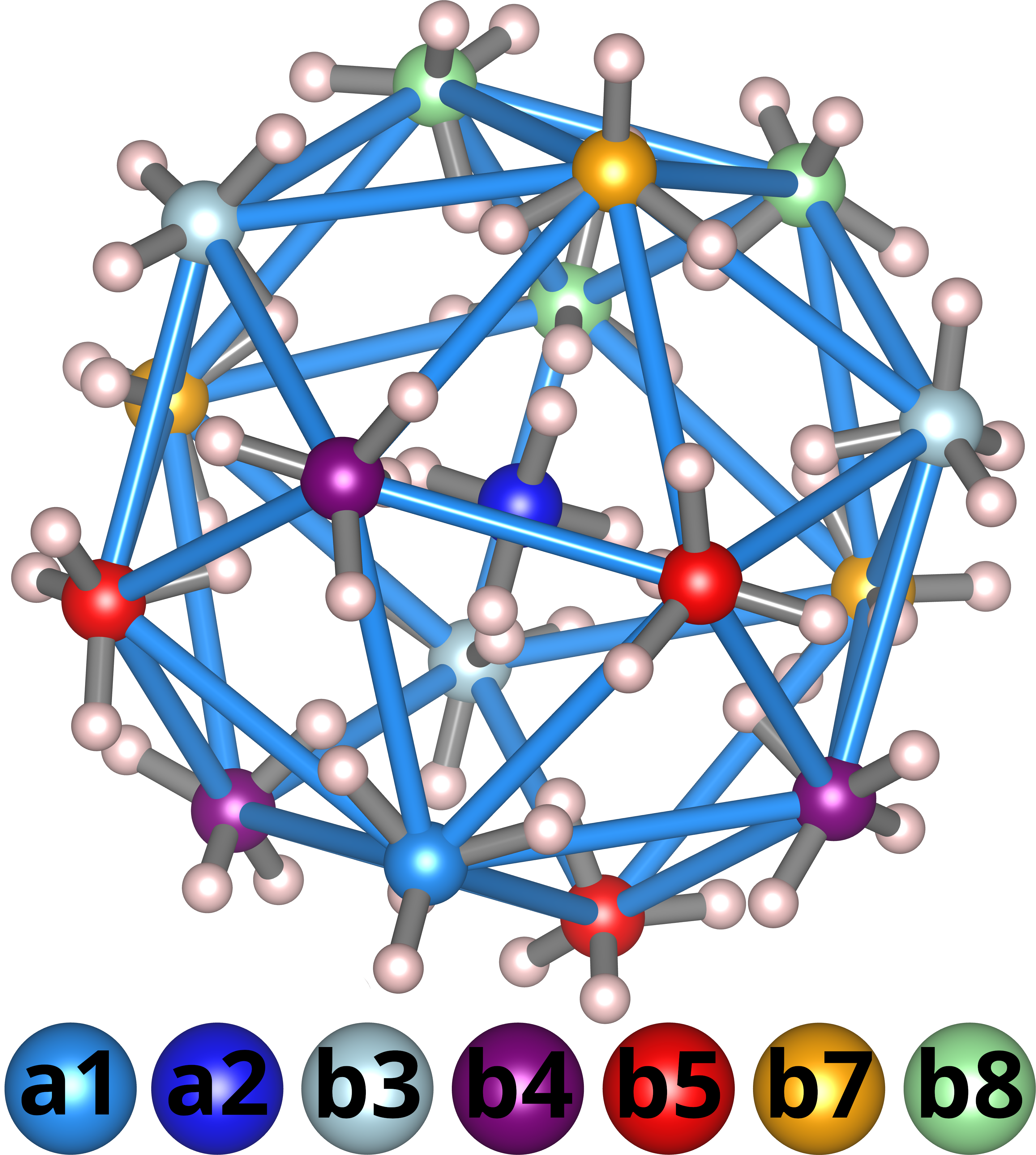}
\put(0,90){\small\textbf{HP}}
\end{overpic}
\end{minipage}\hfill
\begin{minipage}[c]{0.22\textwidth}
\centering
\begin{minipage}[c]{\linewidth}
\begin{minipage}[c]{0.48\linewidth}
\centering
\begin{overpic}[width=\linewidth]{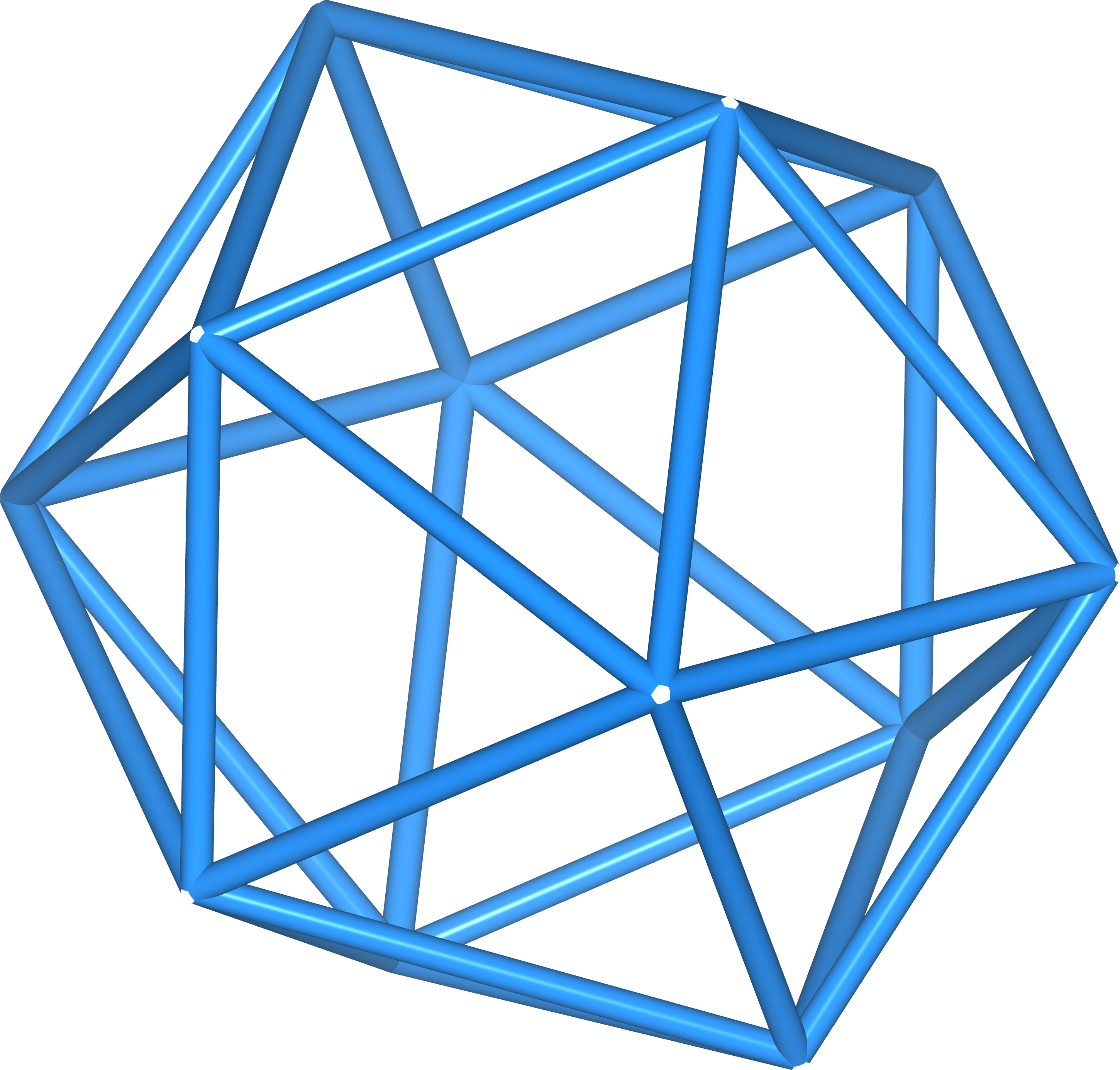}
\put(0,87){\small\textbf{A}}
\end{overpic}
\end{minipage}\hfill
\begin{minipage}[c]{0.48\linewidth}
\centering
\begin{overpic}[width=\linewidth]{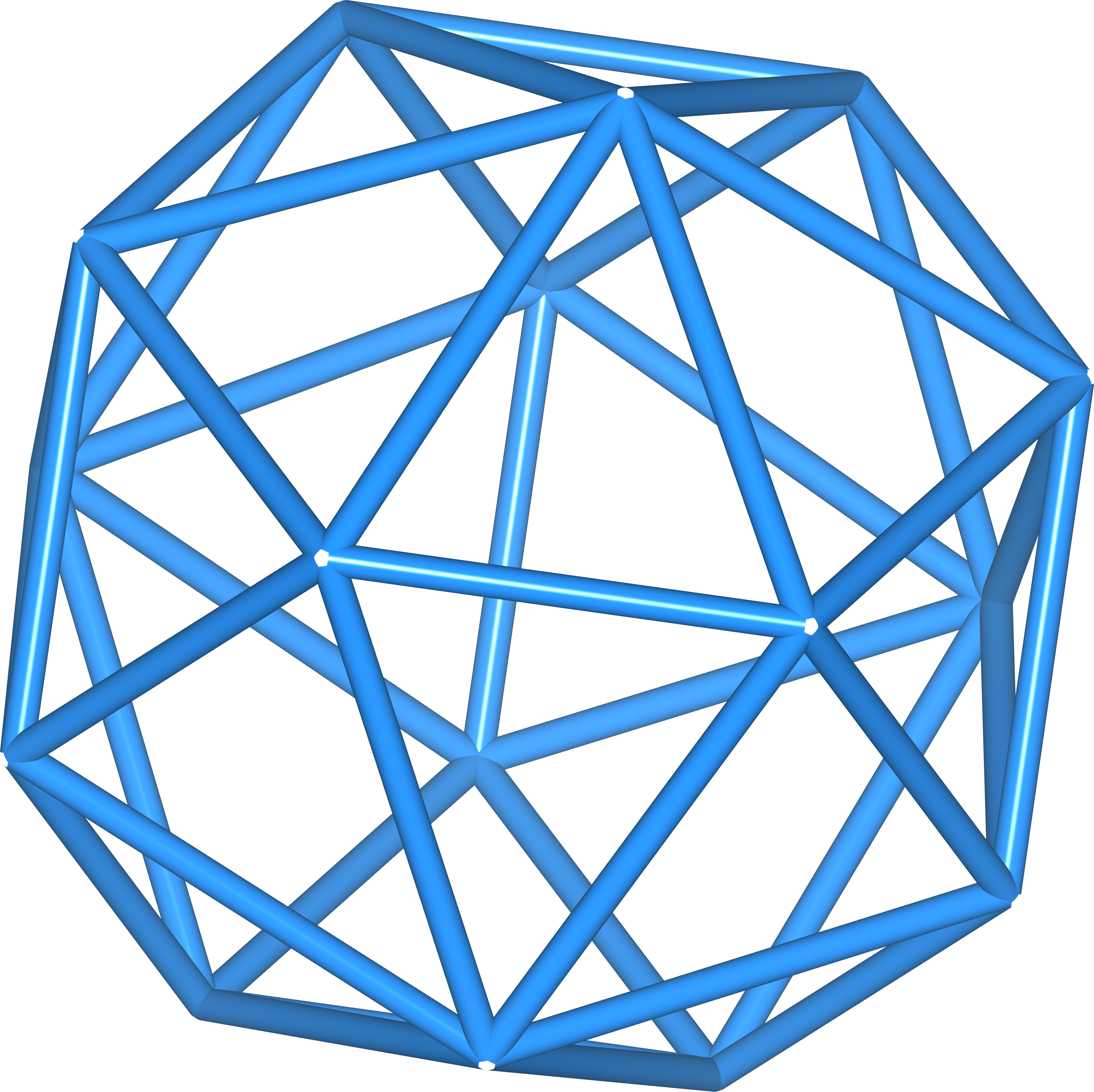}
\put(0,90){\small\textbf{B}}
\end{overpic}
\end{minipage}

\end{minipage}

\begin{minipage}[c]{\linewidth}
\begin{minipage}[c]{0.48\linewidth}
\centering
\vspace{2ex}
\begin{overpic}[width=\linewidth]{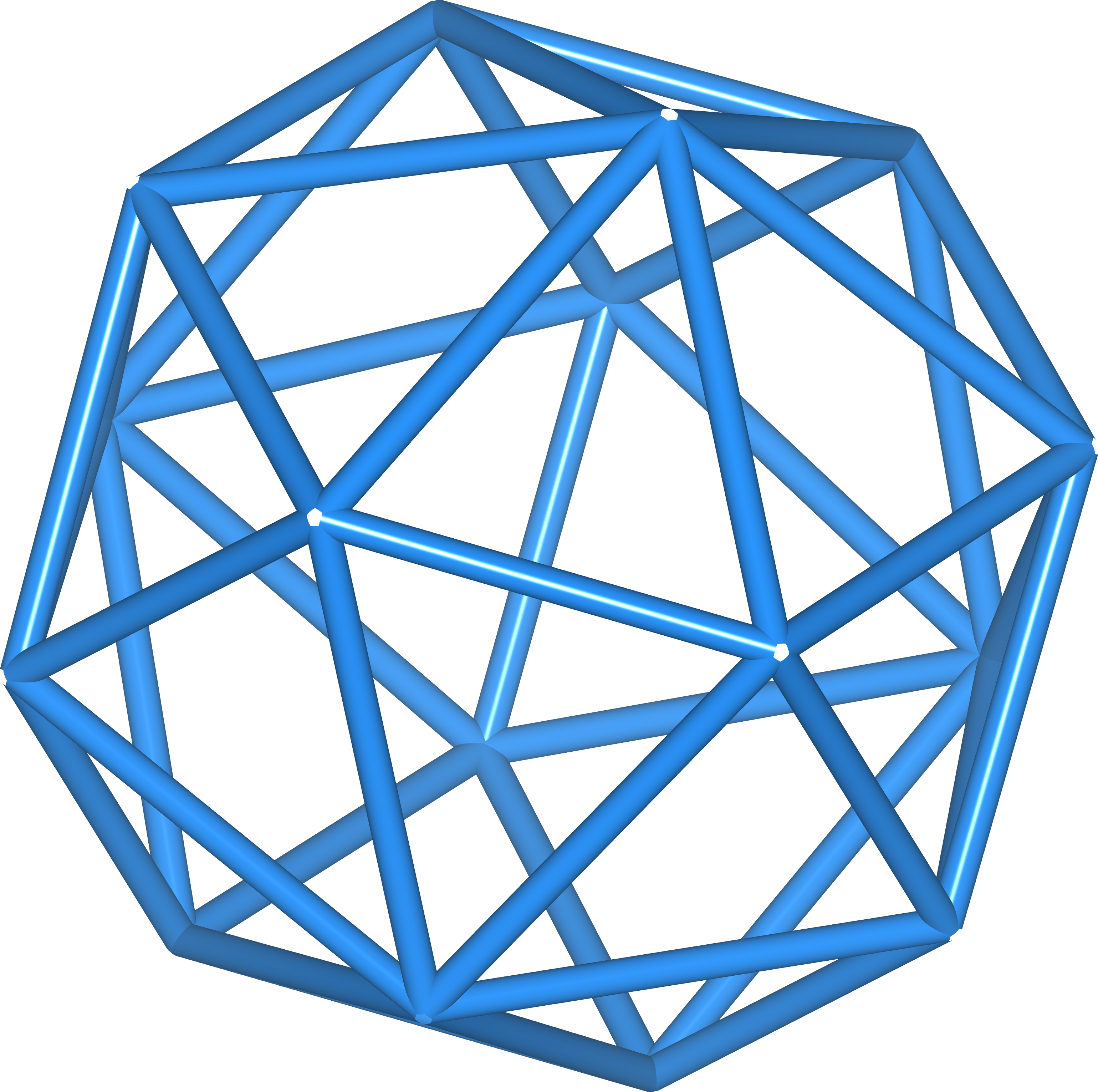}
\put(0,96){\small\textbf{HP}}
\end{overpic}
\end{minipage}
\end{minipage}
\end{minipage}

\caption{The supermolecular clusters labelled with the phase in which they occur. Different colours indicate carbon atoms of different Wyckoff sites. The large green sphere in A indicates that the molecule on the $1b$ site  is rotating.  In A, the molecules are oriented with outward-point C--H bonds.  In B and HP, they are oriented with outward-pointing C--H bonds.  The B and HP Z16 clusters are essentially identical, both in molecular positions and orientations. The differences between these phases arise from rotations of the non-Z16 molecules, which affects symmetry naming/coloring of the Wyckoff sites.  }
\label{fig:supermol}
\end{figure}

The near-spherical supermolecular objects are shown in figure \ref{fig:supermol}.

\subsection{Low Temperature Structures} \label{sec:low_temp}

Extensive structure searches~\cite{gao2010, conway2019, bykov2021structural} identify a set of zero-temperature stable structures that differ from those observed experimentally at finite temperatures. 
These calculations were initially done with PBE, but we have verified that the stability is independent of exchange-correlation functional, and whether the runs are done with or without van der Waals corrections. 

By contrast, the lowest enthalpy structures have much more homogeneity in intermolecular spacing.  In 
$P 2_1 2_1 2_1$ and $P2_1/c$ all the molecules in the unit cell are symmetry-equivalent.
In $P 2_1 2_1 2_1$ 
every molecule has four closest neighbours calculated at 2.743~\AA (50GPa), arranged in a tetrahedron such that the C--H bond points away from the neighbours, but with a variety of dihedral angles.  
By contrast each molecule in $P2_1/c$ has one nearest neighbour at 2.626~\AA, with further neighbours at  2.745,  2.834, 2.839, and 2.910.  The significantly shorter distance in  $P2_1/c$ is associated with eclipsed dihedral angles. This complexity arises from the wide range of enthalpies (bondlengths) for bonds with similar energy, and the impossibility of orienting molecules favourably with respect to all their neighbours.

Static lattice DFT finds lowest enthalpy for a  $P 2_1 2_1 2_1$ structure below 5~GPa
and a   $P2_1/c$ (the SnBr$_4$ structure) between 5-40GPa.
However, lattice dynamics calculations demonstrate that zero point energy is already enough to destabilize the  $P2_1/c$ structure with respect to  $P 2_1 2_1 2_1$ at 25~GPa. At room temperature the HP structure, becomes stable.
Lattice dynamics does not properly describe rotational behaviour, but this indicates the importance of entropy in determining stability.

We have investigated these phases, with further BOMD simulations of $P2_1/c$ and $P 2_1 2_1 2_1$ CH$_4$ supercells with 60 and 64 molecules respectively, as detailed in SM table. In this case, simulations were done at 20~GPa and the MSD of the simulations at $T=300$~K is shown in SM. After 15~ps, we increased the temperature of $P2_1/c$ to 750~K and ran it for an extra 12~ps. The MSD of the final ps is shown in SM.

While the displacement of carbon seems small in both structures, a closer look at the trajectory reveals a change in symmetry in both cases. In the case of $P2_12_12_1$, the average carbon positions symmetrise to a $Pnma$ sublattice (tolerance of 0.05~\AA). The shift in atomic positions is  clearly seen in SM. The larger displacements of carbons in $P2_1/c$ methane do not result in a change to a high symmetry position, although the molecules reorient so as to exchange their hydrogen positions.   On heating still further, to 750~K, the  carbons will adopt a $C2/m$ structure and we interpret this observation as a transition to a phase-I-like plastic phase within the constraints of the chosen BOMD cell.
There are qualitative differences in between $P2_1/c$ and $P 2_1 2_1 2_1$. In both cases, both C and H start vibrating about their initial positions, and it takes about 2~ps for the molecules to reorient significantly. At that point, the monoclinic $P2_1/c$  phase will become plastic in another 2--4~ps, much slower than the rhombohedral $P 2_1 2_1 2_1$.

\section{Discussion and Conclusions}

We have combined extensive static DFT calculations with Born--Oppenheimer molecular dynamics (BOMD).  The central outcome is that methane's ``complex'' high-pressure phases are best understood as entropy-stabilised, frustrated packings of near-spherical \emph{multimolecule building blocks} (supermolecules), rather than as straightforward orientational orderings of single CH$_4$ molecules on a simple lattice.

Our novel ADF/Mollweide analysis reveals a range of molecular motions.  In Phase I the molecules rotate freely, but strongly avoid certain directions.  In Phase A we find a free rotor and rapid rotation between all 12 permutations.  In phase
B we find cases where one bond becomes locked, with rapid rotation around to the remaining three permutations.  Also in Phase B, the $g1$ molecules oscillate between two different orientations, as well as the multiple permutations.

In general, the rotations are facilitated by temperature and suppressed by pressure, giving rise to an onion-ring type phase diagram in good agreement with experiment.  Curiously, the most stable zero-Kelvin DFT structures are completely different from the finite temperature ones.  Calculations of rotational entropy precise and extensive enough to determine phase boundaries proved impossible with BOMD.  The quasiharmonic approximation is clearly inappropriate  for non-oscillating states, however low barriers to rotation are correlated with low curvature of the potential, so a phonon approximation can give qualitatively correct trends.

It is interesting that the observed classical ``exchange'' of hydrogen positions gives the correct structure, which indicates that a full quantum treatment of the protons is not required to understand the structures.  Curiously, classical states have low free energy due to their high entropy, whereas the lowest energy quantum free rotor (J=0) is stabilised by low kinetic energy.

We also reconcile the long-standing mismatch between (i) low-temperature, minimum-enthalpy structures from structure searching and (ii) the experimentally observed finite-temperature phases of crystalline methane.

In Phase~I, BOMD stabilises the experimentally observed $fcc$ carbon lattice via the high rotational entropy:
($S^\mathrm{rot} \approx 0.78$), 
with time-averaged cubic symmetry.
Unusually, rather than hopping between well-defined orientations, the molecule rotates freely but does not access certain forbidden directions.

The  symmetry of the methane molecule does not allow a simple orientational ordering on an $fcc$ lattice, subject to the general rule of  steric hindrance: that hydrogen atoms should avoid one another. This leads to orientational frustration, for which the period-doubling/symmetry-breaking $P2_1 2_1 2_1$ structure appears to be a good compromise which may be stable at low pressure and very low temperatures. 

The complex structures of phases A and B have multiple different methane sites: this occurs to accommodate the wide range of bondlengths arising from different orientations.

Phase A is revealed as having R$\bar{3}$ symmetry enabled by molecular orientation at the inversion centre. The icosahedral supermolecules pack in a simple cubic arrangement, very slightly distort due to the incompatibility of the icosahedral and cubic symmetry.   We demonstrate that this structure is fully consistent with previous experimental data, even though it differs from previous interpretations of that data.

Phase B has cubic symmetry, based on a bcc stacking of Z16 supermolecules, the remaining molecules are in tetrahedral interstitial sites of bcc, disordered between two possible orientations.  The HP phase forms when these `interstitial' molecules order.  

The different sites serve to allow space for some molecules to rotate at T, P conditions where $fcc$ rotation is severely hindered.   Phase stability can thereby be attributed to some molecules having high entropy from rotation, while others have low enthalpy from preferential orientation.  The sequence of transitions $\text{I} \rightarrow \text{A} \rightarrow \text{B} \rightarrow \text{HP} \rightarrow P2_1/c$ with decreasing T can then be understood as reducing the fraction of high entropy sites while increasing the fraction of short, low-enthalpy bonds.  

The observed sluggishness of the transformations is also easily understood, since A and B are both large unit cells with incompatible numbers of atoms per cell.  Under such circumstances, no simple transformation mechanism will exist and the transition can only proceed via nucleation and recrystallisation. This constraint does not seem to apply to the B-HP transition, which involves cessation of rotation in some of the sites in $\alpha-Mn$, but no major rearrangement of the molecular centres beyond a small martensitic distortion along one of the four (111) axes.  While there is no major energy barrier between the phases, it is likely that in an experiment HP will contain many symmetry-related orientational domains.  If these domains are smaller than the X-ray correlation length, they can lead to preferential suppression of the intensity of some diffraction peaks~\cite{crain1993phase} which could be misinterpreted as disorder in the occupation of hydrogen sites.

To summarise, we have demonstrated that the DFT and experimental descriptions of high-pressure methane are consistent, once the weaknesses of each approach are properly accounted for.  Specifically, naive structure searching plus quasiharmonic free energy calculation with DFT neglects the rotational entropy which is key to creating the onion-ring phase diagram.  Conversely, diffraction averages over many unit cells and misses local correlations (e.g. C--H bonds not pointing at each other), and X-ray diffraction is unable to convincingly locate the hydrogen positions.  The complex rearrangements required to transform between phases make determination of phase boundaries challenging: in experiment there is hysteresis and in calculation huge supercells are required to be compatible with two phases.  

The major conceptual conclusion is that high-pressure methane is organised by packings of supermolecules. Phase~A is built from a 13-molecule icosahedral cluster plus an 8-molecule cube-like motif (21=13+8), and Phase~B/HP from a 17-molecule Z16 Frank--Kasper cluster plus 12 additional molecules in tetrahedral interstices (29=17+12 in the primitive setting). In this language, the ``complex'' phases reduce to familiar cubic Strukturbericht archetypes (B2/CsCl-like and A2/bcc-like) of \emph{clusters}. The phase stability is controlled by a balance between site-resolved orientational entropy and accommodation of a range of bondlengths. 

The explanation of the near-cubic structures highlights the advantages of describing the structures in terms of their multimolecular building blocks, in addition to symmetry analysis.  It is likely that such an approach will be fruitful in other systems.

\section*{Supplementary Material}
See the Supplementary Material for additional figures and simulation details.

\begin{acknowledgments}
This work was supported by the ERC Advanced Grant HECATE, and by EPSRC via an eCSE award to MK.  We thank ARCHER2 for access to computational resources. For the purpose of open access, the author has applied a Creative Commons Attribution (CC BY) licence to any Author Accepted Manuscript version arising from this submission.
\end{acknowledgments}

\section*{Author Declarations}

\subsection*{Conflict of Interest}
The authors have no conflicts to disclose.

\subsection*{Data Availability}
The data that support the findings of this study will be deposited in the University of Edinburgh DataShare repository and
made publicly available upon acceptance. During peer review, the data are available from the corresponding authors upon request.

\bibliography{methaneabaimd}

\end{document}